\documentclass[12pt]{article}

\usepackage[allcolors=blue,colorlinks=true]{hyperref}

\usepackage{microtype}
\usepackage{graphicx}
\usepackage{subfigure}
\usepackage{booktabs} 
\usepackage{url}

\usepackage{mathtools}
\usepackage{microtype}
\usepackage{graphicx}
\usepackage{subfigure}
\usepackage{booktabs} 
\usepackage{amsmath,amssymb}
\usepackage{multirow}
\usepackage{multicol}
\usepackage{blindtext}
\usepackage{booktabs}
\usepackage{multirow}
\usepackage{siunitx}
\usepackage{caption}
\usepackage{doi}

\usepackage{paralist}
\usepackage{graphicx}
\usepackage{amsmath,amssymb}
\usepackage{color}
\usepackage{algorithm}
\usepackage{algorithmic}

\usepackage[dvipsnames]{xcolor}
\DeclareMathOperator*{\argmin}{argmin}  
  
\DeclareMathOperator*{\minimize}{minimize}  
\DeclareMathOperator*{\subto}{subject \hspace{2mm} to}  
\DeclareMathOperator*{\prox}{prox}  

\DeclareMathOperator{\bz}{\textbf z}\DeclareMathOperator{\bb}{\textbf b}

\usepackage{thmtools}

\def \curl {\boldsymbol\ell}
\def\V{\mathbf V}
\def\U{\mathbf U}
\def\A{\mathbf A}
\def\B{\mathbf B}
\def\C{\mathbf C}

\DeclareMathOperator{\bLambda}{\boldsymbol \Lambda}

\DeclareMathOperator{\bx}{\textbf x}
\DeclareMathOperator{\bu}{\textbf u}
\DeclareMathOperator{\bv}{\textbf v}
\DeclareMathOperator{\blambda}{\boldsymbol \lambda}

\newcommand{\br}{\textbf r} 
\newcommand{\bzeta}{\boldsymbol \zeta} 
\newcommand{\bw}{\textbf w}

\def\X{\mathbf X}
\def\U{\mathbf U}
\def\M{\mathbf M}

\def\V{\mathbf V}
\def\Z{\mathbf Z}

\def\W{\mathbf W}
\def\D{\mathbf D}

\def\R{\mathbf R}
\def\N{\mathbf N}
\def\bKSI{\mathbf \Psi}

\def\E{\mathbf E}

\usepackage[margin=1in]{geometry}

\usepackage[square,numbers]{natbib}
\usepackage[resetlabels]{multibib}
\usepackage{enumerate}
\usepackage{amsmath,latexsym,amssymb,graphicx,wrapfig,  enumitem, mathtools}
\allowdisplaybreaks[1]

\usepackage{footnote}
\usepackage{thm-restate}

\usepackage[utf8]{inputenc}
\usepackage[english]{babel}

\makesavenoteenv{tabular}
\makesavenoteenv{table}

\makeatletter
\renewcommand*{\@fnsymbol}[1]{\@alph{#1}}
\makeatother

\bibpunct{(}{)}{;}{a}{}{,}

\usepackage{setspace}

\begin{document}


\title{\bf Integrative Generalized Convex Clustering Optimization and Feature Selection for Mixed Multi-View Data}
\author{ Minjie Wang\thanks{Department of Statistics, Rice University, Houston, TX}\hspace{.2cm}
and Genevera I. Allen\thanks{Departments of Electrical and Computer Engineering, Statistics, and Computer Science, Rice University, Houston, TX} \textsuperscript{,}\thanks{Jan and Dan Duncan Neurological Research Institute, Baylor College of Medicine, Houston, TX}}
\date{}
\maketitle

\begin{abstract}
In mixed multi-view data, multiple sets of diverse features are measured on the same set of samples.  By integrating all available data sources, we seek to discover common group structure among the samples that may be hidden in individualistic cluster analyses of a single data-view.  While several techniques for such integrative clustering have been explored, we propose and develop a convex formalization that will inherit the strong statistical, mathematical and empirical properties of increasingly popular convex clustering methods.  Specifically, our Integrative Generalized Convex Clustering Optimization (iGecco) method employs different convex distances, losses, or divergences for each of the different data views with a joint convex fusion penalty that leads to common groups.  Additionally, integrating mixed multi-view data is often challenging when each data source is high-dimensional.  To perform feature selection in such scenarios, we develop an adaptive shifted group-lasso penalty that selects features by shrinking them towards their loss-specific centers.  Our so-called iGecco+ approach selects features from each data-view that are best for determining the groups, often leading to improved integrative clustering.  To fit our model, we develop a new type of generalized multi-block ADMM algorithm using sub-problem approximations that more efficiently fits our model for big data sets.  Through a series of numerical experiments and real data examples on text mining and genomics, we show that iGecco+ achieves superior empirical performance for high-dimensional mixed multi-view data.
\end{abstract}

\noindent%

\newpage


\doublespacing

\section{Introduction}

As the volume and complexity of data grows, statistical data integration has gained increasing attention as it can lead to discoveries which are not evident in analyses of a single data set.  We study a specific data-integration problem where we seek to leverage common samples measured across multiple diverse sets of features that are of different types (e.g., continuous, count-valued, categorical, skewed continuous and etc.).  This type of data is often called mixed, multi-view data \citep{hall1997introduction,acar2011all,lock2013joint,tang2018integrated,baker2019feature}.  
While many techniques have been developed to analyze each individual data type separately, there are currently few methods that can directly analyze mixed multi-view data jointly.  Yet, such data is common in many areas such as electronic health records, integrative genomics, multi-modal imaging, remote sensing, national security, online advertising, and environmental studies. For example in genomics, scientists often study gene regulation by exploring only gene expression data, but other data types, such as short RNA expression and DNA methylation, are all part of the same gene regulatory system.  
Joint analysis of such data can give scientists a more holistic view of the problem they study.  But, this presents a major challenge as each individual data type is high-dimensional (i.e., a larger number of features than samples) with many uninformative features.  Further, each data view contains different data types: expression of genes or short RNAs measured via sequencing is typically count-valued or zero-inflated plus skewed continuous data whereas DNA methylation data is typically proportion-valued.  In this paper, we seek to leverage multiple sources of mixed data to better cluster the common samples as well as select relevant features that distinguish the inherent group structure.   

We propose a convex formulation which integrates mixed types of data with different data-specific losses, clusters common samples with a joint fusion penalty and selects informative features that separate groups. Due to the convex formulation, our methods enjoy strong statistical, mathematical and empirical properties. We make several methodological contributions. First, we consider employing different types of losses for better handling non-Gaussian data with Generalized Convex Clustering Optimization (Gecco), which replaces Euclidean distances in convex clustering with more general convex losses. We show that for different losses, Gecco's fusion penalty forms different types of centroids which we call loss-specific centers. To integrate mixed multi-view data and perform clustering, we incorporate different convex distances, losses, or divergences for each of the different data views with a joint convex fusion penalty that leads to common groups; this gives rise to Integrative Generalized Convex Clustering (iGecco). Further, when dealing with high-dimensional data, practitioners seek interpretability by identifying important features which can separate the groups. To facilitate feature selection in Gecco and iGecco, we develop an adaptive shifted group-lasso penalty that selects features by shrinking them towards their loss-specific centers, leading to Gecco+ and iGecco+ which performs clustering and variable selection simultaneously. 
To solve our methods in a computationally efficient manner, we develop a new general multi-block ADMM algorithm using sub-problem approximations, and make an optimization contribution by proving that this new class of algorithms converge to the global solution.

\subsection{Related Literature}
Our goal is to develop a unified, convex formulation of integrative clustering with feature selection based on increasingly popular convex clustering methods.  
\cite{pelckmans2005convex, lindsten2011just, hocking2011clusterpath} proposed convex clustering which uses a fusion penalty to achieve agglomerative clustering like hierarchical clustering. This convex formulation guarantees a global optimal solution, enjoys strong statistical and mathematical theoretical properties, and often demonstrates superior empirical performance to competing approaches.  Specifically, in literature,  \citet{pelckmans2005convex,chi2017convex} showed it yields stable solutions to small perturbations on the data or tuning parameters; \cite{radchenko2017convex} studied statistical consistency; \cite{tan2015statistical} established its link to  hierarchical clustering as well as prediction consistency; and many others have studied other appealing theoretical properties \citep{zhu2014convex,sui2018convex,chi2019recovering}. Despite these advantages, convex clustering has not yet gained widespread popularity due to its intensive computation. Recently, some proposed fast and efficient algorithms to solve convex clustering and estimate its regularization paths \citep{chi2015splitting,weylandt2019dynamic}. Meanwhile, convex clustering has been extended to biclustering \citep{chi2017convex} and many other applications \citep{chi2018provable,choi2019regularized}.

One potential drawback to convex clustering however, is that thus far, it has only been studied employing Euclidean distances between data points and their corresponding cluster centers. As is well known, the Euclidean metric suffers from poor performance with data that is highly non-Gaussian such as binary, count-valued, skewed data, or with data that has outliers. While  \cite{wang2016robust} studied robust convex clustering and \cite{sui2018convex} investigated convex clustering with metric learning, there has not been a general investigation of convex clustering for non-Gaussian data and data integration on mixed data has not been studied. But, many others have proposed clustering methods for non-Gaussian data in other contexts. One approach is to perform standard clustering procedures on transformed data \citep{anders2010differential,bullard2010evaluation,marioni2008rna,robinson2010scaling}. But, choosing an appropriate transformation that retains the original cluster signal is a challenging problem.   Another popular approach is to use hierarchical clustering with specified distance metrics for non-Gaussian data \citep{choi2010survey,fowlkes1983method}.   Closely related to this, \cite{banerjee2005clustering} studied different clustering algorithms utilizing a large class of loss functions via Bregman divergences. Yet, the proposed methods are all extensions of existing clustering approaches and hence inherit both good and bad properties of those approaches. There has also been work on model-based clustering, which assumes that data are generated by a finite mixture model; for example \citet{banfield1993model,si2013model} propose such a model for the Poisson and negative binomial distributions.  Still these methods have a non-convex formulation and local solutions like all model-based clustering methods.
 We propose to adopt the method similar to \cite{banerjee2005clustering} and study convex clustering using different loss functions; hence our method inherits the desirable properties of convex clustering and handles non-Gaussian data as well. More importantly, there is currently no literature on data integration using convex clustering and we achieve this by integrating different types of general convex losses with a joint fusion penalty.

Integrative clustering, however, has been well-studied in the literature. The most popular approach is to use latent variables to capture the inherent structure of multiple types of data.   This achieves a joint dimension reduction and then clustering is performed on the joint latent variables \citep{shen2009integrative,shen2012integrative,shen2013sparse,mo2013pattern,mo2017fully,meng2015mocluster}. Similar in nature to the latent variables approach, matrix factorization methods assume that the data has an intrinsic low-dimensional representation, with the dimension often corresponding to the number of clusters \citep{lock2013joint,hellton2016integrative,zhang2012discovery,chalise2017integrative,zhang2011novel,yang2015non}. There are a few major drawbacks of latent variable or dimension reduction approaches, however.  First it is often hard to directly interpret latent factors or low-dimensional projections.  Second, many of these approaches are based on non-convex formulations yielding local solutions.  And third, choosing the rank of factors or projections is known to be very challenging in practice and will often impact resulting clustering solutions.  Another approach to integrative clustering is clustering of clusters (COC) which performs cluster analysis on every single data set and then integrates the primary clustering results into final group assignments using consensus clustering \citep{hoadley2014multiplatform,lock2013bayesian,kirk2012bayesian,savage2013identifying,wang2014similarity}. 
This, however, has several potential limitations as each individual data set might not have enough signal to discern joint clusters or the individual cluster assignments are too disparate to reach a meaningful consensus.  
Finally, others have proposed to use distance-based clustering for mixed types of data by first defining an appropriate distance metric for mixed data (for example, the Gower distance by \citealp{gower1971general}) and then applying an existing distance-based clustering algorithm such as hierarchical clustering \citep{ahmad2007k,ji2012fuzzy}. Consequently, this method inherits both good and bad properties of distance-based clustering approaches. Notice that all of these approaches are either two-step approaches or are algorithmic or non-convex problems that yield local solutions.  In practice, such approaches often lead to unreliable and unstable results.

Clustering is known to perform poorly for high-dimensional data as most techniques are highly sensitive to uninformative features. One common approach is to reduce the dimensionality of the data via PCA, NMF, or t-SNE before clustering \citep{ghosh2002mixture,bernardo2003bayesian, tamayo2007metagene}.  A major limitation of such approaches is that the resulting clusters are not directly interpretable in terms of feature importance.  To address this, several have proposed sparse clustering for high-dimensional data. This performs clustering and feature selection simultaneously by iteratively applying clustering techniques to subsets of features selected via regularization \citep{witten2010framework,sun2012regularized,chang2014sparse}.  The approach, however, is non-convex and is highly susceptible to poor local solutions.  Others have proposed penalized model-based clustering that selects features \citep{raftery2006variable,wang2008variable,pan2007penalized}.  Still, these methods inherit several disadvantages of model-based clustering approaches.   Moreover, sparse integrative clustering is relatively under-studied. \citet{shen2013sparse,mo2013pattern} extended iCluster using a penalized latent variable approach to jointly model multiple omics data types. They induce sparsity on the latent variable coefficients via regularization.  As feature selection is performed on the latent variables, however, this is less interpretable in terms of selecting features directly responsible for distinguishing clusters.  
Recently, and most closely related to our own work, \citet{wang2018sparse} proposed sparse convex clustering which adds a group-lasso penalty term on the cluster centers to shrink them towards zero, thus selecting relevant features.
This penalty, however, is only appropriate for Euclidean distances when the data is centered; otherwise, the penalty term shrinks towards the incorrect cluster centers.  For feature selection using different distances and losses, we propose an adaptive shifted group-lasso penalty that will select features by shrinking them towards their appropriate centroid.

\section{Integrative Generalized Convex Clustering with Feature Selection}

In this section, we introduce our new methods, beginning with the Gecco and iGecco and then show how to achieve feature selection via regularization.  We also discuss some practical considerations for applying our methods and develop an adaptive version of our approaches.

\subsection{Generalized Convex Clustering Optimization (Gecco)}  

In many applications, we seek to cluster data that is non-Gaussian. In the literature, most do this using different distance metrics other than Euclidean distances \citep{choi2010survey,fowlkes1983method,de2004clustering}. Some use losses based on exponential family or deviances closely related to Bregman divergences \citep{banerjee2005clustering}.

To account for different types of losses for non-Gaussian data, we propose to replace the Euclidean distances in convex clustering with more general convex losses; this gives rise to Generalized Convex Clustering Optimization (Gecco).
$$
\minimize_{\U} \hspace{2mm}\sum_{i=1}^n  \curl(\X_{i.},\U_{i.}) +  \gamma \sum_{i < i'} w_{ii'} \| \U_{i.} - \U_{i'.}  \|_q
$$ 
Here, our data $\X$ is an $n \times p$ matrix consisting of $n$ observations and $p$ features; $\U$ is an $n \times p$ centroid matrix with the $i^{th}$ row, $\U_{i.}$, the cluster centroid attached to point $\X_{i.}$. The general loss $\curl(\X_{i.},\U_{i.}) $ refers to a general loss metric that measures dissimilarity between the data point $\X_{i.}$ and assigned centroids $\U_{i.}$.
$\|\cdot\|_q$ is the $\ell_q$-norm of a vector and usually $q \in \{1,2,\infty\}$ is considered \citep{hocking2011clusterpath}. Here we prefer using the $\ell_2$-norm in the fusion penalty $(q=2)$ as it encourages the entire rows of similar observations to be fused together simultaneously and is also rotation-invariant; but one could use $\ell_1$ or $\ell_{\infty}$ norms as well. $\gamma$ is a positive tuning constant and $w_{ij}$ is a nonnegative weight.
When $\gamma$ equals zero, each data point occupies a unique cluster. As $\gamma$ increases, the fusion penalty encourages some of the rows of the cluster center $\U$ to be exactly fused, forming clusters. When $\gamma$ becomes sufficiently large, all centroids eventually coalesce to a single cluster centroid, which we define as the loss-specific center associated with $\curl(\cdot)$. Hence $\gamma$ regulates both the cluster assignment and number of clusters, providing a family of clustering solutions. The weight $w_{ij}$ should be specified by the user in advance and is not a tuning parameter; we discuss choices of weights for various convex losses in Section \ref{practical}.  

Going beyond Euclidean distances, we propose to employ convex distance metrics as well as deviances associated with exponential family distributions and Bregman divergences, which are always convex. Interestingly, we show that each of these possible loss functions shrink the cluster centers, $\U$, to different loss-specific centers, instead of the mean-based centroid as in convex clustering with Euclidean distances.  For example, one may want to use least absolute deviations ($\ell_{1}$-norm or Manhattan distances) for skewed data or for data with outliers; with this loss, we show that all observations fuse to the median when $\gamma$ is sufficiently large. We emphasize loss-specific centers here as they will be important in feature selection in the next section. For completeness, we list common distances and deviance-based losses, as well as their loss-specific centers $\tilde x_j$ respectively in Table~\ref{loss-table}. (See Appendix~\ref{centroidcal} for all calculations associated with loss-specific centers, and we provide a formal proof when studying the properties of our approaches in Section \ref{property}.)

\begin{table}[h]
	\vskip 0.15in
	\begin{center}
		\begin{small}
			\scalebox{0.78}{
				\begin{tabular}{ |l|l|l|l| }
					\hline
					\textbf{Data Type} &  \textbf{Loss Type} & \textbf{Loss Function} & \textbf{Loss-specific Center} $\tilde \bx$ \\ \hline
					\multirow{1}{*}{Continuous} & Euclidean ($\ell_2$) & $ \frac{1}{2}  \|\bx_i - \bu_i\|_2^2$ &    $\bar \bx$ \\ \hline
					\multirow{6}{*}{Skewed Continuous} & Manhattan ($\ell_1$) & $\sum_{j=1} |x_{ij} - u_{ij}|$ & median($\bx$) \\ 
					& Minkowski ($\ell_q$) & $\sqrt[q]{\sum_{j=1} |x_{ij} - u_{ij}|^q}$  & no closed form \\
					& Mahalanobis (weighted $\ell_2$) & $(\bx_i-\bu_i)^T \textbf C^{-1}(\bx_i-\bu_i)$ & no closed form \\
					&Chebychev ($\ell_{\infty}$) & $\max_j \{|x_{ij} - u_{ij}|\}$  &no closed form  \\
					&Canberra (weighted $\ell_1$) &  $\sum_{j=1} \frac{|x_{ij} - u_{ij}|}{|x_{ij}| + |u_{ij}|}$  &  no closed form\\
					\hline
					\multirow{4}{*}{Binary} & Bernoulli log-likelihood & $  - x_{ij} u_{ij} + \log(1+e^{u_{ij}})$   & $\text{logit}(\bar \bx)$ \\
					& Binomial Deviance  & $ - x_{ij} \log{u_{ij}} - (1-x_{ij})\log({1-u_{ij}})$ & $\bar \bx$  \\
					& Hinge Loss & $\max(0,1-u_{ij} x_{ij})$     &     mode$(\bx)$\\	& KL divergence & $-x_{ij} \log_2 u_{ij}$ & no closed form     \\ 
					&Hamming ($\ell_0$)  &$ \sum_j \#(x_{ij} \neq u_{ij})/n$   &   mode ($\bx$)  \\
					\hline
					\multirow{6}{*}{Count} & Poisson log-likelihood & $ -x_{ij} u_{ij} + \exp(u_{ij})$ &   $\log (\bar \bx)$  \\
					& Poisson Deviance & $ -x_{ij} \log u_{ij} + u_{ij}$ & $\bar \bx$ \\
					&  Negative Binomial log-likelihood &  $-x_{ij} u_{ij} + (x_{ij} + \frac{1}{\alpha} ) \log (\frac{1}{\alpha} + e^{u_{ij}} ) $  &$\log (\bar \bx)$  \\
					& Negative Binomial Deviance & $ x_{ij} \log(\frac{x_{ij}}{u_{ij}})  - (x_{ij} + \frac{1}{\alpha}) \log (\frac{1+\alpha x_{ij}}{1+\alpha u_{ij} }    ) $   & $\bar \bx$ \\ 
					& Manhattan ($\ell_1$) & $\sum_{j=1} |x_{ij} - u_{ij}|$ & median($\bx$) \\
					& Canberra (weighted $\ell_1$) &  $\sum_{j=1} \frac{|x_{ij} - u_{ij}|}{|x_{ij}| + |u_{ij}|}$  &  no closed form\\
					
					\hline
					\multirow{2}{*}{Categorical} & Multinomial log-likelihood & $  \big\{   \sum_{k=1}^{K}    -x_{ijk} u_{ijk} + \log(\sum_{k=1}^K e^{u_{ijk}}) \big \} $ &  $\text{mlogit}(\bar \bx)$  \\
					& Multinomial Deviance  &  $  \big\{   \sum_{k=1}^{K}    -x_{ijk} \log(u_{ijk})  \big \}$, $\sum \limits_{k=1}^K u_{ijk} = 1$ &  $\bar \bx$ \\
					\hline
			
				\end{tabular}
			}
		\end{small}
	    \caption{Different losses and their loss-specific centers. We provide all calculations associated with loss-specific centers in Appendix~\ref{centroidcal}. Note the Gecco problem with Hamming or Canberra distances is not convex. Though we discuss general convex losses in this paper, we list those non-convex losses for reference. For multinomial log-likelihood and multinomial deviance, we change Gecco formulation slightly to accommodate three indices; we provide a detailed formulation in Appendix~\ref{multin}.}	    	
	    \label{loss-table}
	\end{center}
	\vskip -0.1in
\end{table}

\subsection{Integrative Generalized Convex Clustering (iGecco)} \label{igecco}

In data integration problems, we observe data from multiple sources and would like to get a holistic understanding of the problem by analyzing all the data simultaneously. In our framework, we integrate mixed multi-view data and perform clustering by employing different convex losses for each of the different data views with a joint convex fusion penalty that leads to common groups. Hence we propose Integrative Generalized Convex Clustering (iGecco) which can be formulated as follows:
\begin{align*}
\minimize_{\U^{(k)}} \hspace{2mm} \sum_{k=1}^K \pi_k \curl_k(\X^{(k)},\U^{(k)})  + \gamma \sum_{i < j} w_{ii'}  \sqrt{ \sum_{k=1}^K  \| \U_{i.}^{(k)} - \U_{i'.}^{(k)} \|^2 }
\end{align*}
Here, we have $K$ data sources.  The $k^{th}$ data-view $\X^{(k)}$ is an $n \times p_k$ matrix consisting of $n$ observations and $p_k$ features; $\U^{(k)}$ is also an $n \times p_k$ matrix and the $i^{th}$ row, $\U_{i.}^{(k)}$, is the cluster center associated with the point $\X_{i.}^{(k)}$. And, $\curl_k(\X_{i.}^{(k)},\U_{i.}^{(k)}) $ is the loss function associated with the $k^{th}$ data-view. Each loss function is weighted by $\pi_k$, which is fixed by the user in advance.  We have found that setting $\pi_k$ to be inversely proportional to the null deviance evaluated at the loss-specific center, i.e.,  $\pi_k = \frac{1}{\curl_k(\X^{(k)},\tilde \X^{(k)})}$, performs well in practice. Note that $\tilde \X = \begin{pmatrix} \tilde \X^{(1)} \cdots \tilde \X^{(K)} \end{pmatrix}$ where each $j^{th}$ column of $\tilde \X^{(k)}$ denotes the loss-specific center $\tilde x_j^{(k)}$. We employ this loss function weighting scheme to ensure equal scaling across data sets of different types. Finally, notice that we employ a joint group-lasso penalty on all of the $\U^{(k)}$'s; this incorporates information from each of the data sources and enforces the same group structure amongst the shared observations.  We study this further and prove these properties in Section \ref{property}.

\subsection{Feature Selection: Gecco+ and iGecco+}

In high dimensions, it is important to perform feature selection both for clustering purity and interpretability. 
Recently, \citet{wang2018sparse} proposed sparse convex clustering by imposing a group-lasso-type penalty on the cluster centers which achieves feature selection by shrinking noise features towards zero. This penalty, however, is only appropriate for Euclidean distances when the data is centered; otherwise, the penalty term shrinks towards the incorrect cluster centers.   For example, the median is the cluster center with the $\ell_1$ or Manhattan distances.  Thus, to select features in this scenario, we need to shrink them towards the median, and we should enforce ``sparsity" with respect to the median and not the origin. To address this, we propose adding a shifted group-lasso-type penalty which forces cluster center $\U_{\cdot j}$ to shrink toward the appropriate loss-specific center $\tilde x_j$  for each feature.  Both the cluster fusion penalty and this new shifted-group-lasso-type feature selection penalty will shrink towards the same loss-specific center.

To facilitate feature selection with the adaptive shifted group-lasso penalty for one data type, our Generalized Convex Clustering Optimization with Feature Selection (Gecco+) is formulated as follows:
\begin{align*}
\minimize_{\U} \hspace{2mm}   &\sum_{i=1}^n  \curl(\X_{i.},\U_{i.})  +  \gamma \sum_{  i < i'}^n w_{ii'} \| \U_{i.} - \U_{i'.}  \|_2 \\
&+ \alpha \sum_{j=1}^p   \zeta_j  \| \U_{.j}  - \tilde x_j \cdot \textbf 1_n  \|_2  
\end{align*}
Again, $\U$ is an $n \times p$ matrix and $\tilde x_j$ is the loss-specific center for the $j^{th}$ feature introduced in Table~\ref{loss-table}.  The tuning parameter $\alpha$ controls the number of informative features and the feature weight $\zeta_j$ is a user input which plays an important role to adaptively penalize the features.  (We discuss choices of $\zeta_j$ in Section \ref{adaptive} when we introduce the adaptive version of our method.)  When $\alpha$ is small, all features are selected and contribute to defining the cluster centers.  When $\alpha$ grows sufficiently large, all features coalesce at the same value, the loss-specific center $\tilde{x}_j$, and hence no features are selected and contribute towards determining the clusters.  Another way of interpreting this is that the fusion penalty exactly fuses some of the rows of the cluster center $\U$, hence determining groups of rows.  On the other hand, the shifted group-lasso penalty shrinks whole columns of $\U$ towards their loss-specific centers, thereby essentially removing the effect of uninformative features.  Selected features are then columns of $\U$ that were not shrunken to their loss-specific centers, $\U_{.j} \neq \tilde x_j \cdot \textbf 1_n$.  These selected features, then, exhibit differences across the clusters determined by the fusion penalty.  Clearly, sparse convex clustering of \citet{wang2018sparse} is a special case of Gecco+ using Euclidean distances with centered data.  Our approach using both a row and column penalty is also reminiscent of convex biclustering \citep{chi2017convex} which uses fusion penalties on both the rows and columns to achieve checker-board-like biclusters.

Building upon integrative generalized convex clustering in Section \ref{igecco} and our proposed feature selection penalty above, our Integrative Generalized Convex Clustering Optimization with Feature Selection (iGecco+) is formulated as follows:
\begin{align}  
\minimize_{\U^{(k)}} \hspace{2mm} &\sum_{k=1}^K \pi_k \curl_k(\X^{(k)},\U^{(k)})  + \gamma \sum_{i < i'} w_{ii'}  \sqrt{ \sum_{k=1}^K  \| \U_{i.}^{(k)} - \U_{i'.} ^{(k)} \|^2 }  \nonumber \\
&  + \alpha \sum_{k=1}^K \sum_{j=1}^{p_k}  \zeta_j^{(k)} \| \U_{.j}^{(k)}  - \tilde x_j^{(k)} \cdot \textbf 1_n  \|_2 \label{eq:1}
\end{align} 
Again, $\U^{(k)}$ is an $n \times p_k$ matrix and $\tilde x_j^{(k)}$ is the loss-specific center for the $j^{th}$ feature for $k^{th}$ data type.  By construction, iGecco+ directly clusters mixed multi-view data and selects features from each data view simultaneously. Similarly, adaptive choice of $\zeta_j^{(k)}$ gives rise to adaptive iGecco+ which will be discussed in Section \ref{adaptive}. Detailed discussions on practical choices of tuning parameters and weights can be also found in Section \ref{practical}.

\subsection{Properties}\label{property}
In this section, we develop some properties of our methods, highlighting several advantages of our convex formulation.  The corresponding proofs can be found in Section~\ref{propprove} of the Appendix.

Define the objective function in \eqref{eq:1}  as $F_{\gamma,\alpha} (\U)$  where $\U = \begin{pmatrix} \U^{(1)} \cdots \U^{(K)} \end{pmatrix}$. Then due to convexity, we have the following:

\begin{restatable}{proposition}{prop1}
\label{theorem:diff1}(Global solution)
 If $\curl_k$ is convex for all $k$, then any minimizer of $F_{\gamma,\alpha}(\U)$, $\U^*$, is a global minimizer. If $\curl_k$ is strictly convex for all $k$, then $\U^*$ is unique.
\end{restatable}

\begin{restatable}{proposition}{prop2}
\label{theorem:diff2}(Continuity with respect to data and input parameters)
 The global minimizer $\U^{*}_{\bw, \mathbf \pi, \bzeta,\X}(\gamma, \alpha)$  of iGecco+ exists and depends continuously on the data, $\X$, tuning parameters $\gamma$ and $\alpha$, the weight matrix $\bw$, the loss weight $\pi_k$, and the feature weight $\zeta_j^{(k)}$. 
\end{restatable}

\begin{restatable}{proposition}{prop3}
\label{theorem:diff3}(Loss-specific center)
Define $\tilde \X = \begin{pmatrix} \tilde \X^{(1)} \cdots \tilde \X^{(K)} \end{pmatrix}$ where each $j^{th}$ column of $\tilde \X^{(k)}$ equals the loss-specific center $\tilde x_j^{(k)}$.  Suppose each observation corresponds to a node in a graph with an edge between nodes $i$ and $j$ whenever $w_{ij}>0$. If this graph is fully connected, then $F_{\gamma,\alpha}(\U)$ is minimized by the loss-specific center $\tilde \X$ when $\gamma$ is sufficiently large or $\alpha$ is sufficiently large.
\end{restatable}

\textbf{Remark:} As Gecco, Gecco+ and iGecco are special cases of iGecco+, it is easy to show that all of our properties hold for these methods as well.

These properties illustrate some important advantages of our convex clustering approaches.  Specifically, many other widely used clustering methods are known to suffer from poor local solutions, but any minimizer of our problem will achieve a global solution.  Additionally, we show that iGecco+ is continuous with respect to the data, tuning parameters, and other input parameters.  Together, these two properties are very important in practice and illustrate that the global solution of our method remains stable to small perturbations in the data and input parameters.  Stability is a desirable property in practice as one would question the validity of a clustering result that can change dramatically with small changes to the data or parameters.  
Importantly, most popular clustering methods such as k-means, hierarchical clustering, model-based clustering, or low-rank based clustering, do not enjoy these same stability properties.  

\begin{figure}[ht]
\centering
\includegraphics[scale = 0.75]{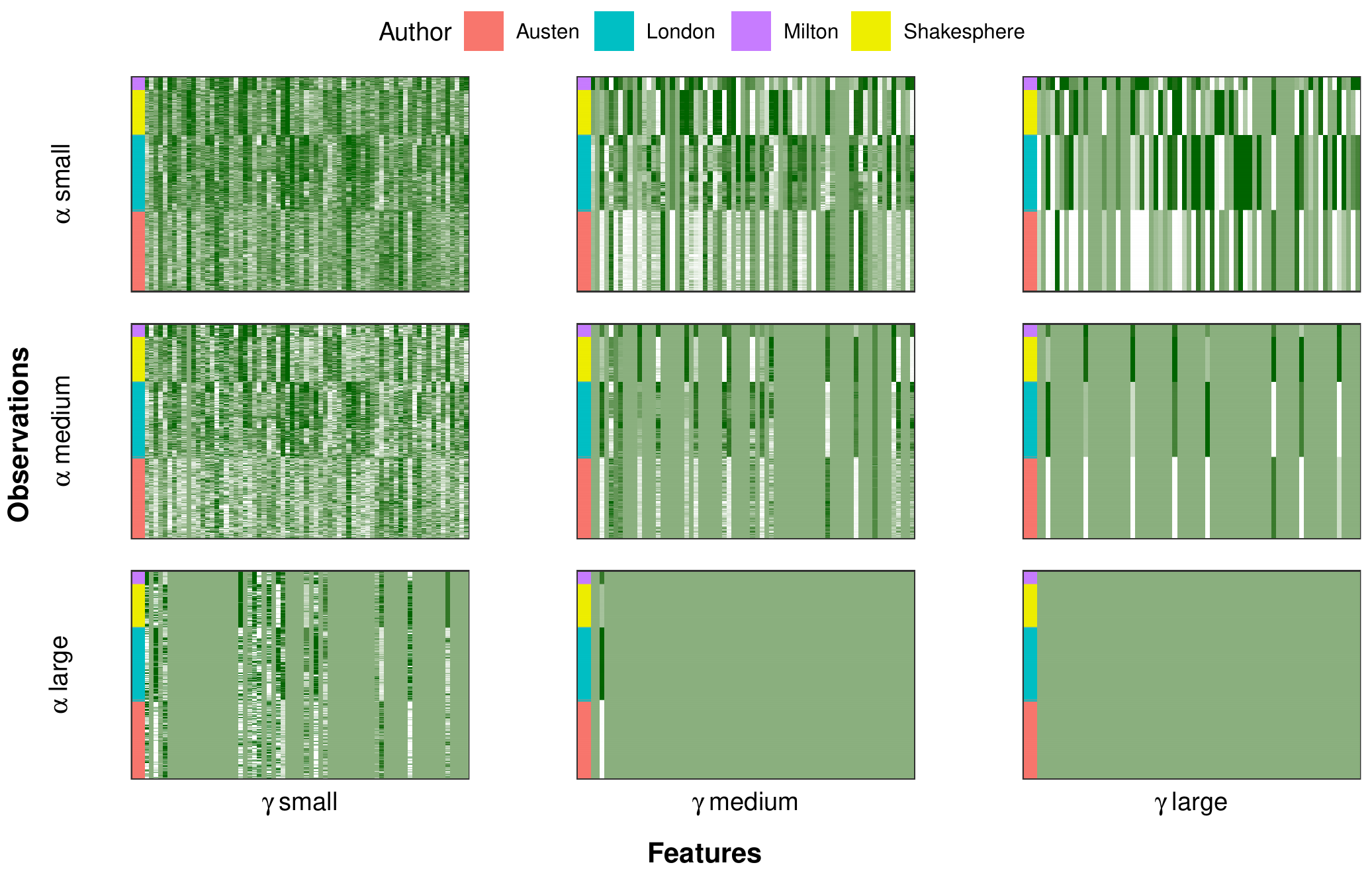}
\caption{Regularization path of Gecco+ solutions $\hat \U({\gamma,\alpha})$ for authors data. From left to right, we increase the parameter for fusion penalty $\gamma$. From top to bottom, we increase the parameter for feature penalty $\alpha$. The interpretation of regularization path is discussed in more detail in Section \ref{property}.}
\label{viz-path-all}
\end{figure}

Finally in Proposition \ref{theorem:diff3}, we verify that when the tuning parameters are sufficiently large, full fusion of all observations to the loss-specific centers is achieved.  Hence, our methods indeed behave as intended, achieving joint clustering of observations.  We illustrate this property in Figure \ref{viz-path-all} where we apply Gecco+ to the authors data set (described fully in Section  \ref{Simulation}).  Here, we illustrate how our solution, $\hat \U({\gamma,\alpha})$, changes as a function of $\gamma$ and $\alpha$.  This so-called ``cluster solution path" begins with each observation as its own cluster center when $\gamma$ is small and stops when all observations are fused to the loss-specific center when $\gamma$ is sufficiently large. In between, we see that observations are fusing together as $\gamma$ increases. Similarly, when $\alpha$ is small, all features are selected and as $\alpha$ increases, some of the features get fused to their loss-specific center.

\subsection{Practical Considerations and Adaptive iGecco+} \label{practical}

In this section, we discuss some practical considerations for applying our method to real data. We discuss choosing user-specific inputs such as weights as well as how to select tuning parameters.  In doing so, we introduce an adaptive version of our method as well.

\subsubsection{Choice of Weights and Tuning Parameters}

In practice, a good choice of fusion weights ($w_{ij}$) has been shown to enhance both computational efficiency and clustering quality of convex clustering \citep{chi2015splitting}.  
It has been empirically demonstrated that using weights inversely proportional to the distances yields superior clustering performance; this approach is widely adopted in practice.  Further, setting many of the weights to zero helps reduce computation cost. Considering these two, the most common weights choice for convex clustering is to use $K$-nearest-neighbors method with a Gaussian kernel. Specifically, the weight between the sample pair ($i,j$) is set as $w_{ij} = I_{ij}^k \exp(- \phi d(\X_{i.} , \X_{j.}))$, where $I_{ij}^k$ equals 1 if observation $j$ is among observation $i$'s $K$ nearest neighbors or vice versa, and 0 otherwise. However, this choice of weights based on Euclidean distances may not work well for non-Gaussian data in Gecco(+) or for mixed data in iGecco(+). To account for different data types and better measure the similarity between observations, we still adopt $K$-nearest-neighbors method with an exponential kernel, but further extend this by employing appropriate distance metrics for specific data types in the exponential kernel.
In particular, for weights in Gecco and Gecco+, we suggest using the same distance functions or deviances in the loss function of Gecco and Gecco+.  For weights in iGecco and iGecco+, the challenge is that we need to employ a distance metric which measures mixed types of data. In this case, the Gower distance, which is a distance metric used to measure the dissimilarity of two observations measured in different data types \citep{gower1971general}, can address our problem. To be specific, the Gower distance between observation $i$ and $i'$ overall can be defined as $d(\X_{i.},\X_{i'.}) = \sum_{k=1}^K \sum_{j=1}^{p_k} d_{ii'j}^{(k)}  \big/ \sum_{k=1}^K p_k$ where $d_{ii'j}^{(k)} =  \frac{ | \X_{ij}^{(k)} - \X_{i'j}^{(k)} |}{R_j^{(k)}}$ refers to the Gower distance between observation $i$ and $i'$ for feature $j$ in data view $k$ and $ R_j^{(k)} = \max_{i , i'} | \X_{ij}^{(k)} - \X_{i'j}^{(k)} |$ is the range of feature $j$ in data view $k$. In the literature, Gower distance has been commonly used as distance metrics for clustering mixed types of data \citep{wangchamhan2017efficient,hummel2017clustering,akay2018clustering} and shown to yield superior performance than other distance metrics \citep{ali2013k,dos2015categorical}.

Alternatively, we also propose and explore using stochastic neighbor embedding weights based on symmetrized conditional probabilities \citep{maaten2008visualizing}.  These have been shown to yield superior performance in high-dimensions and if there are potential outliers.  Specifically, the symmetrized conditional probabilities are defined as $p_{ij} = \frac{p_{j|i} + p_{i|j}}{2n}$, where $p_{j|i} = \frac{\exp(-\phi d(\X_{i.} , \X_{j.}))}{\sum_{k \neq i} \exp(-\phi d(\X_{i.} , \X_{k.}))}$.
We propose to use the weights $w_{ij} = I_{ij}^k \cdot p_{ij}$ where $I_{ij}^k$ still equals 1 if observation $j$ is among observation $i$'s $K$ nearest neighbors or vice versa, and 0 otherwise.
Again, we suggest using distance metrics appropriate for specific data types or the Gower distance for mixed data.  In empirical studies, we experimented with both weight choices and found that stochastic neighbor embedding weights tended to work better in high-dimensional settings and if there are outliers.  Hence, we recommend these and employed them in our empirical investigations in Section \ref{Simulation} and \ref{realdata}.   

Estimating the number of clusters in a data set is always challenging. Current literature for tuning parameter selection mainly focuses on stability selection or consensus clustering \citep{wang2010consistent,fang2012selection} and hold-out validation \citep{chi2017convex}. In this paper, we adopt hold-out validation approach for tuning parameter selection and we follow closely the approach described in \cite{chi2017convex}; we have found that this performs well in empirical studies.  

For the choice of the feature selection tuning parameter, $\alpha$, we find that clustering result is fairly robust to choices of $\alpha$. Hence, we suggest using only a few possibilities for $\alpha$ and to choose the combination of $\alpha$ and $\gamma$ which jointly minimizes hold-out error or within-cluster deviance. In many cases, we know the number of clusters a priori (or have an idea of an appropriate range for the number of clusters) and we can directly choose $\alpha$ which minimizes the hold-out error or within cluster deviance for that number of clusters.

\subsubsection{Adaptive Gecco+ and iGecco+ to Weight Features}\label{adaptive}

Finally, we consider how to specify the feature weights, $\zeta_j$ used in the shifted group-lasso penalty.  While employing these weights are not strictly necessary, we have found, as did \citet{wang2018sparse}, that like the fusion weights, well-specified $\zeta_j$'s can both improve performance and speed computation.  But unlike the fusion weights where we can utilize the pairwise distances, we don't have prior information on which features may be relevant in clustering.  Thus, we propose to use an adaptive scheme that first fits the iGecco+ with no feature weights and uses this initial estimate to define feature importance for use in weights.  This is similar to many adaptive approaches in the literature \citep{zou2006adaptive,wang2018sparse}.  

Our adaptive iGecco+ approach is given in Algorithm~\ref{alg:adaptive-feature}; this  applies to adaptive Gecco+ as a special case as well. We assume that the number of clusters (or a range of the number of clusters) is known a priori.
We begin by fitting iGecco+ with $\alpha = 1$ and uniform feature weights $\zeta_j^{(k)}=1$. We then find the $\gamma$ which gives the desired number of clusters, yielding the initial estimate, $\hat \U^{(k)}$.  (Alternatively, we can use hold-out validation to select $\gamma$.)  Next, we use this initial estimate to adaptively weight features by proposing the following weights: $\zeta_j^{(k)} = 1/\| \hat \U_{.j}^{(k)} - \tilde x_j^{(k)} \cdot \textbf 1_n  \|_2$.   These weights place a large penalty on noise features as $\| \hat \U_{.j}^{(k)} - \tilde x_j^{(k)} \cdot \textbf 1_n  \|_2$ is close to zero in this case. 
We also notice that noise features impact the distances used in the fusion weights as well.  Hence, we suggest updating the distances adaptively by using the selected features to better measure the similarities between observations.  To this end, we propose a new scheme to compute weighted Gower distances.   First, we scale the features within each data view so that informative features in different data views contribute equally and on the same scale.  Then, we employ the inverse of $\pi_k$, i.e., the null deviance, to weight the distances from different data types, resulting in an aggregated and weighted Gower distance, $\hat d(\X_{i.},\X_{i'.})$ as further detailed in Algorithm \ref{alg:adaptive-feature}.  
Note that if the clustering signal from one particular data type is weak and there are few informative features for this data type, then our weighting scheme will down-weight this entire data type in the weighted Gower distance.  In practice, our adaptive iGecco+ scheme works well as evidenced in our empirical investigations in the next sections.

\begin{algorithm}[H]
	\caption{Adaptive iGecco+}
	\label{alg:adaptive-feature}
	\begin{algorithmic}
		\STATE {1. Fit iGecco+ with $\alpha = 1$ and a sequence of $\gamma$}
		\STATE {2. Find $\gamma$ which gives desired number of clusters ; Get the estimate $\hat \U^{(k)}$}\\
		\STATE {3. Update the feature weights $\hat \zeta_j^{(k)} = \frac{1}{1+\|\hat \U_{.j}^{(k)} - \tilde x_j^{(k)} \cdot \textbf 1_n  \|_2  }$ and fusion weights $\hat w_{ij} = I_{ij}^k \cdot \exp(-\phi \hat d(\X_{i.},\X_{i'.}))$ where $\hat d(\X_{i.},\X_{i'.}) = \sum_{k=1}^K \sum_{j=1}^{p_k}
\frac{\|\U_{.j}^{(k)}  - \tilde x_j^{(k)} \cdot \textbf 1_n  \|_2}{\max_j   \|\U_{.j}^{(k)}  - \tilde x_j^{(k)} \cdot \textbf 1_n  \|_2} \cdot \frac{1}{\pi_k}      \cdot d_{ii'j}^{(k)}$.}
		\STATE {4. Fit adaptive iGecco+ with $\hat \bzeta$ and $\tilde \bw$; }
	\end{algorithmic}
\end{algorithm}


\section{iGecco+ Algorithm}

In this section, we introduce our algorithm to solve iGecco+, which can be easily extended to Gecco, Gecco+ and iGecco. We first propose a simple, but rather slow ADMM algorithm as a baseline approach. To save computation cost, we further develop a new multi-block ADMM-type procedure using inexact one-step approximation of the sub-problems. Our algorithm is novel from optimization perspective as we extend the multi-block ADMM to higher number of blocks and combine it with the inexact sub-problem solve ADMM literature, which often results in major computational savings.

\subsection{Full ADMM to Solve iGecco+ (Naive Algorithm)}
Given the shifted group-lasso and fusion penalties along with general losses, developing an optimization routine for iGecco+ method is less straight-forward than convex clustering or sparse convex clustering.  In this section, we propose a simple ADMM algorithm to solve iGecco+ as a baseline algorithm and point out its drawbacks.

The most common approach to solve problems with more than two non-smooth functions is via multi-block ADMM \citep{lin2015global,deng2017parallel}, which decomposes the original problem into several smaller sub-problems and solves them in parallel at each iteration. \cite{chen2016direct} established a sufficient condition for the convergence of three-block ADMM. We develop a multi-block ADMM approach to fit our problem for certain types of losses and prove its convergence.

We first recast iGecco+ problem \eqref{eq:1} as the equivalent constrained optimization problem:
\begin{align*}
&\minimize_{\U^{(k)},\V} \hspace{5mm}   \sum_{k=1}^K \pi_k \curl_k(\X^{(k)},\U^{(k)}) + \gamma \underbrace{ \bigg(\sum_{l \in \mathcal E} w_l \|\V_{l.}\|_2\bigg)}_{P_1(\V;\bw)}+ \alpha \sum_{k=1}^K \sum_{j=1}^{p_k}  \zeta_j^{(k)} \| \U_{.j}^{(k)}  - \tilde x_j^{(k)} \cdot \textbf 1_n  \|_2 \\
& \subto \hspace{5mm}  \D \begin{bmatrix} \U^{(1)} & \cdots & \U^{(K)} \end{bmatrix} - \V = \mathbf 0 
\end{align*}
Recently, \cite{weylandt2019dynamic} derived the ADMM for convex clustering in matrix form and we adopt similar approach. We index a centroid pair by $l = (l_1, l_2)$ with $l_1 < l_2$, define the set of edges over the non-zero weights $\mathcal E = \{l = (l_1,l_2) : w_l > 0\}$, and introduce a new variable $\V = \begin{bmatrix} \V^{(1)} & \cdots & \V^{(K)} \end{bmatrix} \in \mathbb R^{|\mathcal E | \times \sum {p_k}}$ where $\V^{(k)}_{l.} = \U^{(k)}_{l_1.} - \U^{(k)}_{l_2.} $ to account for the difference between the two centroids. Hence $\V^{(k)}$ is a matrix containing the pairwise differences between connected rows of $\U^{(k)}$ and the constraint is equivalent to $\D \U^{(k)} - \V^{(k)} = \mathbf 0$ for all $k$; $\D \in \mathbb R^{ |\mathcal E | \times n} $ is  the  directed  difference  matrix  corresponding  to  the  non-zero  fusion weights. It is clear the $\V$ sub-problem has closed-form solution for each iteration. We give general-form multi-block ADMM (Algorithm~\ref{alg:full-admm}) to solve iGecco+. Here $\text{prox}_{h(\cdot)} (\bx) = \argmin_{\bz} \frac{1}{2} \|\bx - \bz\|_2^2 + h(\bz)$ is the proximal mapping of function $h$.
\begin{algorithm}[h]
	\caption{General Multi-block Algorithm for iGecco+}
	\label{alg:full-admm}
	\begin{algorithmic}
		
		\WHILE{not converged}

		\FOR{all $k = 1,\cdots,K$}
		\STATE 	$\U^{ (k) } =
		\argmin \limits_{\U} \hspace{1mm}     \pi_k \curl_k(\X^{(k)},\U)  + \frac{\rho}{2}   \| \D \U -  \V^{(k)} + \bLambda^{(k)} \|_F^2  +   \alpha  \sum_{j=1}^{p_k}  \zeta_j^{(k)} \| \U_{.j}   - \tilde x_j^{(k)} \cdot \textbf 1_n  \|_2 $

		\ENDFOR
		
		\STATE 		$\V = \text{prox}_{\gamma /\rho P_1(\cdot; \bw )} (\begin{bmatrix} \D \U^{(1)} + \bLambda^{(1)} &  \cdots & \D \U^{(K)} + \bLambda^{(K)}  \end{bmatrix} )$ 
		\STATE $\bLambda^{(k)} =  \bLambda^{(k)} +  ( \D \U^{(k)} - \V^{(k)}  ) $ \hspace{5mm} for all $k$

		\ENDWHILE
		
	\end{algorithmic}
\end{algorithm}

Notice that there is no closed-form solution for the $\U^{(k)}$ sub-problem for general losses. Typically we need to apply an inner optimization routine to solve the $\U^{(k)}$ sub-problem until full convergence. In the next section, we seek to speed up this algorithm by using $\U^{(k)}$ sub-problem approximations.  But, first we propose two different approaches to fully solve the $\U^{(k)}$ sub-problem based on specific loss types and then use these to develop a one-step update to solve the sub-problem approximately with guaranteed convergence. 

\subsection{iGecco+ Algorithm}


We have introduced Algorithm \ref{alg:full-admm}, a simple baseline ADMM approach to solve iGecco+.  In this section, we consider different ways to solve the $\U^{(k)}$ sub-problem in Algorithm \ref{alg:full-admm}. First, based on specific loss types (differentiable and non-differentiable), we propose two different algorithms to solve the $\U^{(k)}$ sub-problem to full convergence. These approaches, however, are rather slow for general losses as there is no closed-form solution which results in nested iterative updates. To address this and in connection with current literature on variants of ADMM with sub-problem approximations, we propose iGecco+ algorithm, a multi-block ADMM which solves the sub-problems approximately by taking a single one-step update. We prove convergence of this general class of algorithms, a novel result in the optimization literature.

\subsubsection{Differentiable Case}
When the loss $\curl_k$ is differentiable, we consider solving the $\U^{(k)}$ sub-problem with proximal gradient descent, which is often used when the objective function can be decomposed into a differentiable and a non-differentiable function.  While there are many other possible optimization routines to solve the $\U^{(k)}$ sub-problem, we choose proximal gradient descent as there is existing literature proving convergence of ADMM algorithms with approximately solved sub-problems using proximal gradient descent \citep{liu2013linearized,lu2016fast}. We will discuss in detail how to approximately solve the sub-problem by taking a one-step approximation in Section \ref{inexact}. Based upon this, we propose Algorithm~\ref{alg:full-diff}, which solves the $\U^{(k)}$ sub-problem by running full iterative proximal gradient descent to convergence. Here $P_2(\tilde  \U^{(k)};\bzeta^{(k)}) = \sum_{j=1}^{p_k}  \zeta_j^{(k)} \|\tilde  \U^{(k)}_{.j}\|_2$.


\begin{algorithm}[h]
	\caption{$\U^{(k)}$ sub-problem for differentiable loss $\curl_{k}$ (Proximal gradient):}
	\label{alg:full-diff}
	\begin{algorithmic}
	\WHILE{not converged}
		\STATE $\U^{(k)} = \prox_{s_{k}  \cdot \alpha P_2(\cdot;\bzeta^{(k)})} \begin{footnotesize} \big( \U^{(k)} - \tilde \X^{(k)}   - s_{k} \cdot [ 
		\pi_k \nabla  \curl_{k}(\X^{(k)},\U^{(k)} ) +  \rho \D^T (\D  \U^{(k)}   -  \V^{(k)} + \bLambda^{(k)} )] \big) + \tilde \X^{(k)} \end{footnotesize} $
		\ENDWHILE
	\end{algorithmic}
\end{algorithm}
In Algorithm~\ref{alg:full-diff} and typically in general (proximal gradient) descent algorithms, we need to choose an appropriate step size $s_k$  to ensure convergence.  Usually we employ a fixed step size  by computing the Lipschitz constant as in the squared error loss case; but in our method, it is hard to compute the Lipschitz constant 
for most of our general losses. Instead, we suggest using backtracking line search procedure proposed by \cite{beck2009gradient,parikh2014proximal}, which is a common way to determine step size with guaranteed convergence in optimization. 
Further, we find decomposing the $\U^{(k)}$ sub-problem to $p_k$ separate  $\U_{.j}^{(k)}$ sub-problems brings several advantages such as (i) better convergence property (than updating $\U^{(k)}$'s all together) due to adaptive step size for each $\U_{.j}^{(k)}$ sub-problem and (ii) less computation cost by solving each in parallel. Hence, in this case, we propose to use proximal gradient for each separate $\U_{.j}^{(k)}$ sub-problem. To achieve this, we assume that the loss is elementwise, which is satisfied by every deviance-based loss. Last, as mentioned, there are many other possible ways to solve the $\U^{(k)}$ sub-problem than proximal gradient, such as ADMM. We find that when the loss is squared Euclidean distances or the hessian of the loss can be upper bounded by a fixed matrix, this method saves more computation.  
We provide all implementation details discussed above in Section~\ref{diffdetail} of the Appendix.

\subsubsection{Non-differentiable Case}
When the loss  $\curl_k$ is non-differentiable, we can no longer adopt the proximal gradient method to solve the $\U^{(k)}$ sub-problem as the objective is now a sum of more than one separable non-smooth function. To address this, as mentioned, we can use multi-block ADMM; in this case, we introduce new blocks for the non-smooth functions and hence develop a full three-block ADMM approach to fit our problem.  

To augment the non-differentiable term, we assume that our loss function can be written as $\curl_k(\X^{(k)},\U^{(k)}) =   f_k(g_k(\X^{(k)},\U^{(k)}))$ where $f_k$ is convex but non-differentiable and $g_k$ is affine. This condition is satisfied by all distance-based losses with $g_k(\X^{(k)},\U^{(k)})  =  \X^{(k)} - \U^{(k)}$; for example, for Manhattan distances, we have $f_k(\Z) = \sum_{j=1}^p \|\bz_j\|_1 = \|\text{vec}(\Z)\|_1$, and $g_k(\X,\U) = \X -\U$. 
The benefit of doing this is that now the $\U^{(k)}$ sub-problem has closed-form solution. Particularly, we can rewrite the $\U^{(k)}$ sub-problem as:
\begin{align*}
&\minimize_{\U^{(k)},\V} \hspace{5mm}   \sum_{k=1}^K \pi_k f_k(\Z^{(k)} ) + \frac{\rho}{2}   \| \D \U -  \V^{(k)} + \bLambda^{(k)} \|_F^2   + \alpha \sum_{k=1}^K \bigg( \underbrace{ \sum_{j=1}^{p_k}  \zeta_j^{(k)} \| \br_j^{(k)} \|_2}_{P_2(\R^{(k)};\bzeta^{(k)})} \bigg) \\
& \subto \hspace{5mm}  \X^{(k)} - \U^{(k)} = \Z^{(k)}  , \hspace{5mm} \U^{(k)}  - \tilde \X^{(k)}  = \R^{(k)}
\end{align*}

where $ \tilde \X^{(k)}$ is an $n \times p_k$ matrix with $j^{th}$ columns equal to scalar $\tilde x_j^{(k)}$.

It is clear that we can use multi-block ADMM to solve the  problem above and each primal variable has simple update with closed-form solution. We propose Algorithm~\ref{alg:full-non-diff}, a full, iterative multi-block ADMM, to solve the $\U^{(k)}$ sub-problem when the loss is a non-differentiable distance-based function.  Algorithm~\ref{alg:full-non-diff} applies to iGecco+ with various distances such as Manhattan, Minkowski and Chebychev distances and details are given in Section~\ref{nondiffdetail} of the Appendix. 

\begin{algorithm}[h]
	\caption{$\U^{(k)}$ sub-problem for non-differentiable distance-based loss $\curl_{k}$ (Multi-block ADMM):}
	\label{alg:full-non-diff}
	\begin{algorithmic}
		\STATE {\bfseries Precompute:}  Difference matrix $\D$, $\M = (\D^T \D + 2 \mathbf I)^{-1}$.
		
	\WHILE{not converged}
		\STATE 	$\U^{(k)} =
		\M(\D^T (\V^{(k)} - \bLambda^{(k)}  ) + \tilde \X^{(k)}   +  \R^{(k)} - \N^{(k)}  + \X^{(k)} - \Z^{(k)} + \bKSI^{(k)} )$ 
		\STATE $\Z^{(k)} = \text{prox}_{\pi_k f_k/\rho} (\X^{(k)} - \U^{(k)} + \bKSI^{(k)} )$ 
		\STATE $\R^{(k)} =  \text{prox}_{\alpha /\rho P_2(\cdot; \bzeta^{(k)} )} (\U^{(k)} - \tilde \X^{(k)}  + \N^{(k)} ) $ 
		\STATE $\bKSI^{(k)} = \bKSI^{(k)} + (\X^{(k)} - \U^{(k)} - \Z^{(k)})  $
		\STATE 	$\N^{(k)} = \N^{(k)} +   ( \U^{(k)} - \tilde \X^{(k)} - \R^{(k)} )   $ 
		\ENDWHILE

	\end{algorithmic}
\end{algorithm}

\subsubsection{iGecco+ Algorithm: Fast ADMM with Inexact One-step Approximation to the Sub-problem}\label{inexact}

Notice that for both Algorithm~\ref{alg:full-diff} and \ref{alg:full-non-diff}, we need to run them iteratively to full convergence in order to  solve the $\U^{(k)}$ sub-problem for each iteration, which is dramatically slow in practice. To address this in literature, many have proposed variants of ADMM with guaranteed convergence that find an inexact, one-step, approximate solution to the sub-problem (without fully solving it); these include the generalized ADMM \citep{deng2016global}, proximal ADMM \citep{shefi2014rate,banert2016fixing} and proximal linearized ADMM \citep{liu2013linearized,lu2016fast}. Thus, we propose to solve the $\U^{(k)}$ sub-problem approximately by taking a single one-step update of the algorithm for both types of losses and prove convergence. For the differentiable loss case, we propose to apply the proximal linearized ADMM approach while for the non-differentiable case, we show that taking a one-step update of Algorithm~\ref{alg:full-non-diff} is equivalent to applying a four-block ADMM to the original problem and we provide a sufficient condition for the convergence of four-block ADMM.  Our algorithm, to the best of our knowledge, is the first to incorporate higher-order multi-block ADMM and inexact ADMM with a one-step update to solve sub-problems for general loss functions.

When the loss is differentiable, 
as mentioned in Algorithm~\ref{alg:full-diff}, one can use full iterative proximal gradient to solve the $\U_{.j}^{(k)}$ sub-problem, which however, is computationally burden-some. 
To avoid this, many proposed variants of ADMM which find approximate solutions to the sub-problems. Specifically, closely related to our problem here, \citet{liu2013linearized,lu2016fast}  proposed proximal linearized ADMM which solves the sub-problems efficiently by linearizing the differentiable part and then applying proximal gradient due to the non-differentiable part. We find their approach fits into our problem and hence develop a proximal linearized 2-block ADMM to solve iGecco+ when the loss $\curl_k$ is differentiable and gradient is Lipschitz continuous. It can be shown that applying proximal linearized 2-block ADMM to Algorithm~\ref{alg:full-admm}  is equivalent to taking 
a one-step update of Algorithm~\ref{alg:full-diff} along with $\V$ and $\bLambda$ update in  Algorithm~\ref{alg:full-admm}. 
In this way, we avoid running full iterative proximal gradient updates to convergence for the $\U^{(k)}$ sub-problem as in Algorithm~\ref{alg:full-diff} and hence save computation cost.

When the loss is non-differentiable, we still seek to take an one-step update to solve the $\U^{(k)}$ sub-problem. We achieve this by noticing that 
taking a one-step update of Algorithm~\ref{alg:full-non-diff} along with $\V$ and $\bLambda$ update in Algorithm~\ref{alg:full-admm} is equivalent to applying multi-block ADMM to the original iGecco+ problem recast as follows (for non-differentiable distance-based loss): 
\begin{align*}
&\minimize_{\U^{(k)},\V} \hspace{5mm}   \sum_{k=1}^K \pi_k f_k(\Z^{(k)} ) + \gamma \underbrace{ \bigg(\sum_{l \in \mathcal E} w_l \|\V_{l.}\|_2\bigg)}_{P_1(\V;\bw)}+ \alpha \sum_{k=1}^K \bigg( \underbrace{ \sum_{j=1}^{p_k}  \zeta_j^{(k)} \| \br_j^{(k)} \|_2}_{P_2(\R^{(k)};\bzeta^{(k)})} \bigg) \\
& \subto \hspace{5mm}  \X^{(k)} - \U^{(k)} = \Z^{(k)}  , \hspace{5mm}      \D \begin{bmatrix} \U^{(1)} & \cdots & \U^{(K)} \end{bmatrix} - \V = \mathbf 0, \hspace{5mm} \U^{(k)}  - \tilde \X^{(k)}  = \R^{(k)}
\end{align*}

Typically, general higher-order multi-block ADMM algorithms do not always converge, even for convex functions \citep{chen2016direct}.  We prove convergence of our algorithm and establish a novel convergence result by casting the iGecco+ with non-differentiable losses as a four-block ADMM, proposing a sufficent condition for convergence of higher-order multi-block ADMMs, and finally showing that our problem satisfies this condition. (Details are given in the proof of Theorem \ref{theorem:inexact-full} in Appendix~\ref{algcovproof}.)  Therefore, taking a one-step update of Algorithm~\ref{alg:full-non-diff} converges for iGecco+ with non-differentiable losses.

So far, we have proposed inexact-solve one-step update approach for both differentiable loss and non-differentiable loss case. For mixed type of losses, we combine those two algorithms and this gives Algorithm~\ref{alg:inexact-admm}, a multi-block ADMM algorithm with inexact one-step approximation to the $\U^{(k)}$ sub-problem to solve iGecco+. We also establish the following convergence result.

\begin{algorithm}[h]
	\caption{iGecco+ Algorithm}
	\label{alg:inexact-admm}
	\begin{algorithmic}
		
		\WHILE{not converged}

		\FOR{all $k = 1,\cdots,K$}
		\STATE 	Update $\U^{(k)}$:
		
		\IF{$\curl_k$ is differentiable}
		\STATE Take a one-step  update of Algorithm~\ref{alg:full-diff}
		\ELSIF{$\curl_k$ is non-differentiable}
		\STATE Take a one-step  update of Algorithm~\ref{alg:full-non-diff}
 	 	\ENDIF
 	 	
		\ENDFOR
		
		\STATE 		$\V = \text{prox}_{\gamma /\rho P_1(\cdot; \bw )} (\begin{bmatrix} \D \U^{(1)} + \bLambda^{(1)} &  \cdots & \D \U^{(K)} + \bLambda^{(K)}  \end{bmatrix} )$ 
		\STATE $\bLambda^{(k)} =  \bLambda^{(k)} +  ( \D \U^{(k)} - \V^{(k)}  ) $ \hspace{5mm} for all $k$

		\ENDWHILE
		
	\end{algorithmic}
\end{algorithm}

\begin{restatable}{theorem}{inexact-full}
\label{theorem:inexact-full}(iGecco+ convergence)
If $\curl_k$ is convex for all $k$, Algorithm~\ref{alg:inexact-admm} converges to a global solution. In addition, if each $\curl_k$ is strictly convex, it converges to the unique global solution.
\end{restatable}

\textbf{Remark:}
Our corresponding Theorem \ref{theorem:inexact-full} establishes a novel convergence result as it is the first to show the convergence of four-block or higher ADMM using approximate sub-problems for both differentiable and non-differentiable losses. 


It is easy to see that Algorithm \ref{alg:inexact-admm} can be applied to solve other Gecco-related methods as special cases. When $K=1$, Algorithm~\ref{alg:inexact-admm} gives the algorithm to solve Gecco+. When $\alpha= 0$, Algorithm~\ref{alg:inexact-admm} gives the algorithm to solve iGecco+. When $K=1$ and $\alpha = 0$, Algorithm~\ref{alg:inexact-admm} gives the algorithm to solve Gecco.

\begin{figure}[ht]
	\vskip 0.2in
	\begin{center}
		\centerline{\includegraphics[scale = 0.67]{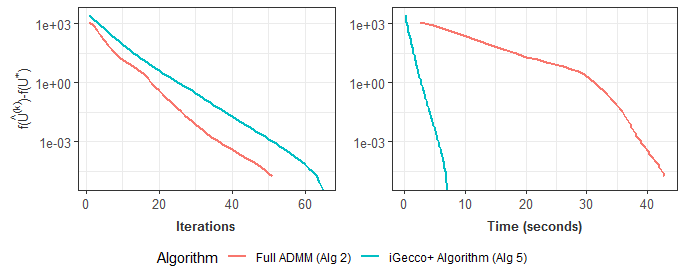}}
		\centerline{\includegraphics[scale = 0.67]{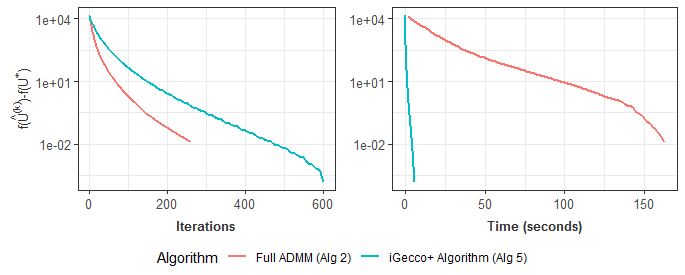}}
		\caption{Comparisons of full ADMM and one-step inexact ADMM algorithm to solve Gecco+ with Poisson log-likelihood (top panel, differentiable loss) and Gecco+ with Manhattan distances (bottom panel, non-differentiable loss). Left plots show the number of iterations needed to converge while right plots show computation time. Algorithm with one-step update to solve the sub-problem saves much more computational time.}
		\label{conv_plot}
	\end{center}
	\vskip -0.2in
\end{figure}

To conclude this section, we compare the convergence results of both full ADMM and inexact ADMM with one-step update in the sub-problem to solve Gecco+ ($n=120$ and $p=210$) in Figure \ref{conv_plot}. The left plots show the number of iterations needed to yield optimization convergence while the right plots show computation time. We see that Algorithm \ref{alg:inexact-admm} (one-step update to solve the sub-problem) saves much more computational time than Algorithm \ref{alg:full-admm} (full updates of the sub-problem). It should be pointed out that though Algorithm \ref{alg:inexact-admm} takes more iterations to converge due to inexact approximation for each iteration,  we still reduce computation time dramatically as the computation time per iteration is much less than the full-solve approach.


\section{Simulation Studies}\label{Simulation}
In this section, we first evaluate performance of Gecco+ against existing methods on non-Gaussian data. Next we compare iGecco+ with other methods on mixed multi-view data. 

\subsection{Non-Gaussian Data} \label{Gecco+_sim}

In this subsection, we evaluate the performance of Gecco and (adaptive) Gecco+ by comparing it with k-means, hierarchical clustering  and sparse convex clustering. For simplicity, we have the following naming convention for all methods: loss type name + Gecco(+). For example, Poisson Deviance Gecco+ refers to Generalized Convex Clustering with Feature Selection using Poisson deviance. 
Sparse CC refers to sparse convex clustering using Euclidean distances. We measure the accuracy of clustering results using adjusted Rand index \citep{hubert1985comparing}.
The adjusted Rand index is the corrected-for-chance version of the Rand index, which is used to measure the agreement between the estimated clustering assignment and the true group label. A larger adjusted Rand index implies a better clustering result. For all methods we consider, we assume oracle number of clusters for fair comparisons.  For our Gecco+, we choose the largest $\alpha$ which minimizes within cluster variance or hold-out error.

Each simulated data set is comprised of $n=120$ observations with 3 clusters. Each cluster has an equal number of observations. Only the first 10 features are informative while the rest are noise. We consider the following simulation scenarios.

\begin{itemize}
	\item S1: Spherical data with outliers
	
	The first 10 informative features in each group are generated from a Gaussian distribution with different $\mu_k$'s  for each class. Specifically, the first 10 features are generated from $N(\mu_k, \mathbf I_{10})$ where $\mu_1= (-2.5\cdot \mathbf 1_{5}^T, \mathbf 0_{5}^T )^T$, $\mu_2= (\mathbf 0_{5}^T , 2.5\cdot \mathbf 1_{5}^T)^T$, $\mu_3= (2.5\cdot \mathbf 1_{5}^T,\mathbf 0_{5}^T)^T$. The outliers in each class are generated from a Gaussian distribution with the same mean centroid $\mu_k$ but with higher variance, i.e, $N(\mu_k, 5 \cdot \mathbf I_{10})$. The remaining noise features are generated from $N(0,1)$. 

	In the first setting (S1A), the number of noise features ranges in $25,50,75,\cdots$ up to 225 with the proportion of the number of outliers fixed ( = 5\%). We also consider the setting when the variance of noise features increases with number of features fixed $p=200$ and number of outliers fixed  (S1B) and high-dimensional setting where $p$ ranges from $250,500,750$ to 1000 (S1C).
	
	\item S2: Non-spherical data with three half moons
	
	Here we consider the standard simulated data of three interlocking half moons as suggested by \cite{chi2015splitting} and \cite{wang2018sparse}. The first 10 features are informative in which each pair makes up two-dimensional three interlocking half moons. We randomly select 5\% of the observations in each group and make them outliers. The remaining noise features are generated from $N(0,1)$. 
	The number of noise features ranges from $25,50,75,\cdots$ up to 225. In both S1 and S2, we compare Manhattan Gecco+ with other existing methods.
	
	\item S3: Count-valued data
	
	The first 10 informative features in each group are generated from a Poisson distribution with different $\mu_k$'s $(i=1,2,3)$ for each class. Specifically, $\mu_1 = 1 \cdot \mathbf 1_{10}$, $\mu_2 = 4  \cdot \mathbf 1_{10}$, $\mu_3 = 7  \cdot \mathbf 1_{10}$. The remaining noise features are generated from a Poisson distribution with the same $\mu$'s which are randomly generated integers from 1 to 10. The number of noise features ranges from $25,50,75,\cdots$ up to 225.
	
\end{itemize}

\begin{figure}[ht]
	\vskip 0.2in
	\begin{center}
		\centerline{\includegraphics[width=\textwidth,height = 11cm]{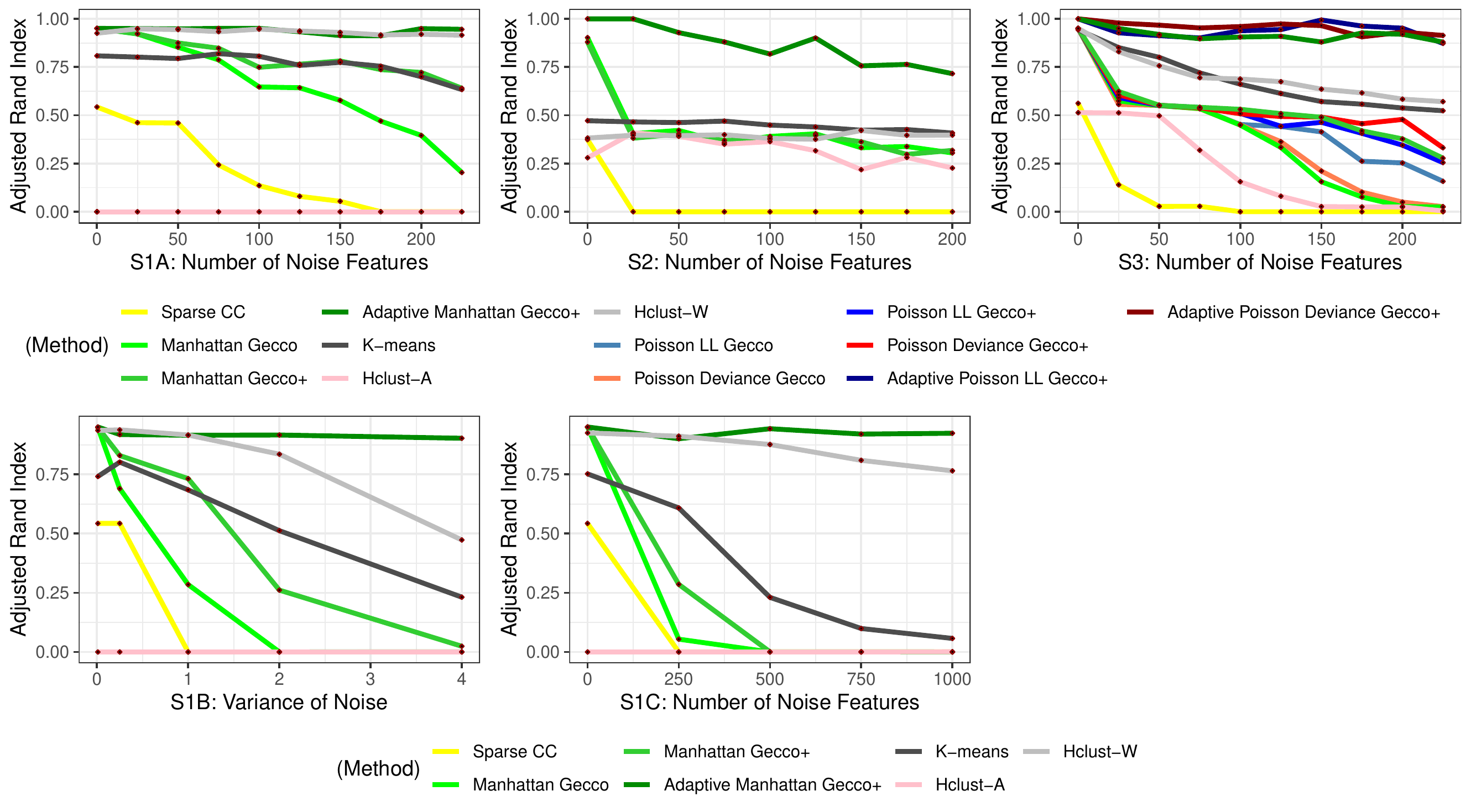}}
		\caption{Simulation results of non-Gaussian data: (S1A) We increase number of noise features for spherical data with outliers; (S2) We increase number of noise features for non-spherical data with outliers; (S3) We increase number of noise features for count-valued data; (S1B) We increase noise level for spherical data with outliers; (S1C) We further increase number of noise features for spherical data with outliers in high dimensions. The adaptive Gecco+ outperforms existing methods in high dimensions.}
		\label{sim_plot}
	\end{center}
	\vskip -0.2in
\end{figure}

We summarize simulation results in Figure~\ref{sim_plot}. 
We find that for spherical data with outliers, adaptive Manhattan Gecco+ performs the best in high dimensions. Manhattan Gecco performs well in low dimensions but poorly as number of noisy features increases. Manhattan Gecco+ performs well as the dimension increases, but adaptive Manhattan Gecco+ outperforms the former as it adaptively penalizes the features, meaning that noisy features quickly get zeroed out in the clustering path and that only the informative features perform important roles in clustering. We see that, without adaptive methods, we do not achieve the full benefit of performing feature selection. As we perform adaptive Gecco+, we show vast improvement in clustering purity as the number of noise features grows where regular Gecco performs poorly. Sparse convex clustering performs the worst as it tends to pick outliers as singleton classes. Our simulation results also show that adaptive Manhattan Gecco+ works well for non-spherical data by selecting the correct features. For count data, all three adaptive Gecco+ methods perform better than k-means, hierarchical clustering and sparse convex clustering. We should point out that there are several linkage options for hierarchical clustering. For visualization purposes, we only show the linkage with the best and worst performance instead of all the linkages. Also we use the appropriate data-specific distance metrics  in hierarchical clustering. Table~\ref{vs-accuracy-F} shows the variable selection accuracy of sparse convex clustering and adaptive Gecco+ in terms of F$_1$ score. In all scenarios, we fix $p=225$. We see that adaptive Gecco+ selects the correct features, whereas sparse convex clustering performs poorly.

\begin{table}[ht]
	\vskip 0.15in
		\begin{center}
  \begin{tabular}{l c c c}
    \toprule
    \multirow{1}{*}{Method} &
      \multicolumn{1}{c}{Scenario 1 (A)} &
      \multicolumn{1}{c}{Scenario 2} &
      \multicolumn{1}{c}{Scenario 3} \\
      \midrule
    Sparse Convex Clustering & 0.37 (3.1e-2) &  0.25 (2.4e-2) &   0.14 (7.2e-3)  \\
    Adaptive Gecco+ &  0.97 (1.9e-2) &  0.99  (1.0e-2) &   0.81 (8.0e-2)\\
    \bottomrule
  \end{tabular}
   \caption{Comparisons of F$_1$ score for adaptive Gecco+ and sparse convex clustering}
 \label{vs-accuracy-F}
  	\end{center}
	\vskip -0.1in
\end{table}

\subsection{Multi-View Data}
In this subsection, we evaluate the performance of iGecco and (adaptive) iGecco+ on mixed multi-view data by comparing it with hierarchical clustering, iCluster+ \citep{mo2013pattern} and Bayesian Consensus Clustering \citep{lock2013bayesian}.  Again, we measure the accuracy of clustering results using the adjusted Rand index \citep{hubert1985comparing}.

As before, each simulated data set is comprised of $n=120$ observations with 3 clusters. Each cluster has an equal number of observations. Only the first 10 features are informative while the rest are noise. We have three data views consisting of continuous data, count-valued data and binary/proportion-valued data.  We investigate different scenarios with different dimensions for each data view and consider the following simulation scenarios: 

\begin{itemize}
	\item S1: Spherical data with $p_1 = p_2 = p_3 = 10$
	
	\item S2: Three half-moon data with $p_1 = p_2 = p_3 = 10$
	
	\item S3: Spherical data with $p_1 = 200$, $p_2= 100$, $p_3 = 50$
	
	\item S4: Three half-moon data with $p_1 = 200$, $p_2= 100$, $p_3 = 50$
	
   \item S5: Spherical data with $p_1 = 50$, $p_2= 200$, $p_3 = 100$
	
	\item S6: Three half-moon data with $p_1 = 50$, $p_2= 200$, $p_3 = 100$
	
\end{itemize}

We employ a similar simulation setup as in Section \ref{Gecco+_sim} to generate each data view. The difference is that here for informative features, we increase the within-cluster variance for Gaussian data and decrease difference of cluster mean centroids $\mu_k$'s for binary and count data so that there is overlap between different clusters. Specifically, for spherical cases, Gaussian data is generated from $N(\mu_k, 3 \cdot \mathbf I_{10})$; count data is generated from Poisson with different $\mu_k$'s ($\mu_1 = 2$, $\mu_2 = 4$, $\mu_3 = 6$, etc); binary data is generated from Bernoulli with different $\mu_k$'s ($\mu_1 = 0.5$, $\mu_2 = 0.2$, $\mu_3 = 0.8$, etc).  For half-moon cases, continuous data is simulated with larger noise and the count and proportion-valued data is generated via a copula transform.  In this manner, we have created a challenging simulation scenario where accurate clustering results cannot be achieved by considering only a single data-view.

Again, for fair comparisons across methods, we assume the oracle number of clusters.
When applying iGecco methods, we employ Euclidean distances for continuous data, Manhattan distances for count-valued data and Bernoulli log-likelihood for binary or proportion-valued data. We use the latter two losses as they perform well compared with counterpart losses in Gecco+ and demonstrate faster computation speed. Again, we choose the largest $\alpha$ that minimizes within-cluster deviance.

Simulation results in Table~\ref{iGecco_rand} and Table~\ref{iGecco+_rand}  show that our methods perform better than existing methods. In low dimensions, iGecco performs comparably with iCluster and Bayesian Consensus Clustering for spherical data. For non-spherical data, iGecco performs much better. For high dimensions, iGecco+ performs better than iGecco while adaptive iGecco+ performs the best as it achieves the full benefit of feature selection. 

\begin{table}[ht]
	\vskip 0.15in
	\begin{center}
			\begin{tabular}{ |l|c|c|}
					\hline
					 Method    &  Scenario 1 & Scenario 2 \\ \hline
					\multirow{1}{*}{Hclust: $\X_1$} & 0.35 (2.9e-2)  & 0.54 (1.3e-2)   \\
					\multirow{1}{*}{Hclust: $\X_2$} & 0.53 (4.6e-2)  &  0.61 (4.0e-2) \\ 
					\multirow{1}{*}{Hclust: $\X_3$} &  0.52 (2.2e-2) &  0.70 (3.0e-2) \\
					\multirow{1}{*}{Hclust: $[\X_1 \X_2 \X_3]$ - Euclidean} & 0.68 (4.7e-2)  & 0.66 (4.4e-2) \\ 
					\multirow{1}{*}{Hclust: $[\X_1 \X_2 \X_3]$ - Gower } & 0.86 (1.5e-2) &  0.84 (4.0e-2) \\ 
					\multirow{1}{*}{iCluster+ with $\lambda = 0$} &  0.90 (1.6e-2)  &  0.70 (8.0e-3) \\ 
					\multirow{1}{*}{Bayesian Consensus Clustering}   &  \textbf{0.95 (1.2e-2)} &   0.63 (1.0e-2)\\ 	
					\multirow{1}{*}{iGecco} &  \textbf{0.93 (4.7e-3)} &  \textbf{1.00 (0.0e-0)} \\ \hline
			\end{tabular}
	\caption{Comparisons of adjusted Rand index for mixed multi-view data}
	\label{iGecco_rand}
	\end{center}
	\vskip -0.1in
\end{table}
\begin{table}[ht]
	\vskip 0.15in
	\begin{center}
		\begin{small}
		\scalebox{0.9}{
			\begin{tabular}{ |l|c|c|c|c|}
					\hline
					    Method &  Scenario 3 & Scenario 4 & Scenario 5  & Scenario 6  \\ \hline
					\multirow{1}{*}{Hclust: $\X_1$} & 0.57 (1.8e-2) &  0.57 (1.4e-2) &  0.44 (2.4e-2)& 0.49 (1.7e-2)\\ 
					\multirow{1}{*}{Hclust: $\X_2$} & 0.22 (1.9e-2) &  0.20 (1.8e-2) & 0.51 (1.7e-2)&  0.51 (2.6e-2) \\ 
					\multirow{1}{*}{Hclust: $\X_3$} & 0.28 (1.1e-2) &  0.25 (2.6e-2) & 0.57 (2.7e-2)&  0.48 (3.3e-2)\\ 
					\multirow{1}{*}{Hclust: $[\X_1 \X_2 \X_3]$ -  Euclidean}    & 0.72 (2.2e-2) & 0.43 (1.9e-2)  & 0.53 (1.9e-2)&   0.56 (2.3e-2) \\ 
					\multirow{1}{*}{Hclust: $[\X_1 \X_2 \X_3]$ - Gower} & 0.78 (1.0e-2) & 0.41 (3.0e-2) & 0.58 (4.2e-2) &  0.64 (2.6e-2) \\ 
					\multirow{1}{*}{iCluster+}   & 0.61 (2.5e-2) &  0.74 (2.8e-2) & 0.62 (1.7e-2)  &  0.61 (1.4e-2)  \\ 
					\multirow{1}{*}{Bayesian Consensus Clustering}  & 0.47 (1.1e-1) &  0.53 (1.0e-2) &  0.60 (1.0e-2) &   0.63 (1.1e-2) \\ 
					\multirow{1}{*}{iGecco}  & 0.14 (8.7e-2) & 0.13 (8.2e-2) & 0.45 (2.9e-2) &  0.42 (4.4e-2)  \\ 
					\multirow{1}{*}{iGecco+}  &  0.37 (6.7e-2) & 0.37 (5.6e-2)  & 0.48 (2.8e-2) &  0.48 (4.7e-2) \\
					\multirow{1}{*}{Adaptive iGecco+}  & \textbf{0.91 (6.1e-3)} & \textbf{0.92 (8.3e-3)}  & \textbf{0.96 (2.4e-2)} &  \textbf{0.94 (4.3e-2)}  \\ \hline					
			\end{tabular}
			}
		\end{small}
		\caption{Comparisons of adjusted Rand index for high-dimensional mixed multi-view data}
	\label{iGecco+_rand}	
	\end{center}
	\vskip -0.1in
\end{table}

Also we show the variable selection results in Table~\ref{iGecco+_vsa} and compare our method to that of iClusterPlus.  Our adaptive iGecco+ outperforms iClusterPlus for all scenarios.
\begin{table}[ht]
	\vskip 0.15in
	\begin{center}
\resizebox{\linewidth}{!}{%
  \begin{tabular}{lcccccccc}
    \toprule
    \multirow{2}{*}{} &
      \multicolumn{2}{c}{Overall} &
      \multicolumn{2}{c}{Gaussian} &
      \multicolumn{2}{c}{Count} & 
      \multicolumn{2}{c}{Binary}  \\
      & {iCluster+} & {A iGecco+} & {iCluster+} & {A iGecco+} & {iCluster+} & {A iGecco+} & {iCluster+} & {A iGecco+} \\
      \midrule
    S3  &  0.88 (1.1e-2)   &  \textbf{0.96 (4.5e-3)}  &   0.96 (9.2e-3)  & 1.00 (0.0e-0)    &  0.81 (1.3e-2)   &  0.89 (5.7e-3)   &  0.87 (1.3e-2)   &   0.98 (7.8e-3)  \\
    S4   & 0.93 (1.5e-2)   & \textbf{0.99 (4.5e-3)}  &  0.97 (2.0e-2)   & 0.99 (7.0e-3)   &   0.93 (1.5e-2) &  1.00 (4.8e-3)  &  0.89 (2.2e-2)   & 0.99 (6.3e-3)  \\
    S5  &  0.95 (3.0e-2) &  \textbf{1.00 (3.3e-3)} &  0.95 (3.3e-2)  & 1.00 (0.0e-0)  & 0.93 (3.6e-2)   &  0.99 (1.0e-2)    &  0.96 (2.2e-2)  & 1.00 (0.0e-0)\\
    S6   & 0.93 (3.1e-2)  &   \textbf{1.00 (1.6e-3)}   & 0.95 (3.3e-2)    & 1.00  (0.0e-0)   & 0.88 (4.4e-2)   &  1.00 (0.0e-0)    &    0.95 (2.5e-2)   &   1.00 (0.0e-0)  \\
    \bottomrule
  \end{tabular}
  }
  \caption{Comparisons of F$_1$ score for adaptive iGecco+ and iClusterPlus}
    \label{iGecco+_vsa}
	\end{center}
	\vskip -0.1in
\end{table}

\section{Real Data Examples}\label{realdata}

In this section, we apply our methods to various real data sets and evaluate our methods against existing ones. We first evaluate the performance of Gecco+ for several real data sets and investigate the features selected by various Gecco+ methods.

\subsection{Authors Data} \label{author}
The authors data set consists of word counts from $n = 841$ chapters written by four famous English-language authors (Austen, London, Shakespeare, and Milton). Each class contains an unbalanced number of observations with 69 features. The features are common ``stop words" like ``a", ``be" and ``the" which are typically removed before text mining analysis.  We use Gecco+ not only to cluster book chapters and compare the clustering assignment with true labels of authors, but also to identify which key words help distinguish the authors. We choose tuning parameters using hold-out validation.

\begin{table}[ht]
\vskip 0.05in
\centering
\begin{tabular}{|l|c|}
\hline
Method &  Adjusted Rand Index    \\\hline
K-means  &  0.73\\
Hierarchical Clustering & 0.73\\
Sparse Convex Clustering & 0.50 \\
Manhattan Gecco+   & 0.96 \\
Poisson LL Gecco+  &   0.96 \\
Poisson Deviance  Gecco+  &  0.96  \\
\hline
\end{tabular}
\caption{Adjusted Rand index of different methods for authors data set}
\label{author-rand}		
\vskip -0.1in
\end{table}

In Table~\ref{author-rand}, we compare Gecco+ with existing methods in terms of clustering quality. For hierarchical clustering, we only show the linkage with the best performance (in this whole section). Our method outperforms k-means and the best hierarchical clustering method.  Secondly, we look at the word texts selected by Gecco+.
As shown in Table~\ref{author-feature}, Jane Austen tended to use the word ``her" more frequently than the other authors; this is expected as the subjects of her novels are typically females. The word ``was" seems to separate Shakespeare and Jack London well. Shakespeare preferred to use present tense more while Jack London preferred to use past tense more. To summarize, our Gecco+ not only has superior clustering performance but also selects interpretable features.
\begin{table}[ht]
	\vskip 0.05in
	\begin{center}
				 \begin{tabular}{lcccr}
					\toprule
					Method &  Features \\
					\midrule
					Manhattan Gecco+   &  \begin{tabular}{@{}c@{}} ``be"  ,``had" ,``her", \\  ``the" ,``to",  ``was" \end{tabular}  \\
					Poisson LL Gecco+    &   ``an"  , ``her" , ``our",  ``your" \\
					Poisson Deviance Gecco+     &   \begin{tabular}{@{}c@{}} ``an",  ``be" , ``had", ``her", \\  ``is",  ``my" , ``the", ``was" \end{tabular}  \\
					\bottomrule
				\end{tabular}
		\caption{Features selected by different Gecco+ methods for authors data set}
	\label{author-feature}	
	\end{center}
	\vskip -0.1in
\end{table}

\subsection{TCGA Breast Cancer Data}
The TCGA data set consists of log-transformed Level III RPKM gene expression levels for 445 breast-cancer patients with 353 features from The Cancer Genome Atlas Network \citep{cancer2012comprehensive}. Five PAM50 breast cancer subtypes are included, i.e, Basal-like, Luminal A, Luminal B, HER2-enriched, and Normal-like. Only 353 genes out of 50,000 with somatic mutations from COSMIC \citep{forbes2010cosmic} are retained. The data is Level III TCGA BRCA RNA-Sequencing gene expression data that have already been pre-processed according to the following steps:
i) reads normalized by RPKM, and ii) corrected for overdispersion by a log-transformation. We remove 7 patients, who belong to the normal-like group and the number of subjects $n$ becomes 438. We also combine Luminal A with Luminal B as they are often considered one aggregate group \citep{choi2014identification}.

\begin{table}[ht]
\centering
\begin{tabular}{|l|c|}
\hline
Method &  Adjusted Rand Index    \\\hline
K-means  &  0.44\\
Hierarchical Clustering & 0.26\\
Sparse Convex Clustering & 0.01 \\
Manhattan  Gecco+ & 0.76 \\
Poisson LL Gecco+   &  0.72  \\
Poisson Deviance Gecco+   &  0.72  \\
\hline
\end{tabular}
\caption{Adjusted Rand index of different methods for TCGA data set}
\label{TCGA-rand}	
\end{table}

From Table~\ref{TCGA-rand}, our method outperforms k-means and the best hierarchical clustering method. Next, we look at the genes selected by Gecco+ in Table~\ref{TCGA-feature}.  FOXA1 is known to be a key gene that characterizes luminal subtypes in DNA microarray analyses \citep{badve2007foxa1}. GATA binding protein 3 (GATA3) is a transcriptional activator highly expressed by the luminal epithelial cells in the breast \citep{mehra2005identification}.  ERBB2 is known to be associated with HER2 subtype and has been well studied in breast cancer \citep{harari2000molecular}. Hence our Gecco+ not only outperforms existing methods  but also selects genes which are relevant to biology and have been implicated in previous scientific studies.

\begin{table}[ht]
	\vskip 0.05in
	\begin{center}
				\begin{tabular}{lcccr}
					\toprule
					Method &  Features \\
					\midrule
					Manhattan Gecco+  &   \begin{tabular}{@{}c@{}}``BCL2" , ``ERBB2" ,``GATA3" \\ ``HMGA1", ``IL6ST" \end{tabular} \\
					Poisson LL Gecco+   &  ``ERBB2" ``FOXA1" ``GATA3"  \\
					Poisson Deviance Gecco+     & \begin{tabular}{@{}c@{}}  ``ERBB2" ,  ``FOXA1",   ``GATA3"  \\ ``RET", ``SLC34A2"\end{tabular} 
					
					\\
					\bottomrule
				\end{tabular}
		\caption{Features selected by different Gecco+ methods for TCGA data set}
	\label{TCGA-feature}	
	\end{center}
	\vskip -0.1in
\end{table}

Next we evaluate the performance of iGecco+ for mixed multi-view data sets and investigate the features selected by iGecco+.

\subsection{Multi-omics Data}\label{omics}

One promising application for integrative clustering for multi-view data lies in integrative cancer genomics. Biologists seek to integrate data from multiple platforms of high-throughput genomic data to gain a more thorough understanding of disease  mechanisms and detect cancer subtypes. In this case study, we seek to integrate four different types of genomic data to study how epigenetics and short RNAs influence the gene regulatory system in breast cancer.

We use the data set from  \cite{cancer2012comprehensive}. \cite{lock2013bayesian} analyzed this data set using integrative methods and we followed the same data pre-processing procedure: i) filter out genes in expression data whose standard deviation is less than 1.5, ii) take square root of methylation data, and iii) take log of miRNA data. 
We end up with a data set of 348 tumor samples including:
\begin{itemize}
    \item RNAseq gene expression (GE) data for 645 genes,
    \item DNA methylation (ME) data for 574 probes,
    \item miRNA expression (miRNA) data for 423 miRNAs, 
    \item Reverse phase protein array (RPPA) data for 171 proteins.
\end{itemize}
The data set contains samples used on each platform with associated subtype calls from each technology platform as well as integrated cluster labels from biologists. We use the integrated labels from biologists as true label. Also we merged the subtypes 3 and 4 in the integrated labels as those two subtypes correspond to Luminal A and Luminal B respectively from the predicted label given by gene expression data (PAM50 mRNA).

Figure \ref{fig:hist} in Appendix~\ref{genohist} gives the distribution of data from different platforms. For our iGecco+ methods, we use Euclidean distances for gene expression data and protein data as the distributions of those two data sets appear gaussian; binomial deviances for Methylation data as the value is between $[0,1]$; Manhattan distances for miRNA data as the data is highly-skewed. We compare our adaptive iGecco+ with other existing methods. From Table \ref{real-data-multi}, we see that our method outperforms all the existing methods.


\begin{table}[ht]
\centering
\begin{tabular}{|l |c|}
\hline
Method & Adjusted Rand   Index \\\hline
Hclust: $\X_1$ GE & 0.51  \\
Hclust: $\X_2$ Meth & 0.39  \\
Hclust: $\X_3$ miRNA & 0.21   \\
Hclust: $\X_4$ Protein & 0.24   \\
Hclust: $[\X_1 \X_2 \X_3 \X_4]$ -  Euclidean & 0.51  \\
Hclust: $[\X_1 \X_2 \X_3 \X_4]$ - Gower & 0.40  \\
iCluster+ & 0.36  \\
Bayesian Consensus Clustering & 0.35\\
Adaptive iGecco+  & \textbf{0.71} \\
\hline
\end{tabular}
\caption{Adjusted Rand index of different methods for multi-omics TCGA data set}
\label{real-data-multi}	
\end{table}


We also investigate the features selected by adaptive iGecco+, shown in Table \ref{TCGA-feature-igecco}, and find that our method is validated as most are known in the breast cancer literature. For example, FOXA1 is known to segregate the luminal subtypes from the others \citep{badve2007foxa1}, and AGR3 is a known biomarker for breast cancer prognosis and early breast cancer detection from blood \citep{garczyk2015agr3}.  Several well-known miRNAs were selected including MIR-135b, which is upregulated in breast cancer and promotes cell growth \citep{hua2016mir} as well as MIR-190 which suppresses breast cancer metastasis \citep{yu2018mir}. Several known proteins were also selected including ERalpha, which is overexpressed in early
stages of breast cancer \citep{hayashi2003expression} and GATA3 which plays an integral role in breast luminal cell differentiation and breast cancer progression \citep{cimino2013gata3}.

\begin{table}[ht]
	\vskip 0.05in
	\begin{center}
				\begin{tabular}{lcccr}
					\toprule
					Data view  &  Features \\
					\midrule
					Gene Expression  &   \begin{tabular}{@{}c@{}}``AGR3", ``FOXA1", ``AGR2", ``ROPN1", \\ ``ROPN1B", ``ESR1", ``C1orf64", ``ART3",``FSIP1" \end{tabular} \\
					miRNA   &    \begin{tabular}{@{}c@{}}``hsa-mir-135b", ``hsa-mir-190b", ``hsa-mir-577", 
					``hsa-mir-934" \end{tabular} \\   
					Methylation     & \begin{tabular}{@{}c@{}} ``cg08047457", ``cg08097882", ``cg00117172", ``cg12265829" \end{tabular}  \\
					Protein   &  ``ER.alpha", ``GATA3", ``AR", ``Cyclin\_E1"  \\
					\bottomrule
				\end{tabular}
		\caption{Features selected by adaptive iGecco+ methods for multi-omics TCGA data set}
	\label{TCGA-feature-igecco}	
	\end{center}
	\vskip -0.1in
\end{table}

\begin{figure}[ht]
\centering
\includegraphics[width=\textwidth,height = 7.8cm]{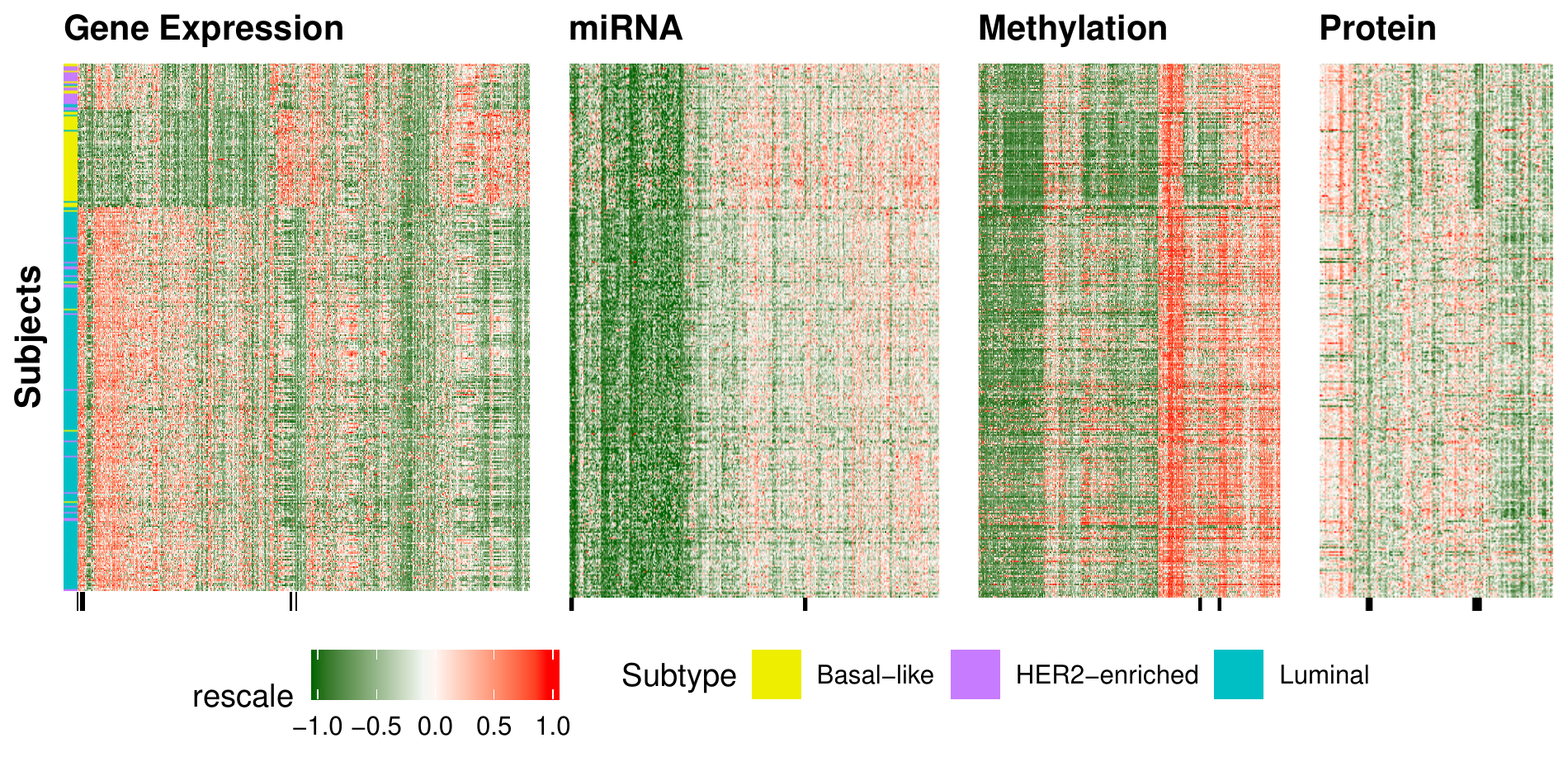}
\caption{Cluster heatmap of multi-omics TCGA data with row orders determined by cluster assignments from iGecco+. The left bar refers to the integrated cluster labels from biologists. The black bars at the bottom of each data view correspond to the selected features. Our adaptive iGecco+ identifies meaningful subtypes.}
\label{viz-omics}
\end{figure}

We also visualize  resulting clusters from adaptive iGecco+. In Figure \ref{viz-omics}, we  see that there is a clear separation between groups and adaptive iGecco+ identifies meaningful subtypes. 

\section{Discussion}

In this paper, we develop a convex formulation of integrative clustering for high-dimensional mixed multi-view data. We propose a unified, elegant methodological solution to two critical issues for clustering and data integration: (i) dealing with mixed types of data and (ii) selecting interpretable features in high-dimensional settings.  
Specifically, we show that clustering for mixed, multi-vew data can be achieved using different data specific convex losses with a joint fusion penalty.  
We also introduce a shifted group-lasso penalty that shrinks noise features to their loss-specific centers, hence selecting features that play important roles in separating groups. In addition, we make an optimization contribution by proposing and proving the convergence of a new general multi-block ADMM algorithm with sub-problem approximations that efficiently solves our problem. Empirical studies show that iGecco+ outperforms existing clustering methods and selects interpretable features in separating clusters.



This paper focuses on the methodological development for integrative clustering and feature selection, but there are many possible avenues for future research related to this work.  For example, we expect in future work to be able to show that our methods inherit the strong theoretical properties of other convex clustering approaches such as clustering consistency and prediction consistency.  An important problem in practice is choosing which loss function is appropriate for a given data set. While this is beyond the scope of this paper, an interesting direction for future research would be to learn the appropriate convex loss function in a data-driven manner. Additionally, many have shown block missing structure is common in mixed data \citep{yu2019optimal,xiang2013multi}. A potentially interesting direction for future work would be to develop an extension of iGecco+ that can appropriately handle block-missing multi-view data. Additionally, \citet{weylandt2019dynamic} developed a fast algorithm to compute the entire convex clustering solution path and used this to visualize the results via a dendrogram and pathwise plot.  In future work, we expect that algorithmic regularization path approaches can also be applied to our method to be able to represent our solution as a dendrogram and employ other dynamic visualizations.  Finally, while we develop an efficient multi-block ADMM algorithm, there may be further room to speed up computation of iGecco+, potentially by using distributed optimization approaches.


In this paper, we demonstrate that our method can be applied to integrative genomics, yet it can be applied to other fields such as multi-modal imaging, national security, online advertising, and environmental studies where practitioners aim to find meaningful clusters and features at the same time. In conclusion, we introduce a principled, unified approach to a challenging problem that demonstrates strong empirical performance and opens many directions for future research.


\section*{Acknowledgements}
The authors would like to thank Michael Weylandt and Tianyi Yao for helpful discussions.  GA and MW also acknowledge support from NSF DMS-1554821, NSF NeuroNex-1707400, and NSF DMS-1264058.

\newpage

\appendix

\singlespacing

\begin{center}
{\bf \LARGE Integrative Generalized Convex Clustering Optimization and Feature Selection for Mixed Multi-View Data: Supplementary Materials} 
\bigskip

{\large Minjie Wang, Genevera I. Allen}
\end{center}

The supplementary materials are organized as follows. In Appendix~\ref{propprove}, we prove properties of our methods discussed in Section~\ref{property}. In Appendix~\ref{algcovproof}, we provide detailed proof for Theorem \ref{theorem:inexact-full}. We provide implementation details for Gecco+ with differentiable losses in Appendix~\ref{diffdetail}. In  Appendix~\ref{nondiffdetail}, we discuss implementation details for  Gecco+ with non-differentiable losses. We introduce multinomial Gecco(+) in Appendix~\ref{multin}. In Appendix~\ref{centroidcal}, we show how to calculate the loss-specific center in Table~\ref{loss-table}. In Appendix~\ref{authorvizone}, we visualize the results of authors data discussed in Section \ref{author}. We show the distribution of data from different platforms in Section~\ref{omics} in Appendix~\ref{genohist}.

\section{Proof of Propositions}\label{propprove}
Proposition \ref{theorem:diff1} and  \ref{theorem:diff2}  are direct extension from Proposition 2.1 in \citet{chi2015splitting}. Notice they proved the solution path depends continuously on the tuning parameter $\gamma$ and the weight matrix $\bw$. It follows that the argument can be also applied to tuning parameter $\alpha$, the loss weight $\pi_k$, and feature weight $\zeta_j^{(k)}$. Also it is obvious that the loss $\curl(\cdot)$ is continuous with respect to the data, $\X$.

We show in detail how to prove Proposition \ref{theorem:diff3} in the following. First we rewrite $ F_{\gamma,\alpha} (\U) $ as:
\begin{small}
\begin{align*}
    F_{\gamma,\alpha} (\U)  &=  \sum_{k=1}^K \pi_k \curl_k(\X^{(k)},\U^{(k)})  + \gamma \sum_{i < i'} w_{ii'}  \sqrt{ \sum_{k=1}^K  \| \U_{i.}^{(k)} - \U_{i'.} ^{(k)} \|^2 }   + \alpha \sum_{k=1}^K \sum_{j=1}^{p_k}  \zeta_j^{(k)} \| \U_{.j}^{(k)}  - \tilde x_j^{(k)} \cdot \textbf 1_n  \|_2 \\
    & = \sum_{k=1}^K \sum_{i=1}^n \pi_k \curl_k(\X_{i.}^{(k)},\U_{i.}^{(k)})  + \gamma \sum_{i < i'} w_{ii'}  \sqrt{ \sum_{k=1}^K  \| \U_{i.}^{(k)} - \U_{i'.} ^{(k)} \|^2 }   + \alpha \sum_{k=1}^K \sum_{j=1}^{p_k}  \zeta_j^{(k)} \| \U_{.j}^{(k)}  - \tilde x_j^{(k)} \cdot \textbf 1_n  \|_2 
\end{align*}
\end{small}

{By definition, loss-specific cluster center is $\tilde \bx^{(k)} = \argmin \limits_{\bu} \sum_{i=1}^n \curl_k(\X_{i.}^{(k)},\bu)$. Since $\curl_k$ is convex, it is equivalent to $\bu$ such that $\partial \sum_i \curl_k(\X_{i.}^{(k)},\bu) = 0$. Hence, $\partial \sum_i \curl_k(\X_{i.}^{(k)},\tilde \bx^{(k)} ) = 0$}

We use the similar proof approach in \cite{chi2015splitting}. A point $\X$ furnishes a global minimum of the convex function $F_Y(\X)$ if and only if all forward directional derivatives $d_{\theta} F_Y(\X)$ at $\X$ are nonnegative. We calculate the directional derivatives:
\begin{align*}
    d_{\theta} F_{\gamma,\alpha} (\tilde \X ) &=  \sum_{k=1}^K \sum_{i=1}^n \pi_k \langle \partial  \curl_k (\X_{i.}^{(k)},\tilde \bx^{(k)}), \Theta_{i.}^{(k)} \rangle +  \gamma \sum_{i < i'} w_{ii'}  \sqrt{ \sum_{k=1}^K  \| \Theta_{i.}^{(k)} - \Theta_{i'.} ^{(k)} \|^2 }   \\ 
    & + \alpha \sum_{k=1}^K \sum_{j=1}^{p_k}  \zeta_j^{(k)} \| \Theta_{.j}^{(k)}  - \tilde x_j^{(k)} \cdot \textbf 1_n  \|_2 
\end{align*}
Note: $\sum_{i=1}^n \langle \partial  \curl_k (\X_{i.}^{(k)},\tilde \bx^{(k)}), \Theta_{i'.}^{(k)} \rangle = 0 $

The generalized Cauchy-Schwartz inequality therefore implies
\begin{align*}
\sum_{i=1}^n \langle \partial  \curl_k (\X_{i.}^{(k)},\tilde \bx^{(k)}), \Theta_{i.}^{(k)} \rangle &= \frac{1}{n} \sum_{i=1}^n  \sum_{i'=1}^n  \langle \partial  \curl_k (\X_{i.}^{(k)},\tilde \bx^{(k)}), \Theta_{i.}^{(k)} \rangle \\
 &= \frac{1}{n} \sum_{i=1}^n  \sum_{i'=1}^n  \langle \partial  \curl_k (\X_{i.}^{(k)},\tilde \bx^{(k)}), \Theta_{i.}^{(k)} - \Theta_{i'.}^{(k)} \rangle \\
 &\geq -\frac{2}{n} \sum_{i < i'} \| \partial  \curl_k (\X_{i.}^{(k)},\tilde \bx^{(k)}) \|_2 \cdot \| \Theta_{i.}^{(k)} - \Theta_{i'.}^{(k)} \|_2 \\
  &\geq -\frac{2}{n} \sum_{i < i'} \| \partial  \curl_k (\X_{i.}^{(k)},\tilde \bx^{(k)}) \|_2 \cdot  \sqrt{ \sum_{k=1}^K  \| \Theta_{i.}^{(k)} - \Theta_{i'.} ^{(k)} \|^2 } 
\end{align*}
Hence 
\begin{align*}
\sum_{k=1}^K \sum_{i=1}^n \pi_k \langle \partial  \curl_k (\X_{i.}^{(k)},\tilde \bx^{(k)}), \Theta_{i.}^{(k)} \rangle \geq -\frac{2}{n} \sum_{k=1}^K \sum_{i < i'}  \pi_k \| \partial  \curl_k (\X_{i.}^{(k)},\tilde \bx^{(k)}) \|_2 \cdot  \sqrt{ \sum_{k=1}^K  \| \Theta_{i.}^{(k)} - \Theta_{i'.} ^{(k)} \|^2 }
\end{align*}

Take $\gamma$ sufficiently large such that:
\begin{align*}
    \gamma \sum_{i < i'} w_{ii'}  \sqrt{ \sum_{k=1}^K  \| \Theta_{i.}^{(k)} - \Theta_{i'.} ^{(k)} \|^2 }   \geq -\frac{2}{n} \sum_{k=1}^K \sum_{i < i'}  \pi_k \| \partial  \curl_k (\X_{i.}^{(k)},\tilde \bx^{(k)}) \|_2 \cdot  \sqrt{ \sum_{k=1}^K  \| \Theta_{i.}^{(k)} - \Theta_{i'.} ^{(k)} \|^2 }
\end{align*}

When all $w_{ii'} > 0$, one can take any $\gamma$ that exceeds 
\begin{align*}
    K \cdot \frac{2}{n} \max_{i,i',k} \frac{ \pi_k \| \partial  \curl_k (\X_{i.}^{(k)},\tilde \bx^{(k)}) \|_2 }{w_{ii'}}
\end{align*}

In general set
\begin{align*}
    \beta = K \cdot \frac{2}{n \min_{w_{ii'} >0} w_{ii'}} \max_{i,i',k}  \pi_k \| \partial  \curl_k (\X_{i.}^{(k)},\tilde \bx^{(k)}) \|_2 
\end{align*}

For any pair $i$ and $i'$ there exists a path $i \to k \to \cdots \to l \to i'$ along which the weights are positive.
It follows that

\begin{align*}
    \frac{2}{n} \sum_{k=1}^K   \pi_k \| \partial  \curl_k (\X_{i.}^{(k)},\tilde \bx^{(k)}) \|_2 \cdot  \sqrt{ \sum_{k=1}^K  \| \Theta_{i.}^{(k)} - \Theta_{i'.} ^{(k)} \|^2 } \leq \beta  \sum_{i < i'}  w_{ii'}  \sqrt{ \sum_{k=1}^K  \| \Theta_{i.}^{(k)} - \Theta_{i'.} ^{(k)} \|^2 }  
\end{align*}
We have
\begin{align*}
    \frac{2}{n} \sum_{k=1}^K  \sum_{i < i'}   \pi_k \| \partial  \curl_k (\X_{i.}^{(k)},\tilde \bx^{(k)}) \|_2 \cdot  \sqrt{ \sum_{k=1}^K  \| \Theta_{i.}^{(k)} - \Theta_{i'.} ^{(k)} \|^2 } \leq \binom{n}{2} \beta  \sum_{i < i'}  w_{ii'}  \sqrt{ \sum_{k=1}^K  \| \Theta_{i.}^{(k)} - \Theta_{i'.} ^{(k)} \|^2 }  
\end{align*}

Hence the forward directional derivative test is satisfied for any $\gamma \geq \binom{n}{2} \beta$.

On the other hand, for fixed $\gamma$, the generalized Cauchy-Schwartz inequality implies
\begin{align*}
\sum_{i=1}^n \langle \partial  \curl_k (\X_{i.}^{(k)},\tilde \bx^{(k)}), \Theta_{i.}^{(k)} \rangle &= \sum_{i=1}^n   \langle \partial  \curl_k (\X_{i.}^{(k)},\tilde \bx^{(k)}), \Theta_{i.}^{(k)}  -  \tilde \bx^{(k)} \rangle \\
& = \sum_{j=1}^{p_k}   \langle \partial  \curl_k (\X_{.j}^{(k)},\tilde x_j^{(k)} \cdot \textbf 1_n    ), \Theta_{.j}^{(k)}  -  \tilde x_j^{(k)} \cdot \textbf 1_n  \rangle \\
&  \geq  -  \sum_{j=1}^{p_k}  \|    \partial  \curl_k (\X_{.j}^{(k)},\tilde x_j^{(k)} \cdot \textbf 1_n    ) \|_2  \cdot \| \Theta_{.j}^{(k)}  -  \tilde x_j^{(k)} \cdot \textbf 1_n  \|_2 
\end{align*}
Hence
\begin{align*}
\sum_{k=1}^K \sum_{i=1}^n \pi_k \langle \partial  \curl_k (\X_{i.}^{(k)},\tilde \bx^{(k)}), \Theta_{i.}^{(k)} \rangle \geq  -  \sum_{k=1}^K \sum_{j=1}^{p_k}  \pi_k \|    \partial  \curl_k (\X_{.j}^{(k)},\tilde x_j^{(k)} \cdot \textbf 1_n    ) \|_2  \cdot \| \Theta_{.j}^{(k)}  -  \tilde x_j^{(k)} \cdot \textbf 1_n  \|_2 
\end{align*}

Take $\alpha$ sufficiently large so that
\begin{align*}
    \alpha \sum_{k=1}^K \sum_{j=1}^{p_k}  \zeta_j^{(k)} \| \Theta_{.j}^{(k)}  - \tilde x_j^{(k)} \cdot \textbf 1_n  \|_2 \geq  -  \sum_{k=1}^K \sum_{j=1}^{p_k}  \pi_k \|    \partial  \curl_k (\X_{.j}^{(k)},\tilde x_j^{(k)} \cdot \textbf 1_n    ) \|_2  \cdot \| \Theta_{.j}^{(k)}  -  \tilde x_j^{(k)} \cdot \textbf 1_n  \|_2 
\end{align*}

When all $\zeta_j^{(k)}  > 0$, one can take any $\alpha$ that exceeds 
\begin{align*}
    \max_{j,k} \frac{ \pi_k \|    \partial  \curl_k (\X_{.j}^{(k)},\tilde x_j^{(k)} \cdot \textbf 1_n    ) \|_2 }{\zeta_j^{(k)}}
\end{align*}

In general, set $\alpha \geq \frac{1}{\min_{\zeta_j^{(k)} > 0} \zeta_j^{(k)}} \cdot \max_{j,k}  \pi_k \|    \partial  \curl_k (\X_{.j}^{(k)},\tilde x_j^{(k)} \cdot \textbf 1_n    ) \|_2  $. It is easy to check the forward directional derivative test is satisfied. $\square$

\section{Proof of Theorem \ref{theorem:inexact-full}}\label{algcovproof}

Recall the iGecco+ problem is:
\begin{align*}
\min_{\U^{(k)}} \hspace{2mm} &\sum_{k=1}^K \pi_k \curl_k(\X^{(k)},\U^{(k)})  + \gamma \sum_{i < i'} w_{ii'}  \sqrt{ \sum_{k=1}^K  \| \U_{i.}^{(k)} - \U_{i'.} ^{(k)} \|^2 }  \\
&  + \alpha \sum_{k=1}^K \sum_{j=1}^{p_k}  \zeta_j^{(k)} \| \U_{.j}^{(k)}  - \tilde x_j^{(k)} \cdot \textbf 1_n  \|_2 
\end{align*}

We can recast the orginal iGecco+ problem as a multi-block ADMM form:
\begin{align}  
&\minimize_{\U^{(k)},\V} \hspace{5mm}   \sum_{k=1}^K \pi_k \curl_k(\X^{(k)},\U^{(k)}) + \gamma \underbrace{ \bigg(\sum_{l \in \mathcal E} w_l \|\V_{l.}\|_2\bigg)}_{P_1(\V;\bw)}  \nonumber  \\
& \hspace{20mm}+ \alpha \sum_{k=1}^K \sum_{j=1}^{p_k}  \zeta_j^{(k)} \| \U_{.j}^{(k)}  - \tilde x_j^{(k)} \cdot \textbf 1_n  \|_2 + \sum_{k'=1}^K \pi_{k'} f_{k'}(\Z^{(k')} )  + \alpha \sum_{k'=1}^K \bigg( \underbrace{ \sum_{j=1}^{p_k'}  \zeta_j^{(k')} \| \br_j^{(k')} \|_2}_{P_2(\R^{(k')};\bzeta^{(k')})} \bigg) \nonumber \\
& \subto \hspace{3mm}  \D \U^{(k)} - \V^{(k)} = 0, \hspace{2mm}  \D \U^{(k')} - \V^{(k')} = 0, \hspace{2mm} \X^{(k')} - \U^{(k')} = \Z^{(k')},  \hspace{2mm} \U^{(k')}  - \tilde \X^{(k')}  = \R^{(k')} \label{eq:2}
\end{align}
where $\curl_k$ refers to the differentiable losses and $\curl_{k'}$ refers to the non-differentiable losses. Hence we have the multi-block ADMM algorithm (Algorithm~\ref{alg:full-igecco+}) to solve the problem above:

\begin{algorithm}[h]
	\caption{Multi-block ADMM to solve iGecco+}
	\label{alg:full-igecco+}
	\begin{algorithmic}
		
	\WHILE{not converged}
	
	\FOR{all $k = 1,\cdots,K$}
		\STATE 	$\U^{(k)} = \argmin \limits_{\U} \hspace{1mm} \pi_k \curl_k(\X^{(k)},\U)   + \frac{\rho}{2}   \| \D \U - \V^{(k)} + \bLambda^{(k)} \|_F^2    +  \alpha  \sum_{j=1}^{p_k} \zeta_j^{(k)} \| \U_{.j}  - \tilde x_j^{(k)} \cdot \textbf 1_n  \|_2  $ 
		
		\STATE 	$\U^{(k')} = \argmin \limits_{\U} \hspace{1mm}  \frac{\rho}{2} \| \X^{(k')} - \U - \Z^{(k')} +  \bKSI^{(k')} \|_F^2 + \frac{\rho}{2} \|  \U - \tilde \X^{(k')} - \R^{(k')} + \N^{(k')}  \|_F^2 + \frac{\rho}{2} \| \D \U - \V^{(k')} + \bLambda^{(k')} \|_F^2 $

		\STATE $\Z^{(k')} = \argmin \limits_{\Z} \hspace{1mm}  \pi_{k'} f_{k'}(\Z ) +  \frac{\rho}{2} \| \X^{(k')} - \U^{(k')} - \Z +  \bKSI^{(k')} \|_F^2 $ 
		\STATE $\R^{(k')} =  \argmin \limits_{\R} \hspace{1mm}  \alpha  \sum_{j=1}^{p_{k'}}  \zeta_j^{(k')} \| \U_{.j}^{(k')}  - \tilde x_j^{(k')} \cdot \textbf 1_n  \|_2 + \frac{\rho}{2} \|  \U^{(k')}  - \tilde \X^{(k')} - \R+ \N^{(k')}  \|_F^2    $ 
		\STATE $\bKSI^{(k')} = \bKSI^{(k')} + (\X^{(k')} - \U^{(k')} - \Z^{(k')})  $
		\STATE 	$\N^{(k')} = \N^{(k')} +   ( \U^{(k')} - \tilde \X^{(k')} - \R^{(k')} )   $ 

			\ENDFOR

		\STATE 		$\V = \argmin \limits_{\V} \hspace{1mm}  \frac{\rho}{2}   \| \D \U^{(k)} - \V^{(k)} + \bLambda^{(k)} \|_F^2   +  \frac{\rho}{2}   \| \D \U^{(k')} - \V^{(k')} + \bLambda^{(k')} \|_F^2 +   \gamma  \bigg(\sum_{l \in \mathcal E} w_l \|\V_{l.}\|_2\bigg) $ 
		\STATE $\bLambda^{(k)} =  \bLambda^{(k)} +  ( \D \U^{(k)} - \V^{(k)}  ) $ \hspace{5mm} for all $k$ and $k'$
		\ENDWHILE

	\end{algorithmic}
\end{algorithm}

To prove convergence of Algorithm~\ref{alg:inexact-admm}, we first show that multi-block ADMM Algorithm~\ref{alg:full-igecco+} converges to a global minimum. Then we show that we can proximal-linearize the sub-problems in the primal updates of Algorithm~\ref{alg:full-igecco+}  with proved convergence  and this is equivalent to Algorithm~\ref{alg:inexact-admm}.

Without loss of generality, we assume we have one differentiable loss $\curl_1(\cdot)$ and one non-differentiable distance-based loss $\curl_2(\cdot)$.

To prove convergence of Algorithm~\ref{alg:full-igecco+}, we first propose a sufficient condition for the convergence of four-block ADMM and prove it holds true. This is an extension of the convergence results in Section 2 of the work by \cite{chen2016direct}. Suppose the convex optimization problem  with linear constraints we want to minimize is
\begin{align}
\min \hspace{2mm} &\theta_1(\bx_1) + \theta_2(\bx_2) + \theta_3(\bx_3) + \theta_4(\bx_4) \nonumber \\
\text{s.t} \hspace{2mm} & \A_1 \bx_1 + \A_2 \bx_2 + \A_3 \bx_3 + \A_4 \bx_4 = \bb \label{eq:3}
\end{align}

The multi-block ADMM has the following form. Note here, the superscript $\bx_i^{(k+1)}$ refers to the $(k+1)^{th}$ iteration in the ADMM updates.
\begin{align}
\begin{cases}\label{eq:4} 
\bx_1^{(k+1)} = \argmin \hspace{1mm}  \{ L_\A(\bx_1,\bx_2^{(k)},\bx_3^{(k)},\bx_4^{(k)},\blambda^{(k)})\}   \\
\bx_2^{(k+1)} = \argmin \hspace{1mm}  \{
L_\A(\bx_1^{(k+1)},\bx_2,\bx_3^{(k)},\bx_4^{(k)},\blambda^{(k)})\}   \\
\bx_3^{(k+1)} = \argmin \hspace{1mm}  \{
L_\A(\bx_1^{(k+1)},\bx_2^{(k+1)},\bx_3,\bx_4^{(k)},\blambda^{(k)})\}  \\
\bx_4^{(k+1)} = \argmin \hspace{1mm}  \{
L_\A(\bx_1^{(k+1)},\bx_2^{(k+1)},\bx_3^{(k+1)},\bx_4,\blambda^{(k)})\}   \\
\blambda^{(k+1)}  = \blambda^{(k)} - (\A_1 \bx_1^{(k+1)}  + \A_2 \bx_2^{(k+1)}  + \A_3 \bx_3^{(k+1)}  + \A_4 \bx_4^{(k+1)}   - \bb     )
 \end{cases} 
\end{align}
where
\begin{align*}
L_\A = \sum_{i=1}^4 \theta_i (\bx_i) - \blambda^T (\A_1 \bx_1  + \A_2 \bx_2  + \A_3 \bx_3  + \A_4 \bx_4   - \bb  ) + \frac{1}{2} \| \A_1 \bx_1  + \A_2 \bx_2  + \A_3 \bx_3  + \A_4 \bx_4   - \bb \|_2^2
\end{align*}

We establish Lemma \ref{theorem:four_block},  a sufficient condition for convergence of four-block ADMM:
\begin{restatable}{lemma}{lemma-four_block}
\label{theorem:four_block}(Sufficient Condition for Convergence of Four-block ADMM)
A sufficient condition ensuring the convergence of \eqref{eq:4} to a global solution of \eqref{eq:3} is: $\A_2^T \A_3 = \mathbf 0$, $\A_2^T \A_4 = \mathbf 0$, $\A_3^T \A_4 = \mathbf 0$.
\end{restatable}

\noindent \textbf{Proof of Lemma \ref{theorem:four_block}}:

According to the first-order optimality conditions of the minimization problems in \eqref{eq:4}, we have:
\begin{footnotesize}
\begin{align*}
\begin{cases}
\theta_1(\bx_1) - \theta_1(\bx_1^{(k+1)}) + (\bx_1 - \bx_1^{(k+1)})^T \big\{   -\A_1^T [ \blambda^{(k)} - (\A_1 \bx_1^{(k+1)}  + \A_2 \bx_2^{(k)}  + \A_3 \bx_3^{(k)}  + \A_4 \bx_4^{(k)}   - \bb)]   \big\}  \geq 0  \\  
\theta_2(\bx_2) - \theta_2(\bx_2^{(k+1)}) + (\bx_2 - \bx_2^{(k+1)})^T \big\{   -\A_2^T [ \blambda^{(k)} - (\A_1 \bx_1^{(k+1)}  + \A_2 \bx_2^{(k+1)}  + \A_3 \bx_3^{(k)}  + \A_4 \bx_4^{(k)}   - \bb)]   \big\}  \geq 0  \\ 
\theta_3(\bx_3) - \theta_3(\bx_3^{(k+1)}) + (\bx_3 - \bx_3^{(k+1)})^T \big\{   -\A_3^T [ \blambda^{(k)} - (\A_1 \bx_1^{(k+1)}  + \A_2 \bx_2^{(k+1)}  + \A_3 \bx_3^{(k+1)}  + \A_4 \bx_4^{(k)}   - \bb)]   \big\}  \geq 0  \\ 
\theta_4(\bx_4) - \theta_4(\bx_4^{(k+1)}) + (\bx_4 - \bx_4^{(k+1)})^T \big\{   -\A_4^T [ \blambda^{(k)} - (\A_1 \bx_1^{(k+1)}  + \A_2 \bx_2^{(k+1)}  + \A_3 \bx_3^{(k+1)}  + \A_4 \bx_4^{(k+1)}   - \bb)]   \big\}  \geq 0  \\ 
\end{cases}
\end{align*}
\end{footnotesize}

Since $\A_2^T \A_3 = \mathbf 0$, $\A_2^T \A_4 = \mathbf 0$, $\A_3^T \A_4 = \mathbf 0$, we have:
\begin{small}
\begin{align*}
\begin{cases}
\theta_1(\bx_1) - \theta_1(\bx_1^{(k+1)}) + (\bx_1 - \bx_1^{(k+1)})^T \big\{   -\A_1^T [ \blambda^{(k)} - (\A_1 \bx_1^{(k+1)}  + \A_2 \bx_2^{(k)}  + \A_3 \bx_3^{(k)}  + \A_4 \bx_4^{(k)}   - \bb)]   \big\}  \geq 0  \\  
\theta_2(\bx_2) - \theta_2(\bx_2^{(k+1)}) + (\bx_2 - \bx_2^{(k+1)})^T \big\{   -\A_2^T [ \blambda^{(k)} - (\A_1 \bx_1^{(k+1)}  + \A_2 \bx_2^{(k+1)}    - \bb)]   \big\}  \geq 0  \\ 
\theta_3(\bx_3) - \theta_3(\bx_3^{(k+1)}) + (\bx_3 - \bx_3^{(k+1)})^T \big\{   -\A_3^T [ \blambda^{(k)} - (\A_1 \bx_1^{(k+1)}    + \A_3 \bx_3^{(k+1)}    - \bb)]   \big\}  \geq 0  \\ 
\theta_4(\bx_4) - \theta_4(\bx_4^{(k+1)}) + (\bx_4 - \bx_4^{(k+1)})^T \big\{   -\A_4^T [ \blambda^{(k)} - (\A_1 \bx_1^{(k+1)}   + \A_4 \bx_4^{(k+1)}   - \bb)]   \big\}  \geq 0  
\end{cases}
\end{align*}
\end{small}
which is also the first-order optimality condition of the scheme:
\begin{align}
\begin{cases} \label{eq:5}
\bx_1^{(k+1)} = \argmin \hspace{1mm}  \{ \theta_1(\bx_1) - (\blambda^{(k)})^T (\A_1 \bx_1)   +   \frac{1}{2} \| \A_1 \bx_1  + \A_2 \bx_2^{(k)}  + \A_3 \bx_3^{(k)}  + \A_4 \bx_4^{(k)}   - \bb \|_2^2          \}    \\
(\bx_2^{(k+1)}, \bx_3^{(k+1)},\bx_4^{(k+1)})  = \argmin \hspace{1mm}  \{ \theta_2(\bx_2) + \theta_3(\bx_3) + \theta_4(\bx_4)   - (\blambda^{(k)})^T (\A_2 \bx_2 + \A_3 \bx_3 + \A_4 \bx_4) \\
\hspace{35mm}  +   \frac{1}{2} \| \A_1 \bx_1^{(k+1)}  + \A_2 \bx_2  + \A_3 \bx_3  + \A_4 \bx_4  - \bb \|_2^2             \}   \\
\blambda^{(k+1)}  = \blambda^{(k)} - (\A_1 \bx_1^{(k+1)}  + \A_2 \bx_2^{(k+1)}  + \A_3 \bx_3^{(k+1)}  + \A_4 \bx_4^{(k+1)}   - \bb     )
\end{cases} 
\end{align}

Clearly, \eqref{eq:5} is a specific application of the original two-block ADMM  to \eqref{eq:3} by regarding $(\bx_2, \bx_3, \bx_4)$ as one variable, $[\A_2,\A_3,\A_4]$ as one matrix and $\theta_2 (\bx_2) + \theta_3(\bx_3) + \theta_4 (\bx_4)$ as one function.  $\square$.

Note that Lemma \ref{theorem:four_block} is stated in vector form and therefore we need to transform the constraints in the original iGecco+ problem   \eqref{eq:2} from matrix form to vector form in order to apply Lemma \ref{theorem:four_block}. Note that $\D \U^{(k)}  = \V^{(k)} \Leftrightarrow {\U^{(k)}}^T \D^T = {\V^{(k)}}^T \Leftrightarrow  (\D \otimes \textbf I_{p_k}) \text{vec} ({\U^{(k)}}^T ) = \text{vec} ({\V^{(k)}}^T )$. 


Hence we can write the constraints in \eqref{eq:2} as:
\begin{align*}
\begin{pmatrix}
\A_1 & \textbf 0 \\  \textbf 0  & \textbf I  \\  \textbf 0 &  \A_2  \\  \textbf 0  & \textbf I   
\end{pmatrix} \bu +  \begin{pmatrix}
 \textbf 0 \\ \textbf I \\ \textbf 0  \\ \textbf 0
\end{pmatrix}  \bz +  \begin{pmatrix}
\textbf 0 \\ \textbf 0   \\  \textbf 0  \\  - \textbf I 
\end{pmatrix}  \br  +   \begin{pmatrix}
- \textbf I & \textbf 0   \\ \textbf 0 & \textbf 0 \\ \textbf 0  &  - \textbf I \\ \textbf 0 & \textbf 0
\end{pmatrix}  \bv = \textbf b  
\end{align*}

where $\bu =  \begin{pmatrix} \bu_1  \\ \bu_2    \end{pmatrix} =  \begin{pmatrix} \text{vec}( {\U^{(1)}} ^T)  \\ \text{vec}({\U^{(2)}}^T)    \end{pmatrix} $. $\A_1 = \D \otimes \textbf I_{p_1}$, $\A_2 = \D \otimes \textbf I_{p_2}$. $\bz = \text{vec}(\Z^T)$, $\br = \text{vec}(\R^T)$.

$\bv = \text{vec} (\V^T) =  \begin{pmatrix} \text{vec}( {\V^{(1)}}^T)  \\ \text{vec}( {\V^{(2)}} ^T)    \end{pmatrix} $, $\textbf b = \begin{pmatrix}
\textbf 0_{p_1 \times |\mathcal E|}  \\ \text{vec}( {\X^{(2)}} ^T) \\ \textbf 0_{p_2 \times |\mathcal E|}  \\  \tilde \bx^{(2)} \\ \vdots \\ \tilde \bx^{(2)}
\end{pmatrix} $, $\tilde \bx \in \mathbb{R}^{p_2}$ is a column vector consisting of all $\tilde x^{(2)}_j$ and is repeated $n$ times in $\textbf b$.

Next we show that the constraint sets in \eqref{eq:2} for our problem satisfies the condition in Lemma \ref{theorem:four_block} and hence the multi-block ADMM Algorithm~\ref{alg:full-igecco+} converges.

By construction, $\E_2 = \begin{pmatrix}
 \textbf 0 \\ \textbf I \\ \textbf 0  \\ \textbf 0
\end{pmatrix}$, $\E_3 =    \begin{pmatrix}
\textbf 0 \\ \textbf 0   \\  \textbf 0  \\  - \textbf I 
\end{pmatrix}$ and $\E_4 =  \begin{pmatrix}
- \textbf I & \textbf 0   \\ \textbf 0 & \textbf 0 \\ \textbf 0  &  - \textbf I \\ \textbf 0 & \textbf 0
\end{pmatrix} $.

It is easy to verify that:

$\E_2^T \E_3 = \textbf 0$, $\E_2^T \E_4 = \textbf 0$, $\E_3^T \E_4 = \textbf 0$.

Hence our setup satisfies the sufficient condition in Lemma \ref{theorem:four_block} and hence the multi-block ADMM Algorithm~\ref{alg:full-igecco+}  converges. 

Next, we see that each primal update in Algorithm~\ref{alg:inexact-admm} is equivalent to the primal update by applying proximal linearized ADMM to the sub-problems in Algorithm~\ref{alg:full-igecco+}. (We will show this in detail in Theorem~\ref{theorem:diff}.)  It is easy to show that those updates with closed-form solutions are special cases of proximal-linearizing the sub-problems. \cite{lu2016fast,liu2013linearized}  showed the convergence of proximal linearized multi-block ADMM.  Hence Algorithm~\ref{alg:inexact-admm} converges to a global minimum if $\curl_k$ is convex for all $k$ and has Lipschitz gradient when it is differentiable. Further, if each $\curl_k$ is strictly convex, it converges to the unique global solution. $\square$.

\section{Gecco+ for Differentiable Losses}\label{diffdetail}
In this section, we propose algorithms to solve Gecco+ when the loss $\curl$ is differentiable and gradient is Lipschitz continuous. In this case, we develop a fast two-block ADMM algorithm  without fully solving the $\U$ sub-problem. Our result is closely related to the proximal linearized ADMM literature \citep{liu2013linearized,lu2016fast}. Also solving the sub-problem approximately is closely connected with the generalized ADMM literature \citep{deng2016global}. 

In the following sections, we discuss algorithms to solve Gecco+ instead of iGecco+ for notation purposes as we would like to include iteration counter indices in the algorithm for illustrating backtracking; but we can easily extend the algorithm to solve iGecco+. To begin with, we clarify different notations in Gecco+ and iGecco+: $\U^{(k)}$ in iGecco+ refers to the $k^{th}$ data source  while $\U^{(k)}$ in Gecco+ refers to the $k^{th}$ iteration counter in the ADMM updates.  We omit iteration counter indices in all iGecco+ algorithm for notation purposes and use the most current iterates.

\subsection{Two-block ADMM in Matrix Form}

Suppose the loss $\curl(\X,\U)$ is differentiable.
Similar to the formulation in convex clustering, we can recast the Gecco+ problem as the equivalent constrained problem:
\begin{align*}
&\minimize_{\U,\V} \hspace{5mm}  \curl(\X,\U) + \gamma \underbrace{ \bigg(\sum_{l \in \mathcal E} w_l \|\V_{l.} \|_2 \bigg)}_{P_1(\V;\bw)} + \alpha \sum_{j=1}^p  \zeta_j \| \U_{.j}   - \tilde x_j \cdot \textbf 1_n  \|_2 \\
& \subto \hspace{5mm} \D \U - \V = 0
\end{align*}

Like in convex clustering \citep{chi2015splitting,weylandt2019dynamic}, we index a centroid pair by $l = (l_1, l_2)$ with $l_1 < l_2$, define the set of edges over the non-zero weights $\mathcal E = \{l = (l_1,l_2) : w_l > 0\}$, and introduce a new variable $\V_{l.} = \U_{l_1.} - \U_{l_2.} $ to account for the difference between the two centroids. Hence $\V$ is a matrix containing the pairwise differences between connected rows of $\U$. $\D$ is the difference matrix defined in the work of \cite{weylandt2019dynamic}.

We can show that the augmented Lagrangian in scaled form is equal to:
\begin{align*}
L(\U,\V,\bLambda) &=  \curl(\X,\U)   + \frac{\rho}{2}  \| \D \U -  \V + \bLambda  \|_F^2    +  \alpha \sum_{j=1}^p \zeta_j \| \U_{.j}  - \tilde x_j \cdot \textbf 1_n  \|_2    + \gamma \sum_{l \in \mathcal E} w_l \|\V_{l.} \|_2 
\end{align*}
where $\D$ is the difference matrix and the dual variable is denoted by $\bLambda$.

To update $\U$, we need to solve the following sub-problem:
\begin{align*}
\minimize_{\U} \hspace{5mm}     \curl(\X,\U)   + \frac{\rho}{2}   \| \D \U -  \V + \bLambda \|_2^2    +  \alpha   \sum_{j=1}^p  \zeta_j \| \U_{.j}  - \tilde x_j \cdot \textbf 1_n  \|_2  
\end{align*}
Let $\tilde \U =  \U  - \tilde \X$. The sub-problem becomes:
\begin{align*}
\minimize_{\tilde \U} \hspace{5mm}    \curl(\X,\tilde \U  +\tilde \X)     + \frac{\rho}{2} \| \D ( \tilde \U  +\tilde \X) - \V + \bLambda \|_2^2 +  \alpha  \sum_{j=1}^p  \zeta_j \|\tilde  \bu_j  \|_2
\end{align*}
where $\tilde  \bu_j$ is the $j^{th}$ column of $\tilde \U$. For each ADMM iterate, we have:
\begin{align*}
\tilde \U^{(k)} = \argmin_{\tilde \U} \hspace{5mm}    \curl(\X,\tilde \U  +\tilde \X)     + \frac{\rho}{2} \| \D ( \tilde \U  +\tilde \X) - \V^{(k-1)} + \bLambda^{(k-1)} \|_2^2 +  \alpha  \underbrace{\sum_{j=1}^p  \zeta_j \|\tilde  \bu_j  \|_2}_{P_2(\tilde  \U;\bzeta)}
\end{align*}
This can be solved by running iterative proximal gradient to full convergence:
\begin{align*}
    \tilde \U^{(k,m)} = \prox_{s_k  \cdot \alpha P_2(\cdot;\bzeta)} \begin{footnotesize} \big( \tilde \U^{(k,m-1)}   - s_k \cdot [ 
		\nabla  \curl(\X,\tilde \U^{(k,m-1)} + \tilde \X ) +  \rho \D^T (\D  (\tilde \U^{(k,m-1)} +  \tilde \X )   -  \V^{(k-1)} + \bLambda^{(k-1)} )] \big) \end{footnotesize} 
\end{align*}

which is equivalent to:
\begin{align*}
    \U^{(k,m)} = \prox_{s_k  \cdot \alpha P_2(\cdot;\bzeta)} \begin{footnotesize} \big( \U^{(k,m-1)} - \tilde \X   - s_k \cdot [ 
		\nabla  \curl(\X,\U^{(k,m-1)} ) +  \rho \D^T (\D  \U^{(k,m-1)}   -  \V^{(k-1)} + \bLambda^{(k-1)} )] \big) + \tilde \X \end{footnotesize} 
\end{align*}

Here $\U^{(k,m)}$ refers to the $m^{th}$ inner iteration counter in the $\U$ sub-problem out of the $k^{th}$ outer iteration counter of the ADMM update. It is straightforward that this is computationally expensive. To address this, we propose to solve the $\U$ sub-problem approximately using just a one-step proximal gradient update and prove convergence in the next section. This approach is based on proximal linearized ADMM \citep{liu2013linearized,lu2016fast}, which solves the sub-problems efficiently by linearizing the differentiable part and then applying proximal gradient due to the non-differentiable part. To ensure convergence, the algorithm requires that gradient shoule be  Lipschitz continuous. The $\V$ and $\bLambda$ updates are just the same as in regular convex clustering.

We adopt such an approach and develop the proximal linearized 2-block ADMM (Algorithm \ref{alg:gecco+_diff_matrix}) to solve Gecco+ when the loss is differentiable and gradient is Lipschitz continuous.

\begin{algorithm}[H]
	\caption{Proximal Linearized 2-block ADMM when the loss is differentiable and gradient is Lipschitz continuous --- Matrix Form}
	\label{alg:gecco+_diff_matrix}
	\begin{algorithmic}
		\WHILE{not converged}

		\STATE $\U^{(k)} = \prox_{s_k  \cdot \alpha P_2(\cdot;\bzeta)} \begin{footnotesize} \big( \U^{(k-1)} - \tilde \X   - s_k \cdot [ 
		\nabla  \curl(\X,\U^{(k-1)} ) +  \rho \D^T (\D  \U^{(k-1)}   -  \V^{(k-1)} + \bLambda^{(k-1)} )] \big) + \tilde \X \end{footnotesize}  $

		\STATE 		$\V^{(k)} = \text{prox}_{\gamma /\rho P_1(\cdot; \bw )} (\D \U^{(k)} + \bLambda^{(k-1)})$ 
		\STATE $\bLambda^{(k)} =  \bLambda^{(k-1)} +  (   \D \U^{(k)}   - \V^{(k)} ) $ 

		\ENDWHILE
	\end{algorithmic} 
\end{algorithm}


Further, if the $\U$ sub-problem can be decomposed to $p$ separate  $\U_{.j}$ sub-problems where the augmented Lagrangian for each now is a sum of a differentiable loss, a quadratic term and a sparse group-lasso penalty, we propose to use proximal gradient descent for each separate $\U_{.j}$ sub-problem. In this way, we yield adaptive step-size for each $\U_{.j}$ sub-problem and hence our algorithm enjoys better convergence property than updating $\U$'s all together. (In the latter case, the step size becomes fairly small as we are moving all $\U$ to some magnitude in the direction of negative gradient.) To achieve this, we assume that the loss is elementwise, which means we can write the loss function as a sum of $p$ terms. (The loss can be written as $\sum_i \curl(\X_{i.},\U_{i.}) = \sum_j \curl(\X_{.j},\U_{.j})  = \sum_i \sum_j q(x_{ij},u_{ij})$ where $q$ is the element-wise version of the loss while $\curl$ is the vector-wise version of the loss.) We see that every deviance-based loss satisfies this assumption. Moreover, by decomposing to  $p$ sub-problems, we can solve each in parallel which saves computation cost. We describe in detail how to solve each $\U_{.j}$ sub-problem in the next subsection.

\subsection{Two-block ADMM in Vector Form in Parallel}

Suppose the $\U$ sub-problem can be decomposed to $p$ separate  $\U_{.j}$ sub-problems mentioned above. The augmented Lagrangian now becomes:
\begin{align*}
L(\U,\V,\bLambda) &= \sum_{j=1}^p  \curl(\X_{.j},\U_{.j})   + \frac{\rho}{2} \sum_{j=1}^p  \| \D \U_{.j} -  \V_{.j} + \bLambda_{.j} \|_2^2  \\
&+  \alpha \sum_{j=1}^p \zeta_j \| \U_{.j}  - \tilde x_j \cdot \textbf 1_n  \|_2    + \gamma \sum_{l \in \mathcal E} w_l \|\V_{l.} \|_2 
\end{align*}
In this way we can perform block-wise minimization. Now minimizing the augmented Lagrangian over $\U$ is equivalent to minimizing over each $\U_{.j}$, $j = 1,\cdots,p$:
\begin{align*}
\minimize_{\U_{.j}} \hspace{5mm}     \curl(\X_{.j},\U_{.j})   + \frac{\rho}{2}   \| \D \U_{.j} -  \V_{.j} + \bLambda_{.j} \|_2^2    +  \alpha  \zeta_j \| \U_{.j}  - \tilde x_j \cdot \textbf 1_n  \|_2  
\end{align*}

Let $\tilde \bu_j =  \U_{.j}  - \tilde x_j \cdot \textbf 1_n$. The problem above becomes:
\begin{align*}
\minimize_{\tilde \bu_j} \hspace{5mm}    \curl(\X_{.j},\tilde \bu_j  +\tilde x_j \cdot \textbf 1_n)     + \frac{\rho}{2} \| \D ( \tilde \bu_j  +\tilde x_j \cdot \textbf 1_n) - \V_{.j} + \bLambda_{.j} \|_2^2 +  \alpha \zeta_j \|\tilde  \bu_j  \|_2
\end{align*}
Similarly, this can be solved by running iterative proximal gradient to full convergence. However, as mentioned above, we propose to solve the $\U$ sub-problem approximately using just a one-step proximal gradient update and prove convergence. Still this approach is based on proximal linearized ADMM \citep{liu2013linearized,lu2016fast}, which solves the sub-problems efficiently by linearizing the differentiable part and then applying proximal gradient due to the non-differentiable part. To ensure convergence, the algorithm requires that gradient shoule be  Lipschitz continuous.

We propose Algorithm~\ref{alg:diff_vec} to solve Gecco+ when $\curl$ is differentiable and gradient is Lipschitz continuous in vector form. Note the $\U$ update in Algorithm~\ref{alg:diff_vec} is  a just a vectorized version of that in Algorithm~\ref{alg:gecco+_diff_matrix} if we use fixed step size $s_k$ for each feature $j$. We use the vector form update here since it enjoys better convergence property mentioned above and we use this form for proof of convergence. Next we prove the convergence of Algorithm~\ref{alg:diff_vec}.

\begin{algorithm}[H]
	\caption{Proximal Linearized 2-block ADMM when the loss is differentiable and gradient is Lipschitz continuous --- Vector Form in parallel}
	\label{alg:diff_vec}
	\begin{algorithmic}
		\STATE {\bfseries Input:} $\mathbf{X}$, $\gamma$, $\mathbf w$, $\alpha$, $\bzeta$
		\STATE {\bfseries Initialize:}  $\U^{(0)},\V^{(0)},\mathbf \Lambda^{(0)}$\\
		\STATE {\bfseries Precompute:} Difference matrix $\D$, $\tilde x_j$
		\WHILE{not converged}
		
		\FOR{$j=1$ {\bfseries to} $p$}
		
		\STATE $\U_{.j}^{(k)} = \prox_{s_k \cdot \alpha \zeta_j \|\cdot \|_2} \big  (\U_{.j}^{(k-1)} - \tilde x_j \cdot \textbf 1_n  - s_k \cdot [ 
		\nabla  \curl(\X_{.j},\U_{.j}^{(k-1)} ) +  \rho \D^T (\D  \U_{.j}^{(k-1)}   -  \V_{.j}^{(k-1)} + \bLambda_{.j}^{(k-1)} )] \big) + \tilde x_j \cdot \textbf 1_n$
		\ENDFOR

		\STATE 		$\V^{(k)} = \text{prox}_{\gamma /\rho P_1(\cdot; \bw )} (\D \U^{(k)} + \bLambda^{(k-1)})$ 
		\STATE $\bLambda^{(k)} =  \bLambda^{(k-1)} +  (   \D \U^{(k)}   - \V^{(k)} ) $ 

		\ENDWHILE
		\STATE {\bfseries Output:} $\U^{(k)}$.
	\end{algorithmic} 
\end{algorithm}

\subsection{Proof of Convergence}

\begin{restatable}{theorem}{theoremdiff}
\label{theorem:diff}
If $\curl$ is convex and differrentiable and  $\nabla \curl$ is Lipschitz continuous,  then Algorithm~\ref{alg:diff_vec} converges to a global solution. Further, if  $\curl$ is strictly convex, it converges to the unique global solution.
\end{restatable}


\noindent \textbf{Proof:}
We will show that the $\U$ sub-problem update in Algorithm~\ref{alg:diff_vec} is equivalent to linearizing the $\U$ sub-problem and then applying a proxmial operator, which is proximal linearized ADMM.

Note that each $\U_{.j}$ sub-problem is:
\begin{align*}
\U_{.j}^{(k+1)} =  \argmin_{\U_{.j}} \hspace{5mm}     \curl(\X_{.j},\U_{.j})   + \frac{\rho}{2}   \| \D \U_{.j} - \V_{.j}^{(k)} + \bLambda_{.j}^{(k)} \|_2^2    +  \alpha  \zeta_j \| \U_{.j}  - \tilde x_j \cdot \textbf 1_n  \|_2  
\end{align*}
For simplicity of notation, we replace $\U_{.j}$ with 
$\bu_j$ in the following.
\begin{align*}
\bu_j^{(k+1)} = \argmin_{\bu_j} \hspace{5mm}    \curl(\X_{.j},  \bu_j )     + \frac{\rho}{2} \| \D  \bu_j  - \V_{.j}^{(k)} + \bLambda_{.j}^{(k)} \|_2^2 +  \alpha \zeta_j  \|  \bu_j  - \tilde x_j \cdot \textbf 1_n \|_2
\end{align*}
Rearranging terms, we have:
\begin{align*}
\bu_j^{(k+1)} = \argmin_{\bu_j}  \hspace{5mm}    \curl(\X_{.j}, \bu_j)     + \frac{\rho}{2} \| \D  \bu_j  - \V_{.j}^{(k)} \|_2^2 +  \rho {\bLambda^{(k)}_{.j}}^T \big(\D   \bu_j   - \V_{.j}^{(k)} \big)  +  \alpha \zeta_j \|   \bu_j  - \tilde x_j \cdot \textbf 1_n   \|_2
\end{align*}
According to the proximal linearized ADMM with parallel splitting algorithm (see Algorithm 3 in \citet{liu2013linearized} and Equation (14) in \citet{lu2016fast}), we can linearize the first two terms and add a quadratic term in the objective:
\begin{align*}
\bu_j^{(k+1)} = \argmin_{\bu_j} \hspace{5mm}    &\curl(\X_{.j},  \bu_j^{(k)} )     + \langle \nabla \curl(\X_{.j}, \bu_j^{(k)}  ) ,   \bu_j -  \bu_j^{(k)} \rangle \\
&+  \langle \rho   \D^T \big( \D    \bu_j^{(k)}  - \V_{.j}^{(k)}  \big)  ,  \bu_j -  \bu_j^{(k)} \rangle  \\
& + \rho  {\bLambda^{(k)}_{.j}}^T \big(\D   \bu_j  - \V_{.j}^{(k)} \big)  +  \alpha \zeta_j \|  \bu_j  - \tilde x_j \cdot \textbf 1_n  \|_2 + \frac{1}{2 s_k} \|  \bu_j -   \bu_j^{(k)}  \|_2^2
\end{align*}
Rearranging terms and removing irrelevant terms, we have:
\begin{align*}
\bu_j^{(k+1)} & = \argmin_{\bu_j} \hspace{5mm} (\nabla g(\bu_j ^{(k)}))^T (\bu_j - \bu_j^{(k)}) + \frac{1}{2 s_k} \|\bu_j - \bu_j^{(k)}\|_2^2 + \alpha \zeta_j  \| \bu_j  - \tilde x_j \cdot \textbf 1_n  \|_2 
\end{align*}
where $\nabla g(\bu_j ^{(k)})  = \nabla \curl(\X_{.j},\bu_j^{(k)}) + \rho \D^T (\D \bu_j^{(k)} -  \V_{.j}^{(k)} + \bLambda_{.j}^{(k)} ) )$.

Let $\hat \bu_j = \bu_j  - \tilde x_j \cdot \textbf 1_n $. We have:
\begin{align*}
\hat \bu_j^{(k+1)} & = \argmin_{\hat \bu_j} \hspace{5mm} (\nabla g(\bu_j ^{(k)}))^T (\hat \bu_j +   \tilde x_j \cdot \textbf 1_n  - \bu_j^{(k)}) + \frac{1}{2 s_k} \|\hat \bu_j +   \tilde x_j \cdot \textbf 1_n - \bu_j^{(k)}\|_2^2 + \alpha  \zeta_j \| \hat \bu_j   \|_2 
\end{align*}

Recall the definition of proximal operator:
\begin{align*}
\bx^{(k+1)} & = \prox_{th} (\bx^{(k)} - t \nabla g(\bx^{(k)}))  \\
&= \argmin_{\bu} \big( h(\bu) + g(\bx^{(k)}) + \nabla g(\bx^{(k)})^T (\bu-\bx^{(k)}) + \frac{1}{2t} \| \bu - \bx^{(k)} \|_2^2 \big)
\end{align*}
Therefore, the $\hat \bu_j$ update is just a  proximal gradient descent update:
\begin{align*}
\hat \bu_j^{(k+1)}  =  \prox_{s_k \cdot \alpha \zeta_j \|\cdot \|_2} \big  (\bu_{j}^{(k)} - \tilde x_j \cdot \textbf 1_n  - s_k \cdot [ 
\nabla \curl(\X_{.j},\bu_{j}^{(k)} ) + \rho \D^T (\D  \bu_{j}^{(k)}  -  \V_{.j}^{(k)} +\bLambda_{.j}^{(k)}  )] \big)
\end{align*}

Now we plug back and get the $\bu_j$, i.e, $(\U_{.j})$ update:
\begin{align*}
\U_{.j}^{(k)}  = \prox_{s_k \cdot \alpha \zeta_j \|\cdot \|_2}&  \big  (\U_{.j}^{(k-1)} - \tilde x_j \cdot \textbf 1_n  - s_k \cdot [ 
\nabla \curl(\X_{.j},\U_{.j}^{(k-1)} ) + \rho \D^T (\D  \U_{.j}^{(k-1)}  - \V_{.j}^{(k-1)} + \bLambda_{.j}^{(k-1)}    )] \big) \\
& + \tilde x_j \cdot \textbf 1_n
\end{align*}

which is equivalent to the $\U_{.j}$ update in Algorithm~\ref{alg:diff_vec}.

The $\V$ and $\bLambda$ update is just the same as the one in convex clustering. Hence Algorithm~\ref{alg:diff_vec} satisfies the condition of proximal linearized ADMM by \cite{liu2013linearized,lu2016fast} and hence converges to a global solution as long as $\nabla \curl$ is Lipschitz continuous. Note that Algorithm~\ref{alg:gecco+_diff_matrix} is equivalent to Algorithm~\ref{alg:diff_vec} with a fixed step size $s_k$. (We can choose $s_k$ to be the minimunm step size $s_k$ over for feature $j$.) Therefore,  Algorithm~\ref{alg:gecco+_diff_matrix} also converges to a global solution as long as $\nabla \curl$ is Lipschitz continuous. Further, if $\curl$ is strictly convex, the optimization problem has unique minimum and hence Algorithm~\ref{alg:gecco+_diff_matrix} and \ref{alg:diff_vec} converges to the global solution. $\square$ 

Note:
\begin{itemize}
    
    \item In \citet{liu2013linearized,lu2016fast}, the algorithm requires that $\nabla \curl$ is Lipschitz continuous to guarantee convergence. We know that $\nabla \curl$ is Lipschitz continuous is equivalent to $\curl$ is strongly smooth. It is easy to show that the hessian of log-likelihood of exponential family and GLM deviance is upper bounded since in (generalized) convex clustering, the value of $\U$ is bounded as it moves along the regularization path from $\X$ to the loss-specific center; also to avoid numerical issues, we add trivial constraint that $u_{ij}>0$ as zero is not defined in the log-likelihood/deviance. Hence the condition for convergence of proximal linearized ADMM is satisfied.
    
    \item To obtain a reasonable step size $s_k$, we need to compute the Lipschitz constant. However, it is non-trivial to calculate the Lipschitz constant for most of our general losses. Instead, we suggest using backtracking line search procedure proposed by \cite{beck2009gradient,parikh2014proximal}, which is a common way to determine step size with guaranteed convergence in optimization. Empirical studies show that choosing step size with backtracking in our framework also ensures convergence. The details for backtracking procedure are discussed below. 
    
    \item For proximal linearized ADMM, \cite{liu2013linearized,lu2016fast} established convergence rate of $O(1/K)$. An interesting future direction might be establishing the linear convergence rate of proximal linearized ADMM when the objective is strongly convex. 
\end{itemize}

\subsection{Backtracking Criterion}
In this section we discuss how to choose the step size $s_k$ in Algorithm~\ref{alg:diff_vec}. As mentioned, usually we employ a fixed step size  by computing the Lipschitz constant as in the squared error loss case; but in our method, it is hard to compute the Lipschitz constant for most of our general losses. Instead, we propose using backtracking line search procedure proposed by \cite{beck2009gradient,parikh2014proximal}, which is a common way to determine step size with guaranteed convergence in optimization. 

Recall the objective function we want to minimize in the $\U$ sub-problem is:
\begin{align*}
f(\tilde \bu_j)  =   \curl(\X_{.j},\tilde \bu_j  +\tilde x_j \cdot \textbf 1_n)     + \frac{\rho}{2} \| \D ( \tilde \bu_j  +\tilde x_j \cdot \textbf 1_n) - \V_{.j} + \bLambda_{.j} \|_2^2 +  \alpha \zeta_j \|\tilde  \bu_j  \|_2
\end{align*}
where $\tilde \bu_j =  \U_{.j}  - \tilde x_j \cdot \textbf 1_n$.

Define:
\begin{align*}
g(\tilde \bu_j)  & =   \curl(\X_{.j},\tilde \bu_j  +\tilde x_j \cdot \textbf 1_n)     + \frac{\rho}{2} \| \D ( \tilde \bu_j  +\tilde x_j \cdot \textbf 1_n) -  \V_{.j} + \bLambda_{.j} \|_2^2  \\
h(\tilde \bu_j)  & =  \alpha \zeta_j \|\tilde  \bu_j  \|_2  \\
G_t(\tilde \bu_j) & = \frac{ \tilde \bu_j - \prox_{t \cdot \alpha \zeta_j \| \cdot \|_2 } ( \tilde \bu_j - t \nabla g (\tilde \bu_j) ) } {t} 
\end{align*}

We adopt the backtracking line search procedure proposed by \cite{beck2009gradient,parikh2014proximal}. At each iteration, while 
\begin{align*}
    g( \tilde \bu_j - t G_t(\tilde \bu_j)) &> g(\tilde \bu_j) - t \nabla g(\tilde \bu_j)^T G_t(\tilde \bu_j) + \frac{t}{2} \| G_t(\tilde \bu_j) \|_2^2 \hspace{10mm} \text{i.e.,}\\
    g( \prox_t ( \tilde \bu_j - t \nabla g (\tilde \bu_j) ) ) &> g(\tilde \bu_j) - \nabla g(\tilde \bu_j)^T  (  \tilde \bu_j - \prox_t ( \tilde \bu_j - t \nabla g (\tilde \bu_j) )  ) \\&+ \frac{1}{2t} \|  \tilde \bu_j - \prox_t ( \tilde \bu_j - t \nabla g (\tilde \bu_j) )  \|_2^2
\end{align*}
shrink $t = \beta t$.

We still adopt the one-step approximation and hence suggest taking a one-step proximal update to solve the $\U$ sub-problem with backtracking. To summarize, we propose Algorithm~\ref{alg:diff-backtracking}, which uses proximal linearized 2-block ADMM with backtracking when the loss is differentiable and gradient is Lipschitz continuous.

\begin{algorithm}[t]
	\caption{Proximal Linearized 2-block ADMM with backtracking when the loss is differentiable and gradient is Lipschitz continuous}
	\label{alg:diff-backtracking}
	\begin{algorithmic}
		\STATE {\bfseries Input:} $\mathbf{X}$, $\gamma$, $\mathbf w$, $\alpha$, $\bzeta$
		\STATE {\bfseries Initialize:}  $\U^{(0)},\V^{(0)},\mathbf \Lambda^{(0)},t$\\
		\STATE {\bfseries Precompute:} Difference matrix $\D$, $\tilde x_j$
		\WHILE{not converged}
		
		\FOR{$j=1$ {\bfseries to} $p$}
		\STATE $t = 1$
		\STATE $\tilde \bu_j^{(k-1)} =  \U_{.j}^{(k-1)}  - \tilde x_j \cdot \textbf 1_n$
		\STATE $\nabla g(\tilde \bu_j^{(k-1)})   =  \nabla \curl(\X_{.j},\tilde \bu_j^{(k-1)}  +\tilde x_j \cdot \textbf 1_n)     + \rho \D^T  \big( \D ( \tilde \bu_j^{(k-1)}  +\tilde x_j \cdot \textbf 1_n) -  \V_{.j}^{(k-1)} + \bLambda_{.j}^{(k-1)} \big) $
		\STATE $\bz = \prox_{t \alpha \zeta_j \|\cdot \|_2} \big(\tilde \bu_j^{(k-1)} - t \nabla g(\tilde \bu_j^{(k-1)})  \big)$
		
		\WHILE{$g(\bz) >g(\tilde \bu_j^{(k-1)}) - \nabla g(\tilde \bu_j^{(k-1)})^T (\tilde \bu_j^{(k-1)} - \bz) + \frac{1}{2t} \| \bz - \tilde \bu_j^{(k-1)}\|_2^2 $}
		
		\STATE $t = \beta t$
		\STATE $\bz = \prox_{t \alpha \zeta_j \|\cdot \|_2} \big(\tilde \bu_j^{(k-1)} - t \nabla g(\tilde \bu_j^{(k-1)})  \big)$
		\ENDWHILE

		\STATE $\U_{.j}^{(k)} = \bz   +  \tilde x_j \cdot \textbf 1_n $
		\ENDFOR

		\STATE 		$\V^{(k)} = \text{prox}_{\gamma /\rho P_1(\cdot; \bw )} (\D \U^{(k)} + \bLambda^{(k-1)})$ 
		\STATE $\bLambda^{(k)} =  \bLambda^{(k-1)} +  (   \D \U^{(k)}   - \V^{(k)} ) $ 

		\ENDWHILE
		\STATE {\bfseries Output:} $\U^{(k)}$.
	\end{algorithmic} 
\end{algorithm}

\subsection{Alternative Algorithm for Differentiable Losses}

It should be pointed out that there are many other methods to solve the $\U$ sub-problem when the loss $\curl$ is differentiable.  We choose to use proximal gradient descent algorithm as there is existing literature on approximately solving the sub-problem using proximal gradient under ADMM with proved convergence \citep{liu2013linearized,lu2016fast}. But there are many other optimization techniques to solve the $\U$ sub-problem such as ADMM. 

In this subsection, we show how to apply ADMM to solve the $\U$ sub-problem and specify under which conditions this method is more favorable. Recall to update $\U$, we need to solve the following sub-problem:
\begin{align*}
    \minimize \limits_{\U}   \hspace{5mm}    \curl(\X,\U)  + \frac{\rho}{2}   \| \D \U -  \V + \bLambda  \|_F^2  +   \alpha  \sum_{j=1}^{p}  \zeta_j \| \U_{.j}   - \tilde x_j \cdot \textbf 1_n  \|_2
\end{align*}
We use ADMM to solve this minimization problem and can now recast the  problem above as the equivalent constrained problem:
\begin{align*}
&\minimize_{\U,\V,\bLambda,\R} \hspace{5mm}     \sum_{j=1}^p \curl(\X_{.j},\U_{.j})   + \frac{\rho}{2}   \| \D \U -  \V + \bLambda \|_F^2  + \alpha \underbrace{ \bigg(\sum_{j=1}^p  \zeta_j \| \br_j  \|_2 \bigg)}_{ P_2(\R;\bzeta)}\\
& \subto \hspace{5mm}    \U  - \tilde \X  = \R 
\end{align*}

The augmented Lagrangian in scaled form is:
\begin{align*}
L(\U,\V,\R,\bLambda,\N) =  & \sum_{j=1}^p \curl(\X_{.j},\U_{.j})    + \frac{\rho}{2}   \| \D \U -  \V + \bLambda \|_F^2 + \frac{\rho}{2}  \|  ( \U  - \tilde \X) -  \R   + \N \|_F^2 \\
&     +  \alpha \sum_{j=1}^p  \zeta_j \| \br_j \|_2 
\end{align*}
where the dual variable for $\V$ is denoted by $\bLambda$;  the dual variable for $\R$ is denoted by $\N$.

The $\U_{.j}$ sub-problem in the inner nested ADMM is:
\begin{align*}
\U_{.j}^{(k)} =  \argmin_{\U_{.j} } \hspace{2mm}   \curl(\X_{.j},\U_{.j})   & + \frac{\rho}{2}   \| \D \U_{.j} -  \V_{.j}^{(k-1)} + \bLambda_{.j}^{(k-1)} \|_F^2 \\
&+ \frac{\rho}{2}  \|  ( \U_{.j}  - \tilde x_j \cdot \textbf 1_n ) -   \br_j^{(k-1)}   + \N_{.j}^{(k-1)} \|_F^2 
\end{align*}
Now, we are minimizing a sum of a differentiable loss $\ell$ and two quadratic terms which are all smooth. Still the $\U_{.j}$ sub-problem does not have closed-form solution for general convex losses and we need to run an iterative descent algorithm (such as gradient descent, Newton method) to full convergence to solve the problem. Similarly, to reduce computation cost, we take a one-step update by applying linearized ADMM \citep{lin2011linearized}  to the $\U_{.j}$ sub-problem.  The $\U_{.j}$ update in the inner ADMM now becomes:
 \begin{align*}
\U_{.j}^{(k)} =  \U_{.j}^{(k-1)}  - s_k \bigg[    \nabla \curl(\X_{.j},\U_{.j}^{(k-1)}) &+ \rho \D^T ( \D \U_{.j}^{(k-1)} -  \V_{.j}^{(k-1)}  + \bLambda_{.j}^{(k-1)} ) \\
&+ \rho  ( \U_{.j}^{(k-1)}  - \tilde x_j \cdot \textbf 1_n  -   \br_j^{(k-1)}   + \N_{.j}^{(k-1)} ) \bigg] 
\end{align*}


In this case, empirical studies show that taking a one-step Newton update is favored than a one-step gradient descent update as the former enjoys better convergence properties and generally avoids backtracking. However, inverting a hessian is computationally burdensome at each iteration when $n$ is large. Exceptions are for Euclidean distances case where there is a closed-form solution for the $\U_{.j}$ update and for Bernoulli log-likelihood case where the hessian of the loss can be upper bounded by a fixed matrix. In the latter case, we propose to pre-compute the inverse of that fixed matrix instead of inverting a hessian matrix at each iteration. To illustrate this, we write out the $\U_{.j}$ sub-problem of Gecco+ with Bernoulli log-likelihood:
\begin{align*}
\U_{.j} =  \argmin_{\U_{.j} } \hspace{2mm}    \bigg( \sum_{i=1}^n   - x_{ij} u_{ij} + \log(1+e^{u_{ij}})  \bigg)  &+ \frac{\rho}{2}   \| \D \U_{.j} -  \V_{.j} + \bLambda_{.j} \|_F^2 \\
&+ \frac{\rho}{2}  \|  ( \U_{.j}  - \tilde x_j \cdot \textbf 1_n ) -   \br_j   + \N_{.j} \|_F^2 
\end{align*}
The hessian is $\text{diag} \bigg \{ \frac{ e^{u_{ij}}}{(1 + e^{u_{ij}})^2} \bigg\} + \rho  \D^T  \D + \rho \mathbf I $
which can be upper bounded by $\frac{1}{4} \textbf I + \rho  \D^T  \D + \rho \mathbf I$. We propose to replace hessian with this fixed matrix in Newton method and use its inverse  as step size. This is closely related to the approximate hessian literature \citep{krishnapuram2005sparse,simon2013blockwise}. In this way, we just pre-compute this inverse matrix instead of inverting the hessian matrix at each iteration, which dramatically saves computation. We give Algorithm \ref{alg:log-likelihood-full} to solve Gecco+ for Bernoulli log-likelihood with a one-step update to solve the $\U$ sub-problem.  Empirical studies show that this is faster than taking the inner nested proximal gradient approach as we generally don't need to perform the backtracking step.
\begin{algorithm}[t]
	\caption{Full-step Multi-block Algorithm for Gecco+ for Bernoulli log-likelihood $\curl_{k}$:}
	\label{alg:log-likelihood-full}
	\begin{algorithmic}
		\STATE {\bfseries Precompute:}  Difference matrix $\D$, $\M_1 = (\frac{1}{4} \textbf I + \rho \D^T \D + \rho \mathbf I)^{-1}$.
		
	\WHILE{not converged}

	\WHILE{not converged}
		\STATE $\U^{(k)} = \U^{(k-1)}  -  \M_1 \bigg[   \nabla \curl(\X,\U^{(k-1)}) + \rho \D^T (\D \U^{(k-1)} -  \V^{(k-1)} + \bLambda^{(k-1)}  ) + \rho(  \U^{(k-1)} - \tilde \X   -  \R^{(k-1)} + \N^{(k-1)}    )      \bigg]$
		\STATE $\R^{(k)} =  \text{prox}_{\alpha /\rho P_2(\cdot; \bzeta )} (\U^{(k)} - \tilde \X  + \N^{(k-1)} ) $ 
		\STATE 	$\N^{(k)} = \N^{(k-1)} +   ( \U^{(k)} - \tilde \X - \R^{(k)} )   $ 
		\ENDWHILE

		\STATE 		$\V^{(k)} = \text{prox}_{\gamma /\rho P_1(\cdot; \bw )} (\D \U^{(k)} + \bLambda^{(k-1)})$ 
		\STATE $\bLambda^{(k)} =  \bLambda^{(k-1)} +  ( \D \U^{(k)} - \V^{(k)}  ) $ 
				\ENDWHILE
	\end{algorithmic}
\end{algorithm}

Yet, Algorithm \ref{alg:log-likelihood-full} is slow as we need to run iterative inner nested ADMM updates to full convergence. To address this, as mentioned, we can take a one-step update of the inner nested iterative ADMM algorithm. To see this, we can recast the original Gecco+ problem as:
\begin{align*}
&\minimize_{\U^{(k)},\V} \hspace{5mm}   \curl(\X,\U) + \gamma \underbrace{ \bigg(\sum_{l \in \mathcal E} w_l \|\V_{l.}\|_2\bigg)}_{P_1(\V;\bw)}+ \alpha \bigg( \underbrace{ \sum_{j=1}^{p}  \zeta_j \| \br_j \|_2}_{P_2(\R;\bzeta)} \bigg) \\
& \subto \hspace{5mm}     \D \U - \V = 0 , \hspace{5mm} \U  - \tilde \X  = \R
\end{align*}

We apply multi-block ADMM to solve this optimization problem and hence get Algorithm \ref{alg:log-likelihood}. As discussed above, we take a one-step update to solve the $\U$ sub-problem with linearized ADMM and use the inverse of fixed approximate hessian as step size.
\begin{algorithm}[ht]
	\caption{One-step Inexact Multi-block Algorithm for Gecco+ for Bernoulli log-likelihood $\curl_{k}$:}
	\label{alg:log-likelihood}
	\begin{algorithmic}
		\STATE {\bfseries Precompute:}  Difference matrix $\D$, $\M_1 = (\frac{1}{4} \textbf I + \rho \D^T \D + \rho \mathbf I)^{-1}$.
		

	\WHILE{not converged}
		\STATE $\U^{(k)} = \U^{(k-1)}  -  \M_1 \bigg[   \nabla \curl(\X,\U^{(k-1)}) + \rho \D^T (\D \U^{(k-1)} -  \V^{(k-1)} + \bLambda^{(k-1)}  ) + \rho(  \U^{(k-1)} - \tilde \X   -  \R^{(k-1)} + \N^{(k-1)}    )      \bigg]$
		\STATE $\R^{(k)} =  \text{prox}_{\alpha /\rho P_2(\cdot; \bzeta )} (\U^{(k)} - \tilde \X  + \N^{(k-1)} ) $ 
		\STATE 	$\N^{(k)} = \N^{(k-1)} +   ( \U^{(k)} - \tilde \X - \R^{(k)} )   $ 

		\STATE 		$\V^{(k)} = \text{prox}_{\gamma /\rho P_1(\cdot; \bw )} (\D \U^{(k)} + \bLambda^{(k-1)})$ 
		\STATE $\bLambda^{(k)} =  \bLambda^{(k-1)} +  ( \D \U^{(k)} - \V^{(k)}  ) $ 
				\ENDWHILE
	\end{algorithmic}
\end{algorithm}

Similarly, we adopt this approach to solve Gecco+ with Euclidean distances (sparse convex clustering). We first recast the original problem as:
\begin{align*}
&\minimize_{\U^{(k)},\V} \hspace{5mm}   \frac{1}{2} \| \X - \U \|_2^2 + \gamma \underbrace{ \bigg(\sum_{l \in \mathcal E} w_l \|\V_{l.}\|_2\bigg)}_{P_1(\V;\bw)}+ \alpha \bigg( \underbrace{ \sum_{j=1}^{p}  \zeta_j \| \br_j \|_2}_{P_2(\R;\bzeta)} \bigg) \\
& \subto \hspace{5mm}     \D \U - \V = 0 , \hspace{5mm} \U  - \tilde \X  = \R
\end{align*}
Still, we use multi-block ADMM to solve this optimization problem and hence get Algorithm \ref{alg:scc}. Note that $\U$ sub-problem now has closed-form solution. 
\begin{algorithm}[ht]
	\caption{One-step Inexact Multi-block Algorithm for Gecco+ for Euclidean distances $\curl_{k}$:}
	\label{alg:scc}
	\begin{algorithmic}
		\STATE {\bfseries Precompute:}  Difference matrix $\D$, $\M_2 = ( \textbf I + \rho \D^T \D + \rho \mathbf I)^{-1}$.
		

	\WHILE{not converged}
		\STATE $\U^{(k)} =  \M_2 \big[  \X + \rho \D^T (  \V^{(k-1)} - \bLambda^{(k-1)}  ) + \rho(   \tilde \X   +  \R^{(k-1)} - \N^{(k-1)}    )      \big]$
		\STATE $\R^{(k)} =  \text{prox}_{\alpha /\rho P_2(\cdot; \bzeta )} (\U^{(k)} - \tilde \X  + \N^{(k-1)} ) $ 
		\STATE 	$\N^{(k)} = \N^{(k-1)} +   ( \U^{(k)} - \tilde \X - \R^{(k)} )   $ 

		\STATE 		$\V^{(k)} = \text{prox}_{\gamma /\rho P_1(\cdot; \bw )} (\D \U^{(k)} + \bLambda^{(k-1)})$ 
		\STATE $\bLambda^{(k)} =  \bLambda^{(k-1)} +  ( \D \U^{(k)} - \V^{(k)}  ) $ 
				\ENDWHILE
	\end{algorithmic}
\end{algorithm}
Typically, we do not have approximate hessian or closed-form solution to the sub-problem for each loss and we have to use one-step gradient descent with backtracking to solve the $\U$ sub-problem. Empirical study shows that this approach converges slower than Algorithm \ref{alg:diff-backtracking} which uses one-step proximal gradient descent with backtracking.

\section{Gecco+  for Non-Differentiable Losses}\label{nondiffdetail}

In this section, we propose algorithm to solve Gecco+ when the loss $\curl$ is non-differentiable. In this case, we develop a multi-block ADMM algorithm to solve Gecco+ and prove its algorithmic convergence.

\subsection{Gecco+ Algorithm for Non-Differentiable Losses}

Suppose the non-differentiable loss $\curl$ can be expressed as $\curl(\X,\U) =  f(g(\X,\U))$ where  $f$ is convex but non-differentiable and $g$ is affine. This expression is reasonable as it satisfies the affine composition of a convex function. For example, for the least absolute loss, $f(\Z) = \sum_{j=1}^p \|\bz_j \|_1 = \| \text{vec} (\Z) \|_1$ and $g(\X,\U) = \X -\U$. We specify the affine function $g$ as we want to augment the non-differentiable term in the loss function $\curl$.

We can rewrite the problem as:
\begin{align*}
\minimize_{\U} \hspace{5mm}    f(g(\X,\U)) + \gamma \sum_{1\leq i < i' \leq n} w_{ii'} \| \U_{i.} - \U_{i'.} \|_2 + \alpha \sum_{j=1}^p  \zeta_j \| \U_{.j}  - \tilde x_j \cdot \textbf 1_n  \|_2
\end{align*}

We can now recast the problem above  as the equivalent constrained problem:
\begin{align*}
&\minimize_{\U,\V,\Z,\R} \hspace{5mm}    f(\Z) + \gamma \underbrace{ \bigg(\sum_{l \in \mathcal E} w_l \|\V_{l.}\|_2\bigg)}_{P_1(\V;\bw)} + \alpha \underbrace{ \bigg(\sum_{j=1}^p  \zeta_j \| \br_j  \|_2 \bigg)}_{ P_2(\R;\bzeta)}\\
& \subto \hspace{5mm}  g(\X,\U) = \Z  \\
& \hspace{25mm} \D \U - \V = 0 \\
& \hspace{25mm}  \U  - \tilde \X  = \R 
\end{align*}
where $ \tilde \X$ is an $n \times p$ matrix with $j^{th}$ columns equal to scalar $\tilde x_j$.

The augmented Lagrangian in scaled form is:
\begin{align*}
L(\U,\V,\Z,\R,\bLambda,\N,\bKSI) =  & \frac{\rho}{2}   \| \D \U -  \V + \bLambda \|_F^2 + \frac{\rho}{2}  \|  ( \U  - \tilde \X) -  \R   + \N \|_F^2 \\
&  + \frac{\rho}{2}  \| g(\X,\U) - \Z + \bKSI \|_F^2 +  f(\Z)    + \gamma \sum_{l \in \mathcal E} w_l \|\V_{l.} \|_2   +  \alpha \sum_{j=1}^p  \zeta_j \| \br_j \|_2 
\end{align*}
where the dual variable for $\V$ is denoted by $\bLambda$;  the dual variable for $\Z$ is denoted by $\bKSI$;  the dual variable for $\R$ is denoted by $\N$.

Since we assume $g$ to be affine, i.e, $g(\X,\U) = \A \X + \B\U + \C$,  the augmented Lagrangian in scaled form can be written as:

\begin{align*}
L(\U,\V,\Z,\R,\bLambda,\N,\bKSI) &=  \frac{\rho}{2}   \| \D \U -  \V + \bLambda \|_F^2 + \frac{\rho}{2}  \|  ( \U  - \tilde \X) -  \R   + \N \|_F^2 \\
&  + \frac{\rho}{2}  \|  \A \X + \B\U + \C - \Z + \bKSI \|_F^2 +  f(\Z)    + \gamma \sum_{l \in \mathcal E} w_l \|\V_{l.} \|_2   +  \alpha \sum_{j=1}^p  \zeta_j \| \br_j \|_2 
\end{align*}

It can be shown that the $\U$ sub-problem has closed-form solution.

Note that, hinge loss is also non-differentiable and we can write $g(\X,\U) = \textbf 1- \U \circ \X$ where $\textbf 1$ is a matrix of all one and ``$\circ$" is the Hadamard product. Now the $\U$ sub-problem does not have closed-form solution and we will discuss how to solve this problem in the next section.

For distance-based losses, the loss function can always be written as: $\curl(\X,\U) = f( \X - \U)$, which means $g(\X,\U)  =  \X - \U$. Then the augmented Lagrangian in scaled form can be simplified as:
\begin{align*}
L(\U,\V,\Z,\R,\bLambda,\N,\bKSI) =  & \frac{\rho}{2}   \| \D \U -  \V + \bLambda \|_F^2 + \frac{\rho}{2}  \|   ( \U  - \tilde \X) -  \R + \N \|_F^2 \\
&  + \frac{\rho}{2}  \| \X  - \U - \Z + \bKSI \|_F^2 +  f(\Z)    + \gamma \sum_{l \in \mathcal E} w_l \|\V_{l.} \|_2   +  \alpha \sum_{j=1}^p  \zeta_j \| \br_j \|_2 
\end{align*}

Now the $\U$ sub-problem has closed-form solution: 		
\begin{align*}
\U^{(k)} = (\D^T \D + 2 \mathbf I)^{-1}
(\D^T (\V^{(k-1)} - \bLambda^{(k-1)}  ) + \tilde \X +  \R^{(k-1)} - \N^{(k-1)}  + \X - \Z^{(k-1)} + \bKSI^{(k-1)} )
\end{align*}
This gives us Algorithm~\ref{alg:non-diff} to solve Gecco+ for non-differentiable distance-based loss.

\begin{algorithm}[H]
	\caption{Multi-block ADMM for Non-differentiable Distance-based Loss}
	\label{alg:non-diff}
	\begin{algorithmic}
		\STATE {\bfseries Precompute:}  Difference matrix $\D$, $\M = (\D^T \D + 2 \mathbf I)^{-1}$.
		
		\WHILE{not converged}
		\STATE 	$\U^{(k)} =
		\M(\D^T (\V^{(k-1)} - \bLambda^{(k-1)}  ) + \tilde \X   +  \R^{(k-1)} - \N^{(k-1)}  + \X - \Z^{(k-1)} + \bKSI^{(k-1)} )$ 
		\STATE $\Z^{(k)} = \text{prox}_{f/\rho} (\X - \U^{(k)} + \bKSI^{(k-1)} )$ 
		\STATE $\R^{(k)} =  \text{prox}_{\alpha /\rho P_2(\cdot; \bzeta )} (\U^{(k)} - \tilde \X  + \N^{(k-1)} ) $ 
		\STATE $\bKSI^{(k)} = \bKSI^{(k-1)} + (\X - \U^{(k)} - \Z^{(k)})  $
		\STATE 	$\N^{(k)} = \N^{(k-1)} +   ( \U^{(k)} - \tilde \X - \R^{(k)} )   $ 
		
		\STATE 		$\V^{(k)} = \text{prox}_{\gamma /\rho P_1(\cdot; \bw )} (\D \U^{(k)} + \bLambda^{(k-1)})$ 
		\STATE $\bLambda^{(k)} =  \bLambda^{(k-1)} +  ( \D \U^{(k)} - \V^{(k)}  ) $

		\ENDWHILE
		
	\end{algorithmic}
\end{algorithm}

Algorithm~\ref{alg:non-diff} can be used to solve Gecco+ problem with various distances such as Manhattan, Minkowski and Chebychev distances by applying the corresponding proximal operator in the $\Z$ update. For example, for Gecco+ with Manhattan distances, the $\Z$ update is just applying element-wise soft-thresholding operator. For Gecco+ with Chebychev distances, the proximal operator in the $\Z$-update can be computed separately across the rows of its argument and reduces to applying row-wise proximal operator of the infinity-norm.  For Gecco+ with Minkowski distances, we similarly apply row-wise proximal operator of the $\ell_q$-norm.

Next we prove the convergence of Algorithm~\ref{alg:non-diff}.

\subsection{Proof of Convergence for Algorithm~\ref{alg:non-diff}}

\begin{restatable}{theorem}{theoremnondiff}
\label{theorem:non-diff}
If $\curl$ is convex, Algorithm~\ref{alg:non-diff} converges to a global minimum.
\end{restatable}


\noindent \textbf{Proof:}
Note we have provided a sufficient condition for the convergence of four-block ADMM in Lemma \ref{theorem:four_block}. Next we show that the constraint set in our problem satisfies the condition in Lemma \ref{theorem:four_block} and hence the multi-block ADMM Algorithm~\ref{alg:non-diff} converges. Recall our problem is:
\begin{align*}
&\minimize_{\U,\V,\Z,\R} \hspace{5mm}    f(\Z) + \gamma  \sum_{l \in \mathcal E} w_l \|\V_{l.}\|_2 + \alpha \sum_{j=1}^p  \zeta_j \| \br_j  \|_2 \\
& \subto \hspace{5mm} \X - \U = \Z  \\
& \hspace{25mm} \D \U - \V = 0 \\
& \hspace{25mm}    \U  - \tilde \X  = \R
\end{align*}

Note that Lemma \ref{theorem:four_block} is stated in vector form. Hence we transform the constraints above from matrix form to vector form. Note that $\D \U  = \V \Leftrightarrow \U^T \D^T = \V^T \Leftrightarrow  (\D \otimes \textbf I_p) \text{vec} (\U^T ) = \text{vec} (\V^T )$. Hence we can write the constraints as:
\begin{align*}
\begin{pmatrix}
\textbf I \\ \A  \\ \textbf I 
\end{pmatrix} \bu +  \begin{pmatrix}
\textbf I \\ \textbf 0  \\ \textbf 0
\end{pmatrix}  \bz +  \begin{pmatrix}
\textbf 0 \\ \textbf 0  \\  - \textbf I 
\end{pmatrix}  \br  +   \begin{pmatrix}
\textbf 0 \\ - \textbf I  \\ \textbf 0
\end{pmatrix}  \bv = \textbf b  
\end{align*}

where $\bu = \text{vec}(\U^T)$, $\A = \D \otimes \textbf I_p$, $\bz = \text{vec}(\Z^T)$, $\br = \text{vec}(\R^T)$, $\bv = \text{vec} (\V^T$), $\textbf b = \begin{pmatrix}
\text{vec}(\X^T) \\ \textbf 0_{p \times |\mathcal E|}  \\  \tilde \bx \\ \vdots \\ \tilde \bx
\end{pmatrix} $, $\tilde \bx \in \mathbb{R}^p$ is a column vector consisting of all $\tilde x_j$ and is repeated $n$ times in $\textbf b$.

By construction, $\A_2 =  \begin{pmatrix}
\textbf I \\ \textbf 0  \\ \textbf 0
\end{pmatrix}$, $\A_3 = \begin{pmatrix}
\textbf 0 \\ \textbf 0  \\  - \textbf I 
\end{pmatrix} $ and $\A_4 = \begin{pmatrix}
\textbf 0 \\ - \textbf I  \\ \textbf 0
\end{pmatrix} $.

It is easy to verify that:

$\A_2^T \A_3 = \mathbf 0$, $\A_2^T \A_4 = \mathbf 0$, $\A_3^T \A_4 = \mathbf 0$.

Hence our setup satisfies the sufficient condition in Lemma \ref{theorem:four_block} and hence the multi-block ADMM Algorithm~\ref{alg:non-diff} converges. $\square$.

\subsection{Special Case: Gecco+ with Hinge Losses}
As mentioned, we cannot directly apply Algorithm~\ref{alg:non-diff} to solve Gecco+ with hinge losses as the function $g(\X,\U)$ in this case is not the same as the one in distance-based losses. Recall Gecco+ with hinge losses  is:
\begin{align*}
\minimize_{\U} \hspace{5mm}  \sum_{i=1}^n \sum_{j=1}^p \max(0,1-u_{ij} x_{ij}) + \gamma \sum_{1\leq i < i' \leq n} w_{ii'} \| \U_{i.} - \U_{i'.} \|_2 + \alpha  \sum_{j=1}^p \zeta_j   \| \U_{.j}  - \tilde x_j \cdot \textbf 1_n \|_2
\end{align*}

Like before,  we can rewrite the problem as:
\begin{align*}
\minimize_{\U} \hspace{5mm}    f(g(\X,\U)) + \gamma \sum_{1\leq i < i' \leq n} w_{ii'} \| \U_{i.} - \U_{i'.} \|_2 + \alpha \sum_{j=1}^p  \zeta_j \| \U_{.j}  - \tilde x_j \cdot \textbf 1_n  \|_2
\end{align*}
We can now recast the  problem above as the equivalent constrained problem:
\begin{align*}
&\minimize_{\U,\V,\Z,\R} \hspace{5mm}    f(\Z) + \gamma \underbrace{ \bigg(\sum_{l \in \mathcal E} w_l \|\V_{l.}\|_2\bigg)}_{P_1(\V;\bw)} + \alpha \underbrace{ \bigg(\sum_{j=1}^p  \zeta_j \| \br_j  \|_2 \bigg)}_{ P_2(\R;\bzeta)}\\
& \subto \hspace{5mm}  \textbf 1- \U \circ \X = \Z  \\
& \hspace{25mm} \D \U - \V = 0 \\
& \hspace{25mm}  \U  - \tilde \X  = \R 
\end{align*}
Here, $f(\Z) = \max(0,\Z)$. With a slight abuse of notation, we refer $f$ to applying element-wise maximum to all entries in the matrix. We set $g(\X,\U) = \textbf 1- \U \circ \X$ where $\textbf 1$ is a matrix of all one and ``$\circ$" is the Hadamard product. $ \tilde \X$ is an $n \times p$ matrix with $j^{th}$ columns equal to scalar $\tilde x_j$.

The augmented Lagrangian in scaled form is:
\begin{align*}
L(\U,\V,\Z,\R,\bLambda,\N,\bKSI) =  & \frac{\rho}{2}   \| \D \U -  \V + \bLambda \|_F^2 + \frac{\rho}{2}  \|  ( \U  - \tilde \X) -  \R   + \N \|_F^2 \\
&  + \frac{\rho}{2}  \| \textbf 1- \U \circ \X  - \Z + \bKSI \|_F^2 +  f(\Z)    + \gamma \sum_{l \in \mathcal E} w_l \|\V_{l.} \|_2   +  \alpha \sum_{j=1}^p  \zeta_j \| \br_j \|_2 
\end{align*}

The $\U$ sub-problem now becomes:
\begin{align*}
\U^{(k+1)} = \argmin_{\U} \hspace{2mm} & \| \D \U -  \V^{(k)} + \bLambda^{(k)} \|_F^2  +   \|  \U  - \tilde \X -  \R^{(k)}   + \N^{(k)} \|_F^2 \\
& +  \| \textbf 1- \U \circ \X  - \Z^{(k)} + \bKSI^{(k)} \|_F^2 
\end{align*}

The first-order optimality condition is:
\begin{align*}
\D^T  ( \D \U -  \V^{(k)} + \bLambda^{(k)} ) +    \U  - \tilde \X -  \R^{(k)}   + \N^{(k)}  +  \X \circ  (  \U \circ \X  + \Z^{(k)} - \textbf 1  -  \bKSI^{(k)} ) = \textbf 0 
\end{align*}
which can be written as:
\begin{align*}
&   (\X \circ \X) \circ \U + \X \circ (\Z^{(k)} - \textbf 1  -  \bKSI^{(k)}  )      + \D^T \D \U  - \D^T (\V^{(k)} -  \bLambda^{(k)} ) +    \U  - \tilde \X -  \R^{(k)}   + \N^{(k)} = \textbf 0 \\
&  (\X \circ \X) \circ \U  + (\D^T \D + \textbf I ) \U = \D^T (\V^{(k)} -  \bLambda^{(k)} ) + \tilde \X +  \R^{(k)}   - \N^{(k)} + \X \circ ( \textbf 1  +  \bKSI^{(k)} - \Z^{(k)} )   
\end{align*}
To solve $\U$ from the above equation, one way is to first find the SVD of the leading coefficient:
\begin{align*}
    \X \circ \X = \sum_k \sigma_k \tilde \bu_k \tilde \bv_k = \sum_k \tilde \bw_k \tilde \bv_k
\end{align*}
From this decomposition we create two sets of diagonal matrices:
\begin{align*}
    \tilde \W_k &= \text{Diag}(\tilde \bw_k)  \\
    \tilde \V_k &= \text{Diag}(\tilde \bv_k)
\end{align*}
The Hadamard product can now be replaced by a sum
\begin{align*}
    \sum_k \tilde \W_k \U \tilde \V_k + (\D^T \D + \textbf I) \U = \C
\end{align*}
where $\C = \D^T (\V^{(k)} -  \bLambda^{(k)} ) + \tilde \X +  \R^{(k)}   - \N^{(k)} + \X \circ ( \textbf 1  +  \bKSI^{(k)} - \Z^{(k)} ) $.

Now we can solve this equation via vectorization:
\begin{align*}
    \text{vec}(\C) &= \big(  \textbf I \otimes (\D^T \D + \textbf I) + \sum_k \tilde \V_k \otimes \tilde \W_k \big) \text{vec}(\U) \\
     \text{vec}(\U) &= \big(  \textbf I \otimes (\D^T \D + \textbf I) + \sum_k \tilde \V_k \otimes \tilde \W_k \big)^{+} \text{vec}(\C) \\
     \U &= \text{Mat} \bigg( \big(  \textbf I \otimes (\D^T \D + \textbf I) + \sum_k \tilde \V_k \otimes \tilde \W_k \big)^{+} \text{vec}(\C) \bigg)
\end{align*}

where $\B^+$ denotes the pseudo-inverse of $\B$, and \text{Mat}() is the inverse of the \text{vec}() operation.

Here we have to compute the pseudo-inverse of a matrix which is computationally expensive in practice. To avoid this, we adopt Generalized ADMM approach proposed by \cite{deng2016global} where the $\U$ sub-problem is augmented by a positive semi-definite quadratic operator. In our case, our modified $\U$ sub-problem becomes:
\begin{align*}
\argmin_{\U} \hspace{2mm} \| \D \U -  \V^{(k)} + \bLambda^{(k)} \|_F^2 & +   \|  ( \U  - \tilde \X) -  \R^{(k)}   + \N^{(k)} \|_F^2 +  \| \textbf 1- \U \circ \X  - \Z^{(k)} + \bKSI^{(k)} \|_F^2 \\
& + \| \big (\textbf 1 -  \X \circ \X  \big)  \circ (\U - \U^{(k)}) \|_F^2 
\end{align*}
The first-order optimality condition now becomes:
\begin{align*}
& \D^T  ( \D \U -  \V^{(k)} + \bLambda^{(k)} ) +    \U  - \tilde \X -  \R^{(k)}   + \N^{(k)}  +  \X \circ  (  \U \circ \X  + \Z^{(k)} - \textbf 1  -  \bKSI^{(k)} ) \\
&+  \big (\textbf 1 -  \X \circ \X  \big)  \circ (\U - \U^{(k)})  = \textbf 0 
\end{align*}
We have:
\begin{align*}
\mathbf H  \U = \D^T  (   \V^{(k)} - \bLambda^{(k)} ) +  \tilde \X +  \R^{(k)}   - \N^{(k)} - \X \circ  (   \Z^{(k)} - \textbf 1  -  \bKSI^{(k)} )  + (\textbf 1 -  \X \circ \X ) \circ \U^{(k)}
\end{align*}
where $\mathbf H = ( \D^T  \D  + \textbf I + \textbf 1 )$. Hence we have analytical update:
\begin{align*}
 \U^{(k+1)} = \mathbf H^{-1} \big( \D^T  (   \V^{(k)} - \bLambda^{(k)} ) +  \tilde \X +  \R^{(k)}   - \N^{(k)} - \X \circ  (   \Z^{(k)} - \textbf 1  -  \bKSI^{(k)} )  + (\textbf 1 -  \X \circ \X ) \circ \U^{(k)} \big)
\end{align*}
It is easy to see that the $\V$, $\Z$ and $\R$ updates all have closed-form solutions.

\section{Multinomial Gecco+}\label{multin}
In this section, we briefly demonstrate how Gecco+ with multinomial losses is formulated, which is slightly different from the original Gecco+ problems. Suppose we observe categorical data as follows ($K=3$):
\begin{align*}
\X_{n \times p} = \begin{pmatrix}
1 & 1 & 1 \\
2 & 2 & 2 \\
3 & 3 & 3 
\end{pmatrix}
\end{align*}
We can get the indicator matrix $\X^{(k)}$ for each class $k$ as:
\begin{align*}
\X^{(1)} = \begin{pmatrix}
1 & 1 & 1 \\
0 & 0 & 0 \\
0 & 0 & 0 
\end{pmatrix}
\hspace{5mm} \X^{(2)} = \begin{pmatrix}
0 & 0 & 0 \\
1 & 1 & 1 \\
0 & 0 & 0 
\end{pmatrix}
\hspace{5mm}  \X^{(3)} = \begin{pmatrix}
0 & 0 & 0 \\
0 & 0 & 0 \\
1 & 1 & 1 
\end{pmatrix}
\end{align*}

Then we concatenate $\X^{(1)},\X^{(2)},\X^{(3)}$ and get $\hat \X_{n \times (p*K)} = \begin{pmatrix} \X^{(1)} &  \X^{(2)} & \X^{(3)} 
\end{pmatrix} $. This is equivalent to the dummy coding of the categorical matrix $\tilde \X$ after some row/column shuffle.
\begin{align*}
\tilde \X =  \begin{pmatrix}
1 & 0 & 0 & 1 & 0 & 0 & 1 & 0 & 0 \\
0 & 1 & 0 & 0 & 1 & 0 & 0 & 1 & 0 \\
0 & 0 & 1 & 0 & 0 & 1 & 0 & 0 & 1  
\end{pmatrix}_{n \times (p*K)}
\end{align*}
It is obvious that measuring the difference of two observations by comparing rows of $\hat \X$ is better than simply comparing the Euclidean distances of rows of original data matrix $\X$. Also parameterizing in $\hat \X$ is beneficial for computing the multinomial log-likelihood or deviance. Hence we concatenate all columns of $\X^{(k)}$ as input data. Similarly, we concatenate all columns of the corresponding $\U^{(k)}$ and then fuse $\hat \U$ in row-wise way.

\subsection{Gecco with Multinomial Log-likelihood} 
Gecco with multinomial log-likelihood can be formulated as:
\begin{align*}
\minimize_{\U} \hspace{5mm} \sum_i \sum_j  \bigg\{   \sum_{k=1}^{K}    -x_{ijk} u_{ijk} + \log(\sum_{k=1}^K e^{u_{ijk}}) \bigg \} + \gamma \sum_{i < i'} w_{ii'} \bigg \|  \begin{bmatrix} \U_{i.}^{(1)} \\ \vdots \\ \U_{i.}^{(K)} \end{bmatrix} -  \begin{bmatrix} \U_{i'.}^{(1)} \\ \vdots \\ \U_{i'.}^{(K)}  \end{bmatrix}   \bigg \|_2
\end{align*}
where $x_{ijk}$ refers to the elements of indicator matrix $\X_{ij}^{(k)}$ discussed previously and $\U_{i.}^{(k)} = \begin{bmatrix} u_{i1k} \\ u_{i2k}  \\ \vdots \\ u_{ipk} \end{bmatrix}$.

\subsection{Gecco+ with Multinomial Log-likelihood}
Gecco+ with multinomial log-likelihood can be formulated as:
\begin{align*}
\minimize_{\U} \hspace{5mm} \sum_i \sum_j  &\bigg\{   \sum_{k=1}^{K}    -x_{ijk} u_{ijk} + \log(\sum_{k=1}^K e^{u_{ijk}}) \bigg \} + \gamma \sum_{i < i'} w_{ii'} \bigg \|  \begin{bmatrix} \U_{i.}^{(1)} \\ \vdots \\ \U_{i.}^{(K)} \end{bmatrix} -  \begin{bmatrix} \U_{i'.}^{(1)} \\ \vdots \\ \U_{i'.}^{(K)}  \end{bmatrix}   \bigg \|_2 \\
&+ \alpha \sum_{j=1}^p \sum_{k=1}^K   \| \U_{.j}^{(k)}   - \tilde x_j^{(k)} \cdot \textbf 1_n  \|_2
\end{align*}
where $\U_{i.}^{(k)} = \begin{bmatrix} u_{i1k} \\ u_{i2k}  \\ \vdots \\ u_{ipk} \end{bmatrix}$, $\U_{.j}^{(k)} = \begin{bmatrix} u_{1jk} \\ u_{2jk}  \\ \vdots \\ u_{njk} \end{bmatrix}$ and $\tilde x_j^{(k)}$ is the loss-specific center for $j$ variable in $k^{th}$ class.

\section{Loss-specific Center Calculation}\label{centroidcal}

In this section, we show how to calculate the loss-specific center in Table~\ref{loss-table}.

\subsection{Continuous Data}
For continuous data, we consider Gecco with Euclidean distances.
\subsubsection{Euclidean Distance}
\begin{align*}
\minimize_{\U} \hspace{5mm} \frac{1}{2} \sum_{i=1}^n \|\X_{i.} - \U_{i.}\|_2^2 + \gamma \sum_{i < i'} w_{ii'} \|\U_{i.} - \U_{i'.} \|_2
\end{align*}
When total fusion, $\U_{i.}= \U_{i'.}$, $\forall i \neq i'$. Let $\U_{i.} = \U_{i'.} = \bu$. The problem above becomes:
\begin{align*}
\minimize_{\bu} \hspace{5mm} \frac{1}{2} \sum_{i=1}^n \|\X_{i.} - \bu\|_2^2 
\end{align*}
Taking derivative, we get:
\begin{align*}
 \sum_{i=1}^n (\X_{i.} - \bu) = 0   \Rightarrow \bu = \bar \bx
\end{align*}

\subsection{Count Data}
For count-valued data, we consider Gecco with Poisson log-likelihood/deviance, negative binomial log-likelihood/deviance and Manhattan distances.
\subsubsection{Poisson Log-likelihood}

\begin{align*}
\minimize_{\U} \hspace{5mm} \sum_i^n \sum_j^p -x_{ij} u_{ij} + \exp(u_{ij}) + \gamma \sum_{i < i'} w_{ii'} \|\U_{i.} - \U_{i'.} \|_2
\end{align*}
When total fusion, $\bu_{ij} = \bu_{i'j}$, $\forall i \neq i'
$. Let $\bu$ be the fusion vector and $\bu = (u_1,\cdots,u_p)$. The problem above becomes:
\begin{align*}
\minimize_{\U} \hspace{5mm} \sum_i^n \sum_j^p -x_{ij} u_{j} + \exp(u_{j})
\end{align*}
Taking derivative, we get:
\begin{align*}
\sum_{i=1}^n -x_{ij} + n \exp(u_{j}) = 0   \Rightarrow \exp(u_{j}) = \bar x_j \Rightarrow u_{j} = \log(\bar x_j) \Rightarrow \bu = \log(\bar \bx)
\end{align*}

\subsubsection{Poisson Deviance}
\begin{align*}
\minimize_{\U} \hspace{5mm}  \sum_i^n \sum_j^p -x_{ij} \log u_{ij} + u_{ij}  + \gamma \sum_{i < i'} w_{ii'} \|\U_{i.} - \U_{i'.}\|_2
\end{align*}

Let $\bu$ be the fusion vector and $\bu = (u_1,\cdots,u_p)$. The problem above becomes:
\begin{align*}
\minimize_{\U} \hspace{5mm} \sum_i^n \sum_j^p -x_{ij} \log u_{j} + u_{j}
\end{align*}
Taking derivative, we get:
\begin{align*}
\sum_{i=1}^n \frac{-x_{ij}}{u_j} + n  = 0   \Rightarrow u_j = \bar x_j \Rightarrow \bu = \bar \bx
\end{align*}

\subsubsection{Negative Binomial Log-likelihood}

\begin{align*}
\minimize_{\U} \hspace{5mm} \sum_i^n \sum_j^p  - x_{ij} u_{ij} + (x_{ij} + \frac{1}{\alpha} ) \log (\frac{1}{\alpha} + e^{u_{ij}} ) + \gamma \sum_{i < i'} w_{ii'} \|\U_{i.} - \U_{i'.}\|_2
\end{align*}
When total fusion, $\bu_{ij} = \bu_{i'j}$, $\forall i \neq i
$. Let $\bu$ be the fusion vector and $\bu = (u_1,\cdots,u_p)$. The problem above becomes:
\begin{align*}
\minimize_{\U} \hspace{5mm} \sum_i^n \sum_j^p  - x_{ij} u_{j} + (x_{ij} + \frac{1}{\alpha} ) \log (\frac{1}{\alpha} + e^{u_{j}} ) 
\end{align*}
Taking derivative, we get:
\begin{align*}
\sum_{i=1}^n -x_{ij} + (x_{ij} + \frac{1}{\alpha} ) \frac{e^{u_{j}}}{\frac{1}{\alpha} + e^{u_{j}}  } &= 0\\
\sum_{i=1}^n  x_{ij} &= \sum_{i=1}^n  (x_{ij} + \frac{1}{\alpha} ) \frac{e^{u_{j}}}{\frac{1}{\alpha} + e^{u_{j}}  } \\
\sum_{i=1}^n  x_{ij}  \cdot \frac{1}{\alpha}  + \sum_{i=1}^n  x_{ij}  \cdot   e^{u_{j}}  & =  \sum_{i=1}^n  x_{ij} \cdot e^{u_{j}}   + \frac{n}{\alpha} e^{u_{j}}  \\
e^{u_{j}} &= \sum_{i=1}^n  x_{ij}  /n \\
\exp(u_{j}) &= \bar x_j \Rightarrow u_{j} = \log(\bar x_j) \Rightarrow \bu = \log(\bar \bx)
\end{align*}

\subsubsection{Negative Binomial Deviance}
\begin{align*}
\minimize_{\U} \hspace{5mm}  \sum_i^n \sum_j^p  x_{ij} \log(\frac{x_{ij}}{u_{ij}})  - (x_{ij} + \frac{1}{\alpha}) \log (\frac{1+\alpha x_{ij}}{1+\alpha u_{ij} }    )  + \gamma \sum_{i < i'} w_{ii'} \|\U_{i.} - \U_{i'.} \|_2
\end{align*}
The formulation above is equivalent to 
\begin{align*}
\minimize_{\U} \hspace{5mm}  \sum_i^n \sum_j^p  - x_{ij} \log u_{ij}  + (x_{ij} + \frac{1}{\alpha}) \log (1+\alpha u_{ij}    )  + \gamma \sum_{i < i'} w_{ii'} \|\U_{i.} - \U_{i'.}\|_2
\end{align*}

Let $\bu$ be the fusion vector and $\bu = (u_1,\cdots,u_p)$. The problem above becomes:
\begin{align*}
\minimize_{\U} \hspace{5mm} \sum_i^n \sum_j^p  - x_{ij} \log u_{j}  + (x_{ij} + \frac{1}{\alpha}) \log (1+\alpha u_{j}    )
\end{align*}
Taking derivative, we get:
\begin{align*}
&\sum_{i=1}^n \frac{-x_{ij}}{u_j} + \frac{ x_{ij} + \frac{1}{\alpha}}{1+\alpha u_{j} } \cdot \alpha  = 0    \\
& \sum_{i=1}^n  x_{ij} + \alpha u_{j} \sum_{i=1}^n  x_{ij} = n  u_{j} + \alpha  u_{j}  \sum_{i=1}^n  x_{ij}  \\
& u_j = \bar x_j \Rightarrow \bu = \bar \bx
\end{align*}

\subsubsection{Manhattan Distance}
\begin{align*}
\minimize_{\U} \hspace{5mm} \sum_{i=1}^n \|\bx_i - \bu_i\|_1 + \gamma \sum_{i < i'} w_{ii'} \|\U_{i.} - \U_{i'.}\|_2
\end{align*}
When total fusion, $\U_{i.} = \U_{i'.}$, $\forall i \neq i'$. Let $\U_{i.} = \U_{i'.} = \bu$. We have:
\begin{align*}
\minimize_{\bu} \hspace{5mm}  \sum_{i=1}^n \|\bx_i - \bu\|_1
\end{align*}
For each $j$, we have:
\begin{align*}
\minimize_{u_j} \hspace{5mm}   \sum_{i=1}^n \|x_{ij} - u_j\|_1
\end{align*}
We know that $u_j$ is just the median of $x_{ij}$ for each $j$.

\subsection{Binary Data}
For binary data, we consider Gecco with Bernoulli log-likelihood, binomial deviance and hinge loss.
\subsubsection{Bernoulli Log-likelihood}
\begin{align*}
\minimize_{\U} \hspace{5mm} \sum_i^n \sum_j^p  - x_{ij} u_{ij} + \log(1+e^{u_{ij}}) + \gamma \sum_{i < i'} w_{ii'} \|\U_{i.} - \U_{i'.}\|_2
\end{align*}
When total fusion, $\bu_{ij} = \bu_{i'j}$, $\forall i \neq i'
$. Let $\bu$ be the fusion vector and $\bu = (u_1,\cdots,u_p)$. The problem above becomes:
\begin{align*}
\minimize_{\U} \hspace{5mm} \sum_i^n \sum_j^p -x_{ij} u_{j} + \log(1+e^{u_{j}})
\end{align*}
Taking derivative, we get:
\begin{align*}
\sum_{i=1}^n -x_{ij} + n \frac{\exp(u_{j})}{1+\exp(u_{j})} = 0   &\Rightarrow \frac{\exp(u_{j})}{1+\exp(u_{j})}  = \bar x_j \\ &\Rightarrow \text{logit}^{-1} (u_{j}) = \bar x_j  \Rightarrow  u_{j} = \text{logit} (\bar x_j) \Rightarrow  \bu = \text{logit} (\bx)
\end{align*}

\subsubsection{Binomial Deviance}
\begin{align*}
\minimize_{\U} \hspace{5mm} \sum_i^n \sum_j^p - x_{ij} \log{u_{ij}} - (1-x_{ij})\log({1-u_{ij}})  + \gamma \sum_{i < i'} w_{ii'} \|\U_{i.} - \U_{i'.}\|_2
\end{align*}

Let $\bu$ be the fusion vector and $\bu = (u_1,\cdots,u_p)$. The problem above becomes:
\begin{align*}
\minimize_{\U} \hspace{5mm} \sum_i^n \sum_j^p - x_{ij} \log{u_{j}} - (1-x_{ij})\log({1-u_{j}})
\end{align*}
Taking derivative, we get:
\begin{align*}
\sum_{i=1}^n \frac{-x_{ij}}{u_j} + \frac{1-x_{ij}}{1-u_{j}}  = 0   &\Rightarrow  \sum_i^n - x_{ij} (1-u_j)  + (1-x_{ij})   u_j = 0 \\ &\Rightarrow \sum_i^n - x_{ij} + u_j = 0 \Rightarrow u_j = \bar x_j \Rightarrow \bu = \bar\bx
\end{align*}

\subsubsection{Hinge Loss}
\begin{align*}
\minimize_{\U} \hspace{5mm} \sum_{i=1}^n \sum_{j=1}^p \max(0,1-u_{ij} x_{ij})  + \gamma \sum_{i < i'} w_{ii'} \|\U_{i.} - \U_{i'.} \|_2
\end{align*}

Let $\bu$ be the fusion vector and $\bu = (u_1,\cdots,u_p)$. The problem above becomes:
\begin{align*}
\minimize_{\U} \hspace{5mm} \sum_{i=1}^n \sum_{j=1}^p \max(0,1-u_{j} x_{ij}) 
\end{align*}
For each feature $j$, the problem becomes:
\begin{align*}
\minimize_{u_j} \hspace{5mm} \sum_{i=1}^n  \max(0,1-u_{j} x_{ij}) 
\end{align*}
Note in hinge loss, $x_{ij} \in \{-1,1\}$. Suppose we have $n_1$ observations for class ``1" and $n_2$ observations for class ``-1". The problem now becomes
\begin{align*}
\minimize_{u_j} \hspace{5mm}   n_1 \max(0,1-u_{j} ) + n_2 \max(0,1+u_{j} )   
\end{align*}
Define $h(t) =  n_1 \max(0,1-t ) + n_2 \max(0,1+t )$. We have:
\begin{align*}
    h(t) = \begin{cases}
    n_1 (1-t)  & \text{if} \hspace{2mm} t \leq -1  \\
    n_1 (1-t)  + n_2 (1+t)  & \text{if} \hspace{2mm}  -1 < t < 1  \\
    n_2 (1+t) & \text{if} \hspace{2mm}  t \geq 1
    \end{cases}
\end{align*}
Clearly, if $n_2 > n_1$ (more ``-1"), $h(t)$ is minimized by $t=-1$; if $n_1 > n_2$ (more ``1"), $h(t)$ is minimized by $t=1$; if $n_1 = n_2$, $h(t)$ is minimized by any $t$ between $[-1,1]$. Therefore, $u_j$ should be the mode of all observations for feature $j$:
\begin{align*}
u_j = \text{mode}_i (x_{ij})
\end{align*}

\subsection{Categorical Data}
For categorical data, we consider Gecco with multinomial log-likelihood and deviance.

\subsubsection{Multinomial Log-likelihood}

\begin{align*}
\minimize_{\U} \hspace{5mm} \sum_i^n \sum_j^p  \big\{   \sum_{k=1}^{K}    -x_{ijk} u_{ijk} + \log(\sum_{k=1}^K e^{u_{ijk}}) \big \} + \gamma \sum_{i < i'} w_{ii'} \bigg \|  \begin{bmatrix} \U_{i.}^{(1)} \\ \vdots \\ \U_{i.}^{(K)} \end{bmatrix} -  \begin{bmatrix} \U_{i'.}^{(1)} \\ \vdots \\ \U_{i'.}^{(K)}  \end{bmatrix}   \bigg \|_2
\end{align*}
When total fusion, $u_{ijk} = u_{i'jk}$, $\forall i \neq i'$. Let $\bu$ be the fusion vector and $\bu_k = (u_{1k},\cdots,u_{pk})$. The problem above becomes:
\begin{align*}
\minimize_{\U} \hspace{5mm} \sum_i^n \sum_j^p \big\{ \sum_{k}^K -x_{ijk} u_{jk} + \log( \sum_{k}^K  e^{u_{jk}}) \big\}
\end{align*}

Taking derivative with respect to $u_{jk}$, we get:
\begin{align*}
\sum_{i=1}^n -x_{ijk} + n \frac{\exp(u_{jk})}{ \sum_k^K \exp(u_{jk})} &= 0 \\  
u_{jk} =  \log \frac{ \bar x_{.jk}} {\sum_k \bar x_{.jk} } & =  \text{mlogit}( \bar x_{.jk} )
\end{align*}

\subsubsection{Multinomial Deviance}
\begin{align*}
&\minimize_{\U} \hspace{5mm} \sum_i^n \sum_j^p \big\{   \sum_{k=1}^{K}    -x_{ijk} \log(u_{ijk})  \big \}  + \gamma \sum_{i < i'} w_{ii'} \bigg \|  \begin{bmatrix} \U_{i.}^{(1)} \\ \vdots \\ \U_{i.}^{(K)} \end{bmatrix} -  \begin{bmatrix} \U_{i'.}^{(1)} \\ \vdots \\ \U_{i'.}^{(K)}  \end{bmatrix}   \bigg \|_2  \\
& \text{subject to} \hspace{5mm} \sum \limits_{k=1}^K u_{ijk} = 1
\end{align*}

When total fusion, $\bu_{ijk} = \bu_{i'jk}$, $\forall i \neq i
$. Let $\bu$ be the fusion vector and $\bu_k = (u_{1k},\cdots,u_{pk})$. The problem above becomes:
\begin{align*}
& \minimize_{\U} \hspace{5mm} \sum_i^n \sum_j^p   \sum_{k}^K -x_{ijk} \log u_{jk}  \\
& \text{subject to} \hspace{5mm} \sum \limits_{k=1}^K u_{jk} = 1
\end{align*}

We can write the constraint in Lagrangian form:
\begin{align*}
& \minimize_{\U} \hspace{5mm} \sum_i^n \sum_j^p   \sum_{k}^K -x_{ijk} \log u_{jk}  +  \lambda( \sum \limits_{k=1}^K u_{jk} - 1)
\end{align*}

Taking derivative with respect to $u_{jk}$, we get:
\begin{align*}
\sum_{i=1}^n \frac{-x_{ijk}}{u_{jk}} + \lambda  &= 0 \\  
\sum_{i=1}^n x_{ijk} = \lambda u_{jk} 
\end{align*}
We have:
\begin{align*}
n = \sum_{i=1}^n \sum_{k=1}^K   x_{ijk} = \sum_{k=1}^K  \sum_{i=1}^n x_{ijk} = \sum_{k=1}^K  \lambda u_{jk}  = \lambda
\end{align*}
Therefore,
\begin{align*}
u_{jk} = \sum_{i=1}^n x_{ijk} / n 
\end{align*}

\section{Visualization of Gecco+ for Authors Data}\label{authorvizone}
\begin{figure}[H]
\centering
\includegraphics[scale = 0.6]{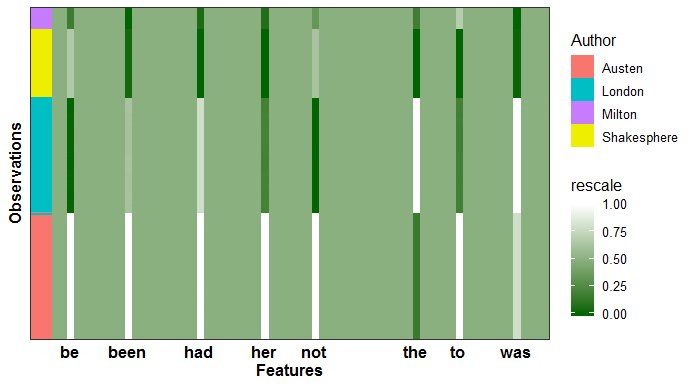}
\caption{Cluster heatmap of Gecco+ solution $\hat \U({\gamma,\alpha})$  for authors data set  with $\alpha=15$ and $\gamma = 10^5$. The left bar refers to the true author label. We highlight selected features at the bottom. Gecco+ selects informative features that separate groups.}
\label{viz-selected-feature}
\end{figure}

Figure \ref{viz-selected-feature} illustrates selected features and cluster assignment for authors data set with one combination of $\alpha$ and $\gamma$. We select meaningful features and achieve satisfactory clustering results. We have already discussed the results and interpretation in detail in Section \ref{author}.

\section{Multi-omics Data}\label{genohist}
In this section, we show the distribution of data from different platforms in Section~\ref{omics}. We see that both gene expression data and protein data appear gaussian; Methylation data is between $[0,1]$; miRNA data is highly-skewed.

\begin{figure}[H]
    \centering
    \includegraphics[width=12cm]{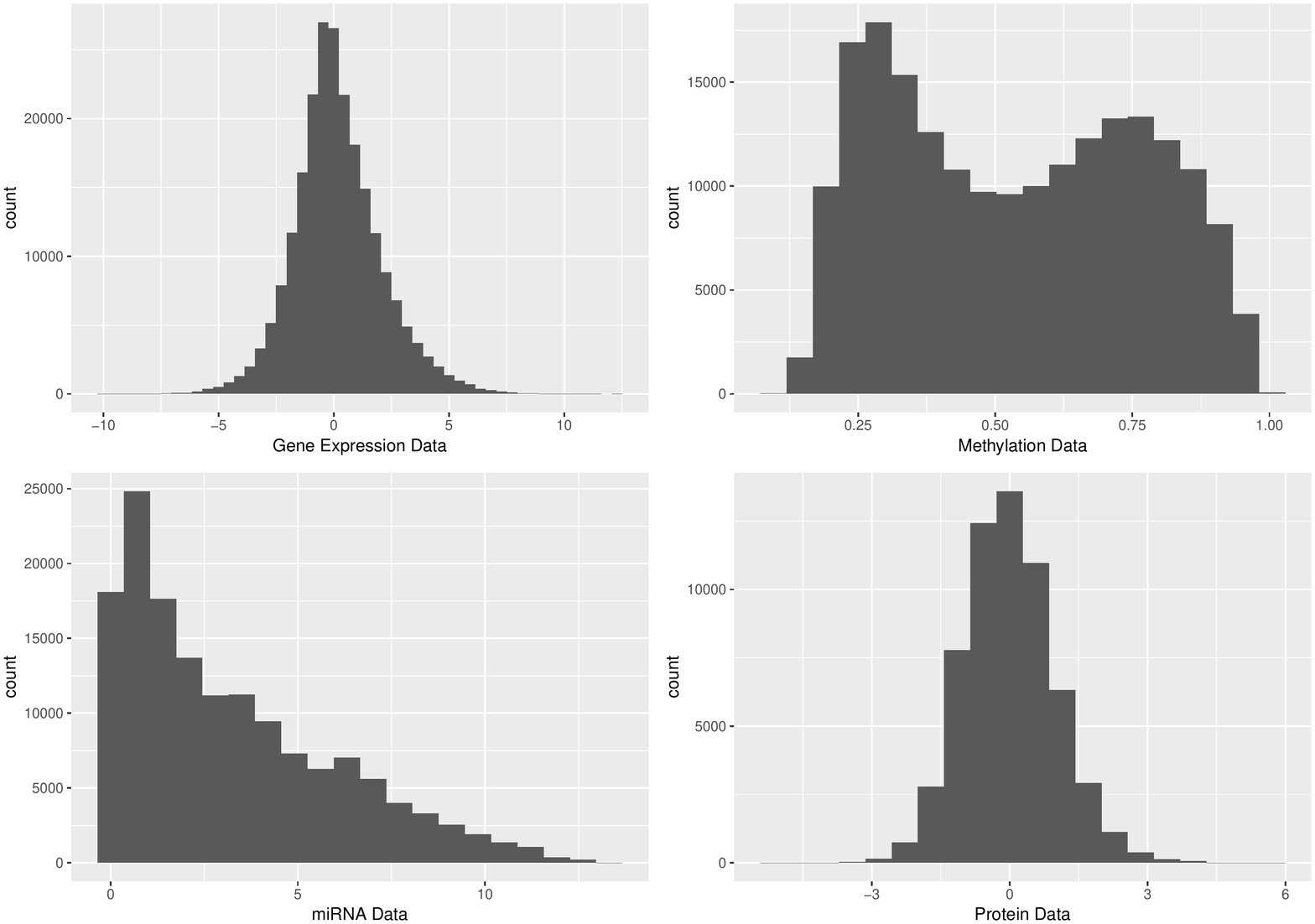}
    \caption{Histograms of data from different platforms for multi-omics TCGA data set. Both gene expression data and protein data appear gaussian; Methylation data is proportion-valued; miRNA data is highly-skewed.}
	\label{fig:hist}
\end{figure}

\bibliographystyle{abbrvnat}
\bibliography{main.bib}

\end{document}